\documentclass[12pt,preprint]{aastex}

\newcommand{\grav}{log($g$)}
\newcommand{\nstars}{122}
\newcommand{\teff}{$T_{\mathrm{\scriptsize eff}}$}
\newcommand{\kms}{km~s$^{-1}$}
\newcommand{\eqw}{$W_{\lambda}$}

\begin{document}

\title{Normal and Outlying Populations of the
  Milky Way Stellar Halo at [Fe/H] $< -2$ \altaffilmark{1}}

\author{Judith G. Cohen\altaffilmark{2}, 
Norbert Christlieb\altaffilmark{3},
Ian Thompson\altaffilmark{4},
Andrew McWilliam\altaffilmark{4}, Stephen Shectman\altaffilmark{4},
Dieter Reimers\altaffilmark{5}, Lutz Wisotzki\altaffilmark{6}
\& Evan Kirby\altaffilmark{7,8}
}

\altaffiltext{1}{Based in part on observations obtained in part at the
W.M. Keck Observatory, which is operated jointly by the California 
Institute of Technology, the University of California, and the
National Aeronautics and Space Administration.}

\altaffiltext{2}{Palomar Observatory, Mail Stop 249-17,
California Institute of Technology, Pasadena, Ca., 91125, 
jlc@astro.caltech.edu}

\altaffiltext{3}{Zentrum f{\"u}r Astronomie der Universit{\"a}t Heidelberg,
Landessternwarte, K{\"o}nigstuhl 12, 69117, Heidelberg,
Germany (N.Christlieb@lsw.uni-heidelberg.de)}

\altaffiltext{4}{Carnegie Observatories,
813 Santa Barbara Street, Pasadena, Ca. 91101,
ian(shec,andy)@obs.carnegiescience.edu}

\altaffiltext{5}
{Hamburger Sternwarte, Gojenbergsweg 112, 21029 Hamburg, Germany,
dreimers@hs.uni-hamburg.de}

\altaffiltext{6}{Leibniz-Institut f\"ur Astrophysik Potsdam, 
An der Sternwarte 16, 14482 Potsdam, Germany, lutz@aip.de}

\altaffiltext{7}{University of California, Department of Physics
and Astronomy, 4129 Frederick Reines Hall, Irvine, Ca. 92697,
ekirby@uci.edu}

\altaffiltext{8}{Southern California Center for Galaxy Evolution Fellow}

\begin{abstract}

From detailed abundance analysis of $>$100 Hamburg/ESO candidate extremely metal-poor (EMP)
stars we find 45 with [Fe/H] $<$ $-3.0$~dex.  
We identify a heretofore unidentified group: Ca-deficient stars, with sub-solar [Ca/Fe]
ratios and the lowest neutron-capture abundances; the Ca-deficient group comprises
$\sim$10\% of the sample, excluding Carbon stars.
Our radial velocity distribution shows that the carbon-enhanced stars with no
$s$-process enhancements, CEMP-no, and which do not
show C$_2$ bands are not preferentially binary systems. Ignoring Carbon stars,
approximately 15\% of our sample are strong (${\geq}5\sigma$) outliers in one, or more, elements
between Mg and Ni; this rises
to $\sim$19\% if very strong (${\geq}10\sigma$) outliers for Sr and Ba are included.  
Examples include:  HE0305$-$0554 with the lowest [Ba/H]
known; HE1012$-$1540 and HE2323$-$0256, two (non-velocity variable) C-rich stars with very strong
[Mg,Al/Fe] enhancements; and HE1226$-$1149 an extremely $r$-process rich star.

\end{abstract}


\section{Introduction \label{section_intro} }

Extremely metal poor stars were presumably among the first
stars formed in the Galaxy, and hence represent, in effect,
a local high-redshift population. It is from this paradigm that the
term ``galactic archeology'' arises.   Such stars
provide important clues to the chemical
history of our Galaxy, the nature of early SN, the
mode of star formation in the proto-Milky Way, and the formation
of the Galactic halo.  The number of extremely metal poor (EMP)
stars, i.e. those with Fe abundance\footnote{The 
standard nomenclature is adopted; the abundance of
element $X$ is given by $\epsilon(X) = N(X)/N(H)$ on a scale where
$N(H) = 10^{12}$ H atoms.  Then
[X/H] = log$_{10}$[N(X)/N(H)] $-$ log$_{10}$[N(X)/N(H)]$_{\odot}$, and similarly
for [X/Fe].}
 [Fe/H] $< -3.0~$dex,
known as of 2005 is summarized by \cite{beers05};
it has grown slowly since that time, but the sample is
still quite small. The goal of the 0Z Survey, which began in 2001, is twofold: (1) to 
increase this sample
substantially so that statistical studies 
of the Galactic EMP star population become feasible and (2) to study at
high spectral dispersion those new EMP stars we find to determine
trends in abundance ratios, deduce implications regarding SN and
nucleosynthesis in the young Milky Way, and derive constraints on 
how well the ISM in the forming Galaxy was mixed.

Our project, the 0Z Survey, focuses on candidate EMP stars from the
Hamburg/ESO Survey  (henceforth the HES) \citep{wis00}.  These stars represent  
an in-situ probe of the Galactic halo.
After completion of the Keck Pilot Project
\citep{keck_pilot_1,keck_pilot_2}, we have  published many papers over
the past decade describing individual EMP stars, or small groups of such stars,
of particular interest found in our survey, 
i.e. \cite{cohen03},  \cite{cohen04}, \cite{feh_cstars},
\cite{cohen06}, \cite{cohen_1424}, and \cite{cohen08};
those papers include a total of 74 individual stars.  At this time the 0Z project
has ended, and we now have  122 candidate
EMP stars mostly from the HES with high resolution spectra, many of which have not appeared
in any of our earlier publications.
This sample is the largest sample of EMP candidates with high dispersion
spectra from a single survey analyzed by a single person.  It is substantially
larger than that of the First Stars Survey at the VLT led by
R. ~Cayrel \citep[see, e.g.][]{cayrel04}.  A large component  of the
sample of EMP stars discussed
by  \cite{yong13} is in fact from our earlier publications.

In this paper we focus on the 0Z sample as a whole, and in particular on the
issue of outliers in various properties related to the chemical inventory
of these stars.  After defining the sample in \S\ref{section_sample},
we give details of our analysis procedures in the next three
sections, with \S\ref{section_tailored} discussing the use of model
atmospheres tailored to the individual abundance distribution within
a specific star and \S\ref{section_comp} reporting on the overall
quite good comparison for stars with independent abundance determinations
by multiple groups.

Discussion of the outliers  begins in \S\ref{section_out},
with linear fits to abundance ratios presented in
\S\ref{section_fits}.  The largest family of outliers, the stars
with sub-solar [Ca/Fe],
is presented in \S\ref{section_lowca},
with the next section discussing the other smaller outlier families.
Carbon-rich EMP stars are discussed in \S\ref{section_cstars};
a presentation of $r$-process rich stars follow.  Distance
effects are examined in \S\ref{section_distance}.  Comments
on nucleosynthesis and related issues are given in 
\S\ref{section_nucleosyn}, and a brief summary concludes the paper.

\section{High Spectral Resolution Sample \label{section_sample} }

\cite{christlieb08} describe the procedure for selection
 of EMP candidates suspected to have [Fe/H] $< -3.0$~dex 
 from the database of digitized spectra of the Hamburg/ESO Survey 
 \citep{wis00},
 an objective prism survey whose primary goal was to find bright quasars. 
The original EMP candidate lists were generated from the HES
database by N.~Christlieb
and a subset of these
were transmitted to the Caltech and Carnegie 0Z core project
members in Aug. 2001.  As will be discussed in \cite{christlieb13},
because the fraction
of higher-metallicity interlopers was unknown, but feared to be substantial, these
 candidates were followed up with moderate resolution spectroscopy using
 the Double Spectrograph \citep{dbsp} at the Hale Telescope or the
Boller and Chivens spectrograph
at the 6-m Magellan Clay Telescope.  Approximately 1540 candidates
 were observed.  From each moderate resolution spectrum, a
 metallicity denoted [Fe/H](HES) was generated using a variant
 of the algorithm of  \cite{beers98}, which relies on the strength of
  H$\delta$  as an estimator of \teff.  The strength of
 the 3933~\AA\ Ca~II absorption line then provides an estimate 
 of the metallicity.  The mod-res followup program and results will be 
 discussed in detail in  \cite{christlieb13}.
 
The sample selected for high dispersion observations from our mod-res follow up
spectra of HES EMP candidates consisted of those stars with the lowest values of 
[Fe/H](HES).
Observations for the sample discussed here are primarily from
HIRES \citep{vogt94} at the Keck~I Telescope
with a substantial contribution of spectra
from MIKE \citep{bernstein_mike} at the Magellan~II ~ Clay Telescope.  The
stellar sample for the Keck Pilot Project \citep{keck_pilot_1,keck_pilot_2,
keck_pilot_3} was assembled by N.~Christlieb.
It consisted of candidates with mod-res spectroscopy
from other telescopes, which were subsequently observed with HIRES in Sep. 2000.
Christlieb and T.~Beers also contributed some of the candidates
for the 2001 observing season at HIRES since at that time 
our 0Z project had not yet built up
its own set of candidates vetted by moderate resolution spectroscopy.
MIKE/Magellan observations for the 0Z project did not begin until 2003,
by which time our 0Z project had built up its own set of vetted candidate EMP stars.

As high dispersion observations and preliminary detailed abundance analyses
proceeded over the years, it became clear that there 
was a 
problem with the \cite{beers98} algorithm in that it underestimated [Fe/H]
for extremely C-rich stars \citep{feh_cstars}.  In later versions of this code, 
no value was assigned to [Fe/H](HES)
for such stars.  Because it took a few years to isolate this bias
and to understand its origin,
which is discussed in detail in
\cite{feh_cstars}, such C-rich stars
are over-represented in our high resolution sample.  Furthermore, as the
metallicity assignment algorithm evolved and hopefully improved with time, the sample of
stars of most interest, i.e. those with [Fe/H](HES) $< -3.0$~dex, also evolved, 
with more stars
being added, and a few dropping out.  Major reviews, merging of
the DBSP/Palomar and B\&C/Magellan lists, and summaries
of all the moderate resolution results were carried out in 2005
and in 2007 to inform subsequent high resolution observations.

\section{High Dispersion Observations of EMP Candidates}

HIRES observations commenced in Sep. 2000 at the Keck~1 Telescope
and continued at a rapid pace
through late 2007, extending at a slower pace until early 2012.
The initial observations  used the original 2048x2048 pixel CCD detector
at spectral resolution 36,000 (1.1 x 7 arcsec slit) with 
4.2 pixels per spectral resolution element.  Because of the
limited size of
the detector, the spectral coverage was restricted to 3805 to 5325~\AA.
This instrumental configuration was adopted since good coverage at blue
wavelengths is crucial given the overall weakness of the 
absorption lines in EMP candidates and the dearth of red absorption lines.
In the late spring of 2004 the HIRES detector was upgraded
to a mosaic of three CCDs 
optimized for the blue, green, and
red spectral regions respectively.  This enabled wider spectral coverage
with smaller pixels (6.7 per spectral resolution element) and with
lower readout noise.  The
spectra taken after this are divided into two classes; those
that used the blue HIRES cross disperser had spectral coverage 
from about 3250 to 5990~\AA.
Most subsequent observations used the red HIRES cross disperser with spectral
coverage from 3880 to 8360~\AA.
There are small gaps between the three CCDs that form the
mosaic, sometimes resulting in the  loss of  part or  in a few cases all of a 
single echelle order.  Initially the HIRES observations were reduced using
echelle routines  written by J.~McCarthy \citep{mccarthy88}
within the Figaro package \citep{figaro}.
Beginning in 2003, we used
the pipeline package MAKEE\footnote{MAKEE was developed
by T.A. Barlow specifically for reduction of Keck HIRES data.  It is
freely available on the world wide web at the
Keck Observatory home page, http://www2.keck.hawaii.edu/inst/hires.}
for this purpose.


Most of the HIRES observations from 2010 to 2012 were repeat observations 
of stars
whose prior high dispersion spectrum either had 
a signal-to-noise ratio (SNR)
lower than desired or whose wavelength
coverage needed to be extended further to the red for the reasons to be
discussed in \S\ref{section_out}.  

MIKE observations began  in 2003 and continued through 2008.  Although
there were a number of detector upgrades on the blue and red arms during
this period, the pixel size and  format of the detectors
remained unchanged; only the performance of the detectors
was improved.  
MIKE observations  have a spectral resolution of 
42,000 in the blue camera and 32,000 in the red camera 
(corresponding to a 0.7~arcsec wide slit) with full spectral
coverage, and are Nyquist sampled, with (typically) 2 pixels per
spectral resolution element.
The MIKE pipeline of \cite{kelson03} was used to reduce the spectra.

The total sample of EMP candidates with high dispersion spectra
and detailed abundance analyses
presented here, including those from the Keck Pilot Project \citep{keck_pilot_1,keck_pilot_2}
and a few stars selected from the HK Survey \citep{hk1,hk2}, is \nstars\ stars.
The sample  and a log of the HIRES/Keck and Mike/Magellan observations
are presented in Table~\ref{table_sample}.  The total integration time
for this set of spectra is $\sim$210 hours.

\begin{figure}
\epsscale{1.05}
\plotone{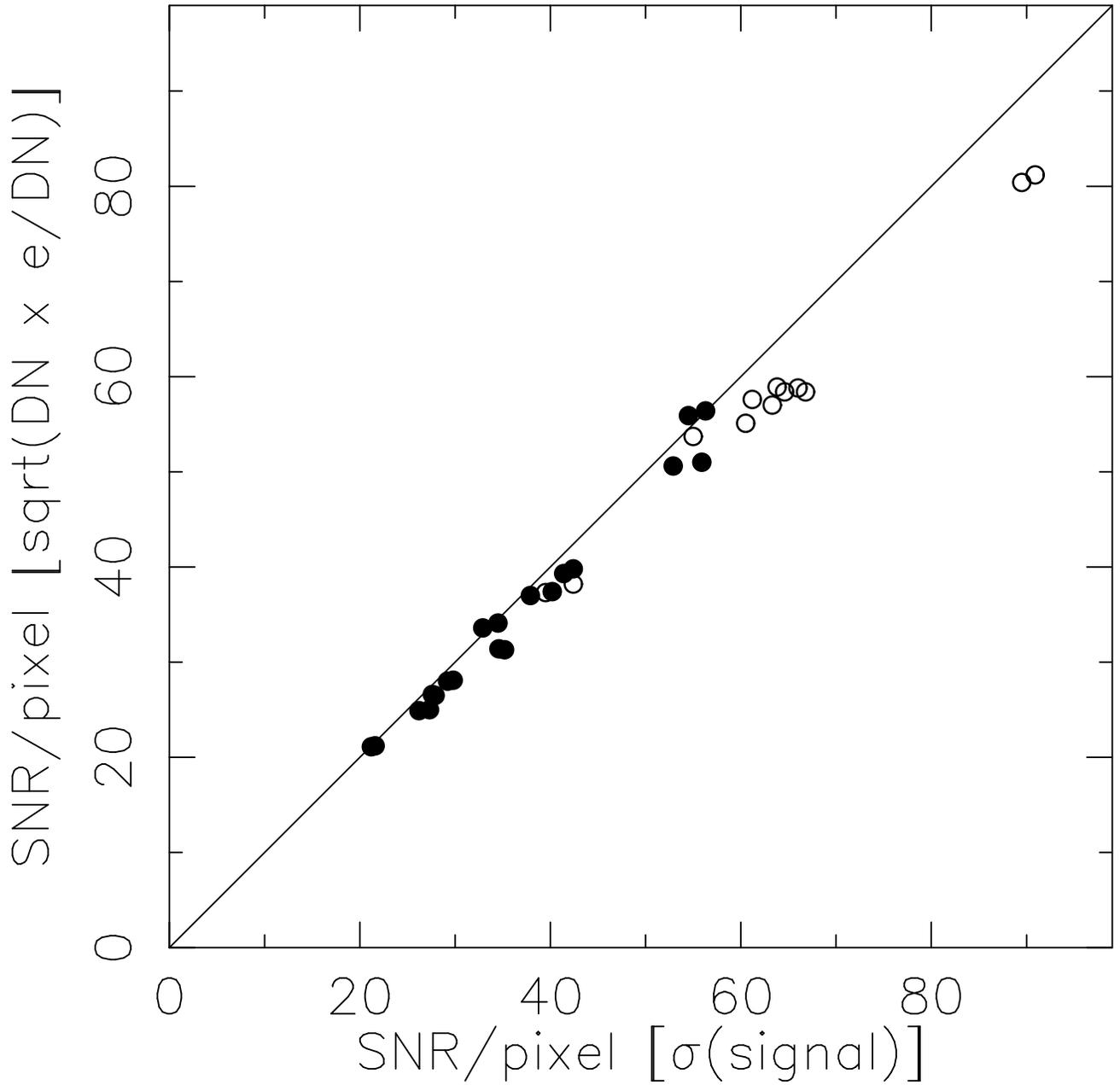}
\caption[]{A comparison of SNR calculated by looking at the dispersion in DN 
within line free regions of spectra of the most metal-poor stars 
(horizontal axis) vs that calculated
 in a simplified way (vertical axis) shows good agreement between the
 two measures.  Filled circles are from HIRES-B spectra, open
 circles from HIRES-R spectra, all using the green or red CCDs 
 of the mosaic HIRES detector.
\label{figure_snr_comp} }
\end{figure}

The SNR per spectral resolution element in the continuum
at the center of the order near 5300~\AA\ is given
in Table~\ref{table_sample}.  The spectral resolution is taken as 36,000 (that of
a 1.1~arcsec wide slit) for HIRES.   For MIKE, the 0.7~arcsec wide
slit corresponds to 2 pixels at the detectors,
to a spectral resolution of 42,000 (32,000) in the blue (red)
camera\footnote{A small number of MIKE and HIRES
spectra were taken at even higher spectral resolution, but the SNR
are quoted for the values given above.}.  For HIRES spectra, the SNR is
calculated from the
observed detected electrons/spectral pixel in the final
1D spectra (DN), ignoring such issues as cosmic rays, the contribution of
sky background to the noise, etc.  We have checked that this 
``quick and dirty'' value  (i.e. SNR(quick) 
= $\sqrt{{\rm DN} \times (e^{-}/{\rm DN}) }$~)
provides
an accurate determination of SNR by selecting line free regions 
approximately 30~\AA\ wide in the
spectra of the most metal-poor stars and calculating the dispersion
in DN across these intervals  in segments of 20 pixels length.  The
short length
removes any curvature in the continuum
arising from the use of an echelle grating.  We then use
the mean of the resulting $\sigma$ values  to determine
a more realistic value for SNR without making any simplifying
assumptions.
This is compared with the SNR(quick) in Fig.~\ref{figure_snr_comp}.  
The agreement,
shown for 32 test cases using the mosaic HIRES detector, is extremely good.  
The slight drop below
equality seen in Fig.~\ref{figure_snr_comp}, most obvious at the highest 
DN levels, may be due to
the limited precision of the conversion factor e$^-$/DN 
in the HIRES on-line documentation.


\section{Equivalent Widths and Radial Velocities \label{section_eqw_vr} }

The pipeline MAKEE when applied to HIRES spectra outputs a 1D spectrum for each
echelle order whose wavelength scale has the heliocentric velocity
correction applied.  For the MIKE spectra 
the heliocentric corrections were applied later. 

Equivalent widths were  
automatically measured from the 1D spectra using the code EWDET2 which is
based on the code EWDET described in \cite{ramirez_m71_fe}. This code carries out
a continuum fit with an algorithm similar to that used in the IRAF script
``continuum'', then searches for absorption features.  For each
feature judged to be real, the continuum fit at that wavelength
is evaluated followed by
a Gaussian fit to the absorption feature.  The output is a list
of the central wavelength, equivalent width, and half width of each feature, 
and an estimate of the uncertainty
in the equivalent width. 
Note that this list, together with a list
of laboratory lines, is also used to determine the radial velocity of the star.
EWDET2 works well on the HIRES spectra, but is somewhat less reliable
for the MIKE spectra, where the spectra are Nyquist sampled.  The smaller
number of pixels per spectral resolution element in the MIKE spectra
makes the Gaussian fitting
routine used by EWDET2 sometimes output spurious results.

Extensive hand checking of equivalent widths had to be done for the
MIKE spectra. For the HIRES spectra, some hand checking was done
particularly to add lines too weak
for the automatic routine to find.  $W_{\lambda}$ for the strongest
lines were always checked by hand to ensure that the full extent of the 
damped wings was included.
 
A master line list of features we adopted for radial velocity
determination was constructed.  It contains 38  reasonably isolated
strong features from 3815 to 6643~\AA\ that one might expect to be
present and reasonably
strong even in EMP stars.  The laboratory wavelengths for these features
were taken from the NIST database as of 2002.  Lines falling within the
strongest CH and CN bands are omitted if the star under consideration
is C-rich.

An initial guess
of the radial velocity was made  by searching the EWDET2 output by hand
for the Mg triplet lines.  Then a code searches
for these 38 features within the list of features found
by EWDET2, rejecting observed lines with  $W_{\lambda} < 50$~m\AA,
and calculates a radial velocity for each of the laboratory lines.
The individual $v_r$ values for lines which occur in two different 
echelle orders, and hence have independent measurements, are averaged.
Finally an average $v_r$  and a rms dispersion are calculated for the
subset of the 38 lab lines that are actually detected in a given spectrum.
This is repeated 
after a 3$\sigma$ clip in $v_r$ to determine the adopted $v_r$ for each star.
At most one laboratory line is eliminated by applying this 3$\sigma$ clip.

The master line list also contains a list of small corrections in wavelength
for each individual line used that was developed in late 2001 
by comparing the line-by-line
$v_r$ measurements of a reasonable number of stars. Most of them
correspond to less than 1~\kms.  This correction file has been
used throughout.  After the HIRES upgrade, additional lines
were added to the master list
at redder wavelengths to form the final list of 38 lines.
All the added lines are  redder than the Mg triplet and have their
correction value set to zero.  

For spectra from HIRES taken after the 2004 detector upgrade
with reasonable SNR of stars which are not highly enhanced
in carbon, this procedure
gives a rms dispersion for the set of $v_r$ lines
around the mean $v_r$ of less than 0.15~km/sec.
Fewer lines from the master list for $v_r$ determinations
can be picked up in the spectra of the cool C-rich stars, and there may
be some blending with molecular features.  Hence the dispersion around
the final $v_r$ is often larger for C-rich stars, sometimes reaching as much as 1.5~km/sec.
The  uncertainty in the adopted $v_r$ is then nominally 
$\sigma$/$\sqrt{N({\rm{lines}})}$.

However, there may be a systematic contribution to the uncertainty
in $v_r$ arising from placement of the star within the slit.  If the
slit is ``wide'' compared to the seeing, this becomes a concern.
The worst case among our high resolution spectra is that of HIRES, 
where a 1.1~arcsec slit was used
for essentially all of the spectra, in an attempt to avoid light loss
while maintaining a sufficiently high spectral dispersion.
But when the seeing is very good (i.e. 0.6~arcsec or less), 
one is centering the star within the slit with only a few percent
of the total light of the star not going through the slit.
If guiding is done using another star within the guider field
once the object of interest
is placed within the spectrograph slit, the location of the
star within the slit should be preserved
throughout the exposure assuming perfect correction for field
rotation.  Thus centering within the slit 
may be thus be slightly different from
one star to another.  

Simulations suggest that a 0.1~arcsec
difference in placing the star within the width of the slit
near its center will produce a shift in measured $v_r$
of less than 0.6~km~s$^{-1}$. 
We therefore assume that all HIRES spectra have a 0.6~km~s$^{-1}$
contribution to the total $v_r$ uncertainty.  MIKE spectra
use narrower slits, so this value should be a firm upper limit to
the $v_r$ offset due to placement of the star in the slit.

In addition, we did not take other steps necessary to secure
the highest possible radial velocity accuracy from HIRES.  The
configuration (usually the setting of the 
cross dispersor) of HIRES varied 
from run to run depending on the goals for the HES and
for whatever other observations were planned.  To
test the $v_r$ accuracy we actually achieved, we 
used the terrestrial atmospheric band near 6860~\AA.  The measured
wavelengths of selected isolated lines in this band should be 
constant in the final 1D spectra, but they are not.  They
show differences from run to run corresponding to to a maximum
of 3~\kms, although the variations within a given run
(i.e. with a fixed HIRES configuration) were
generally closer to 1~\kms. We emphasize that these
variations most likely do
not arise in the spectrograph itself  but rather from the idiosyncrasies of
wavelength fitting within the reduction pipeline.  The largest
apparent wavelength shifts occurred when this atmospheric band
shifted from the first order of the red detector in the
three CCD mosaic to the last order of the green detector,
with one or both being only partial orders due to gaps
between the CCDs in the mosaic.

G.~W.~Preston (private communication, 2001) kindly supplied
a compilation of his monitoring of metal-poor red giants using the
echelle spectrograph at the DuPont 2.5~m Telescope of the 
Las Campanas Observatory.
Those with 8 or more independent measurements which had
a low rms dispersion for the set of measured $v_r$ 
(typically less than 0.5~km~s$^{-1}$) were adopted
by our 0Z Project as radial velocity standard stars.
One or more of these stars was observed during each HIRES run for the
first 4 years of the 0Z project.  
With one exception, all our
observations of stars from Preston's list agree with his $v_r$
to within the observational uncertainties.

A significant fraction ($\sim 25$\%) of the stars in the 0Z sample have 
been observed more than once at high spectral dispersion during the course
of the 0Z project, with
a typical separation of several years; see Table~\ref{table_mult_vr}.  
This offers a good opportunity
to test for binarity. \cite{gunn79} found that a
typical $v_r$ jitter in metal-poor red giants due to
atmospheric effects is $\sim$1~\kms.
Bearing this in mind and given the issues for $v_r$ discussed
above, we assume that velocity differences between the various epochs of observation
for a particular star exceeding 5~km s$^{-1}$ imply that the star is
a binary.  Stars with multiple spectra whose velocity differences are 
less than 3~km s$^{-1}$ are taken to be consistent with no variation
in radial velocity and hence no evidence that the
star is a member of a binary system, while stars with
$3 < \Delta(v_r) < 5$~km s$^{-1}$ are assigned as possible
binaries.
Previously known binaries are
indicated in Table~\ref{table_mult_vr}.  This table includes a few shorter HIRES exposures,
not listed in Table~\ref{table_sample}, acquired
only to check $v_r$, as well as $v_r$ measurements from the literature
for our sample stars when available from other high dispersion studies.

We find (see Table~\ref{table_mult_vr})
that 7 of the total of 8 C-stars 
(i.e. C-rich stars with detectable C$_2$ bands) in this table with
high [Ba/Fe] are binaries or probable binaries as defined above.
The high fraction of binaries is in agreement with an earlier study by
\cite{lucatello_bin} for a sample of comparable size
of C-rich Ba-rich metal-poor stars.  This is 
not surprising given that the
standard explanation of $s$-process enrichment
is mass transfer in a system where the primary
has gone through the AGB, see e.g. \cite{busso_araa}. 
{{Two of the four C-stars with normal [Ba/Fe] are confirmed binaries with
periods, one (a hot dwarf with three epochs of HIRES spectra) does not
appear to be a binary, but might now be or have been one, and the fourth has only one epoch
of observation.  Thus it appears that even the C-stars with normal
[Ba/Fe] are mostly binaries.}}

The picture changes for the 5 C-rich stars which do not show C$_2$ bands.
(Note that the taxonomy and characteristics of C-rich stars are discussed in
\S\ref{section_cstars}.)
The only one with high [Ba/Fe] (the CEMP-$rs$ star HE2148$-$1247) appears to be a binary.  
One of the
other four with normal [Ba/Fe] may be a binary, but there is no
evidence to support binarity for the remaining three stars.


Of the 18 C-normal stars with multiple $v_r$ determinations, 3
are definite binaries  and 3 more are 
probable binaries, i.e. a minimum binary fraction of 17 $\pm$9\%,
possibly being as large as 33\%.  The minimum estimated binary
fraction for our sample of EMP stars is consistent with that
reported by 
\cite{latham02}, who present the results of an extensive radial velocity
monitoring program for a large sample of proper motion stars
carried out over a time span of 18 years.  They find that 
15\% of halo field stars in
the Solar neighborhood are binaries, and that this fraction
is identical to within the uncertainties to that of a similar sized
sample of Galactic disk stars.

In order to investigate more quantitatively 
whether the radial velocity distributions differ between
C-normal, C-rich with C$_2$ bands, and C-rich
without C$_2$ bands,
we computed the velocity range, $\Delta v_r = v_{\rm max}$ $-$ $v_{\rm min}$, for each
star with multiple epoch spectra.  For the 18 stars in the C-normal group $\Delta v_r$ ranges from
0.2 to 5.5~\kms.  For the group of 11 C-rich stars with C$_2$ bands there are four 
with $\Delta v_r$ above 10~\kms and another at 6.8~\kms, all well above the range for the C-normal
group.  The 5 C-rich stars without C$_2$ bands range from 0.8 to 5.7~\kms, very 
similar to the C-normal distribution of $\Delta v_r$ values.
A K--S test comparing the velocities of the C-normal and C-rich stars without C$_2$ indicates
a probability of only 0.3\%  that the two distributions differ.  On the other hand the K--S test
indicates an 80\% probability that the velocities of the C-rich stars with C$_2$
 bands differ from the C-normal velocities.

Thus, the velocities are consistent with the idea that most of the C-rich stars 
with C$_2$ bands are binaries, while the binary fraction of the C-rich stars without C$_2$ 
is similar to the C-normal stars.  Except for HE2148$-$1247, our C-rich stars without C$_2$ 
bands are not particularly enhanced in Ba: they are CEMP-no stars, i.e., without $s$-process 
enhancements.

As we noted in Cohen et al. (2006) the CEMP-no stars show very similar characteristics to
population~I early R-type carbon stars: both show carbon enhancements but normal
$s$-process abundances \citep{dominy84}, and both have low $^{12}$C/$^{13}$C ratios.  Here we find that the CEMP-no stars also have low binary fractions, qualitatively similar to the conclusion
of \cite{mcclure97} for R-type carbon stars.

Thus, our velocities lend further support to the idea that CEMP-no stars
without C$_2$ bands are the
 population~II
equivalent of the population~I early R-type carbon stars.

\section{Stellar Parameters}

We follow the procedures developed earlier by J.~Cohen and collaborators
and described in \cite{cohen05}.  We rely on infrared magnitudes
$J$ and $K_s$ from 2MASS \citep{2mass1,2mass2} combined
with optical photometry from a variety of sources, including
our Andicam queue photometry program at CTIO for $V,I$ described in
\cite{christlieb13}.
If a 0Z sample star with a high resolution spectrum was not 
observed by Andicam
or by the SDSS \citep{york_sdss},
we fell back on our small photometry program at the 40-inch Swope telescope
at the Las Campanas Observatory in 2001, or, in the four cases where nothing
else was available, relied on the HES $V$ mag.  A comparison of
the HES $V$ mags with those of higher precision from the
SDSS will be given in \cite{christlieb13}; for 51 stars from the 0Z Survey in common,
the mean difference in $V$(Andi) vs $V$(SDSS) (the conversion formulae
of \cite{sdss_convert} are used to go from SDSS $gri$ to  $V,I$)
is 0.00~mag, while the rms deviation is
0.07~mag.

From these we form the colors
$V-I$, $V-J$, and $V-K_s$.   We match these using the predicted
color grid of \cite{houdashelt00}, which we have earlier shown to
be essentially identical to those from the Kurucz
\citep{kurucz93} and MARCS models \citep{marcs}.
We chose these specific colors to have good discrimination for
effective temperature,
\teff\, while minimizing their sensitivity to metallicity
and surface gravity.  The all-sky uniform 2MASS colors were
extremely useful, but the faintest 0Z sample stars
are approaching the faint limit for 2MASS; they
have rather large uncertainties for their IR colors, particularly
for $K_s$.  We used the reddening maps of  \cite{schlegel}
assuming that the EMP candidate was beyond the  absorbing gas clouds,
which are confined fairly tightly to the Galactic plane. 
The first pass at \teff\ used [Fe/H](HES) as the stellar metallicity.

Once a first pass at \teff\ was made, we calculated the
surface gravity, log($g$), by combining a 
 bolometric correction adopted from the \cite{houdashelt00} color grid
with an  $\alpha$-enhanced EMP isochrone from \cite{yi03} and
dereddened $V$ mag.  We assume a fixed stellar mass of 0.8~$M_{\odot}$. 
We then repeated this procedure using
the log($g$) from the first iteration and the [Fe/H] from the first
pass abundance analysis to obtain the nominal stellar parameters.
Note that these are set without any reference to the spectral features, except
for a rough value of [Fe/H], which is not crucial over the range
relevant here, since at such low metallicities the contribution of
atomic lines to the total  opacity is not significant and at these low
[Fe/H] the electron density is dominated by ionization of H, not of the metals.

\begin{figure}
\epsscale{1.05}
\plotone{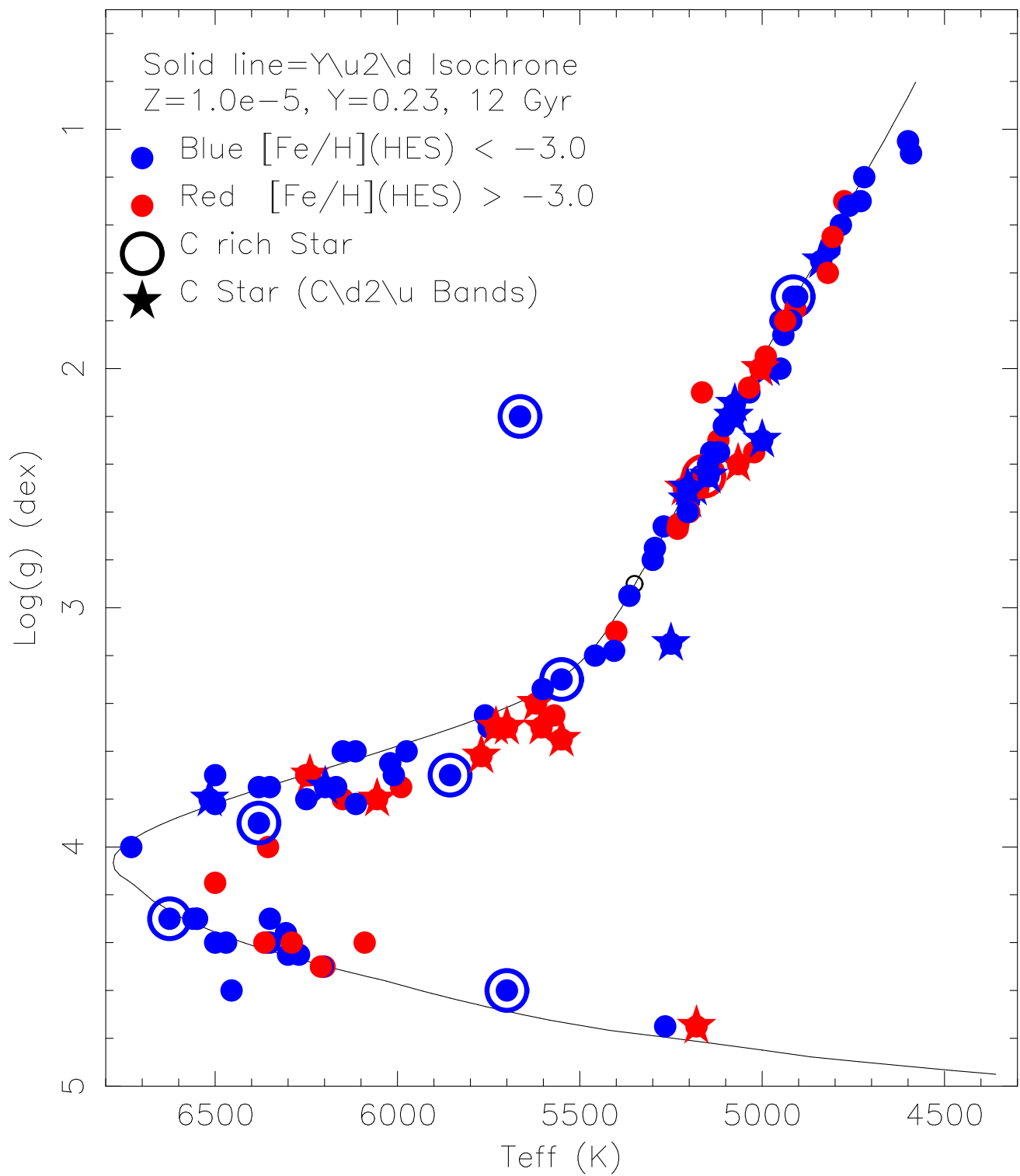}
\caption[]{The adopted $T_{eff}$ vs log($g$) is shown for the 122 stars in
the high resolution
abundance sample studied here.  There is one HB star.  The colors denote
the metallicity, while the symbols indicate the degree of C-enhancement;
C-stars are shown by stars while C-rich stars with no C$_2$ are circled.
(The filled and open circles denote the metallicity in the B/W version
of this figure.)
\label{figure_cmd} }
\end{figure}

Table~\ref{table_param}  lists the nominal and adopted stellar 
parameters for our sample, with the final [Fe/H] values from our detailed
abundance analyses in the last column. 
The uncertainty in \teff\ is taken as the rms $\sigma$ over the
three colors we use to determine \teff.  Stars
with \teff\ near that of the main sequence turnoff can be either
above it (i.e. subgiants) or lower luminosity dwarfs.  
In the absence of a distance, distinguishing
between these two choices can only be made by analyzing the spectra,
i.e. by the Fe ionization equilibrium.
Our 0Z high dispersion sample is shown in the \teff\ -- log($g$) plane
in Fig.~\ref{figure_cmd}.

There are some stars with
very large uncertainties in their nominal \teff\ (see the
third column of Table~\ref{table_param}). In early 2012, in an effort to understand
the origin of these few very large $\sigma$, 
those stars with $\sigma$(\teff) $> 200$~K were searched
for in DR8  of the SDSS \citep{sdss_dr8}.  Several were located, but the SDSS photometry,
transformed into $V$ and $I$ using the equations of \cite{sdss_convert},
was in all cases very close
to that from Andicam. 
Most of the high $\sigma$(\teff) stars are 
C-stars; presumably the strong CH, CN,  and C$_2$ molecular bands are distorting
the colors.  The few other stars with high $\sigma({T_{eff}})$
are at the faint end of the 2MASS sample
with large uncertainties at $K_s$. There are only four stars
for which we had to rely on $V$(HES) as no
accurate $V$ could be located nor were these four stars included
in our own photometry programs, but these are not among those
with unusually large $\sigma({T_{eff}})$.
In a few other cases the optical photometry may not 
be as 
accurate as claimed.  But, as can be seen in Table~\ref{table_param}, the
dispersion around the nominal \teff\ of the three contributing
colors is not unreasonably large for most stars 
and is in most cases consistent with the 
uncertainties of each color from their original sources.
The quartiles in the distribution of dispersion in \teff\
from the three colors are 42, 67, and 133~K, and  
68\% (that of 1$\sigma$ for a Gaussian distribution)
of the total set of dispersions is 
less than 100~K and 94\% (equivalent to 2$\sigma$
for a Gaussian distribution) is less than 200~K. 
Only 13\% have $\sigma({T_{eff}}) > 150$~K.
Thus adopting 100~K as our nominal \teff\ uncertainty seems appropriate.
With this \teff\ uncertainty, the slope of the red giant
branch in low metallicity  isochrones from \cite{yi03}
suggests 0.25~dex as the appropriate uncertainty in log($g$).


\section{Abundance Analysis Procedures \label{section_abund_proc} }

Our detailed abundance analyses begin with the stellar parameters
derived as described above.  We use plane-parallel LTE model atmospheres
from the grid of \cite{kurucz93} with no convective overshooting.
We use the abundance determination code MOOG \citep{moog} as of 2002.
Recently an improved version of MOOG was presented by  \cite{sobeck_moog}. 
This update
incorporates a number of improvements that are important
for EMP stars. In particular, it includes a better treatment of coherent
isotropic scattering, which in the 2002 version is treated as pure absorption.
This provides a proper treatment of Rayleigh scattering, which can
become an important
opacity source in the blue  and UV in EMP stars.
The difference in deduced abundances
between MOOG-2002 and MOOG-2010-SCAT is thus largest
for the coolest EMP giants, and also is a function of the wavelength
of the line, becoming significant only at $\lambda < 4000$~\AA, as was
pointed out by \cite{sobeck_moog}.
The five stars in the HES sample with $T_{eff} < 4800$~K, as well as 
the 15 stars with
$4800 < T_{eff} < 5120$~K with spectra extending well below 4000~\AA,
were reanalyzed  in early 2012 with the updated code of \cite{sobeck_moog}.
These stars are indicated by the letter S after their ID in column 1 of
Table~\ref{table_param}.  In general the changes were small;
$v_t$ tended to be reduced by 0.1~km/sec, and the largest difference
in abundance was a decrease of 0.3~dex for the UV features in the
coolest sample stars.

The
determination of stellar parameters, measurement of equivalent widths, and detailed
abundance analyses were all carried out by J.~Cohen over the past decade,
with the exception of 9 of the 11 stars\footnote{Supplementary
observations with HIRES were subsequently obtained
by J.~Cohen for two of the 11 stars included in the
Keck Pilot Project and she then reanalyzed them.}
included in the Keck Pilot Project \citep{keck_pilot_1,keck_pilot_2}. 
During that time
the $gf$ values for several species have been updated.  In particular, 
we are now using the current NIST (version 4.0) \citep{nist} values 
for Mg and Ca, which are not the same as those we have used in the past.
In late 2011 we therefore went through the abundance analyses of all the
HES stars (except those from the Keck Pilot Project which were not subsequently 
reobserved) 
to homogenize the $gf$ values used for all the detected absorption lines 
in the abundance analyses of each of the sample stars
to those on  J.~Cohen's current master
list.  The largest impact in deduced abundance was for Mg as the
updates to the $gf$ values for several of the Mg~I lines were large,
and there are only a few detectable Mg~I lines in EMP stars.  Note that
the $gf$ values for the Mg triplet lines did not change.
The updated Mg~I $gf$ values are clearly better; they significantly reduce
the dispersion in Mg abundance determined from a typical set of Mg~I lines.
In addition to reducing $\sigma$,
depending on which specific
lines were detected in a particular star, a decrease in
[Mg/H] over our older values of up to 0.12~dex resulted.

The impact
of our update of the Ca~I $gf$ values was smaller as there
are usually more detectable Ca~I lines, only some of which 
had their transition probabilities
changed.  For a given star, smaller changes in [Ca/H], ranging from 
+0.01 to +0.04~dex,
occurred as a result of these changes.

Because of the uncertainty in the stellar parameters, we felt free
to make small adjustments in the adopted \teff\ and/or log($g$) to
improve the results of the abundance analyses (usually the ionization
equilibrium)
 if necessary.  These changes, which were always less than the
dispersion in \teff\ values from the various available colors,
are indicated in Table~\ref{table_param}, where both the
purely photometric stellar parameters and the final adopted
values are given for our 122 star sample.  The adopted stellar parameters
of only 17  of the {\nstars} sample stars differ from their photometric values by
more than 20~K in \teff\ or more than 0.1~dex in log($g$).

\begin{figure}
\epsscale{1.05}
\plotone{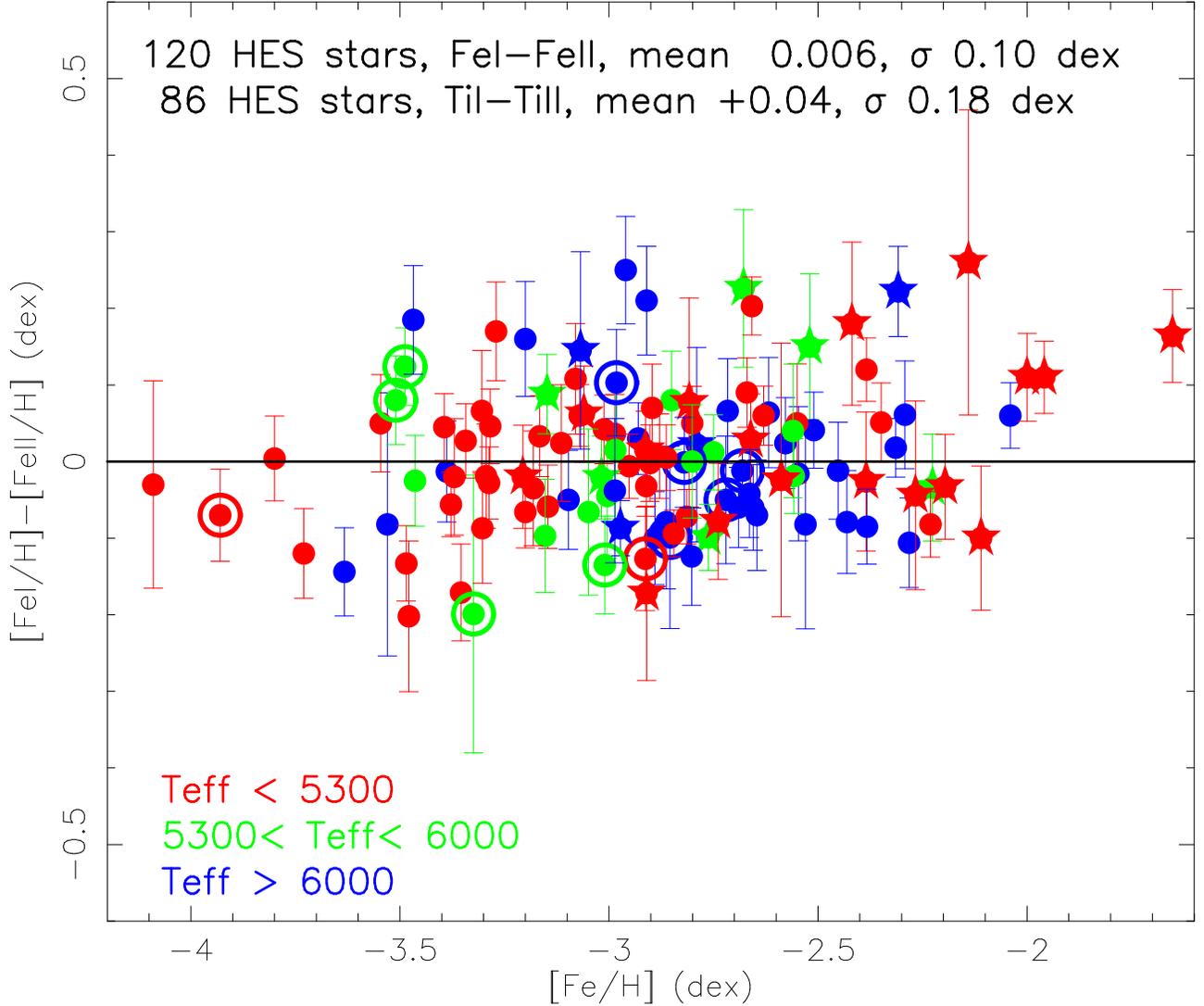}
\caption[]{The ionization equilibrium for neutral vs singly ionized
Fe is shown as a function of [Fe/H].      Statistics for the Fe
and Ti ionization equilibria are given in the text at the top of the figure.
The symbol colors denote ranges of $T_{eff}$, red:  $T_{eff} < 5300$~K,
green: $5300 < T_{eff} < 6000$~K, blue:  $T_{eff} > 6000$~K.
The star symbols are C-stars, while the circled
points are C-rich but do not show C$_2$ bands.  For the B/W version,
the temperature ranges, from cooler to hotter, are indicated by
open circles, filled grey circles, and black filled circles. This
is the set of symbols used to define $T_{eff}$ and degree of
C enhancement for the rest of the figures in this paper,
with the exception of 
Figs.~\ref{figure_ca_distance} and \ref{figure_sr_distance}. 
\label{figure_ioneq} }
\end{figure}

Standard tests of the validity of our procedures include whether 
we were able to achieve good ionization equilibrium, and consistency
of abundances from the set of Fe~I lines observed in each star
with $\chi$, $W_{\lambda}$, and $\lambda$.  The results are very
encouraging.  For 120 stars (a few of the EMP stars have
no detected Fe~II lines), we find a mean difference between
Fe abundance as determined from the neutral versus the ionized
species of 0.00~dex, with $\sigma$ of only 0.10~dex.
The adopted \teff\ uncertainty of 100~K
corresponds to a shift in the ionization equilibrium between
Fe~I and Fe~II of 0.09~dex, and the predicted
1$\sigma$ uncertainty in [FeI/FeII] for giants 
including both stellar parameter and analysis uncertainties is 0.14~dex
(see Table~\ref{table_unc_ratio_giants},
with values for main sequence turnoff region stars in
Table~\ref{table_unc_dwarfs}).  The good
agreement between these values suggests that our adopted
uncertainty in \teff\ is reasonable.
For 86 stars with both Ti~I and Ti~II detected, the mean difference between the 
deduced Ti abundance was only 0.04~dex, with a larger $\sigma$
of 0.18~dex; the corresponding entry in  Table~\ref{table_unc_ratio_giants} is 
$\sigma = 0.12$~dex, fairly close to that observed.
The ionization equilibrium for neutral vs singly
ionized Fe is shown for our sample in Fig.~\ref{figure_ioneq}.

\begin{figure}
\epsscale{0.7}
\plotone{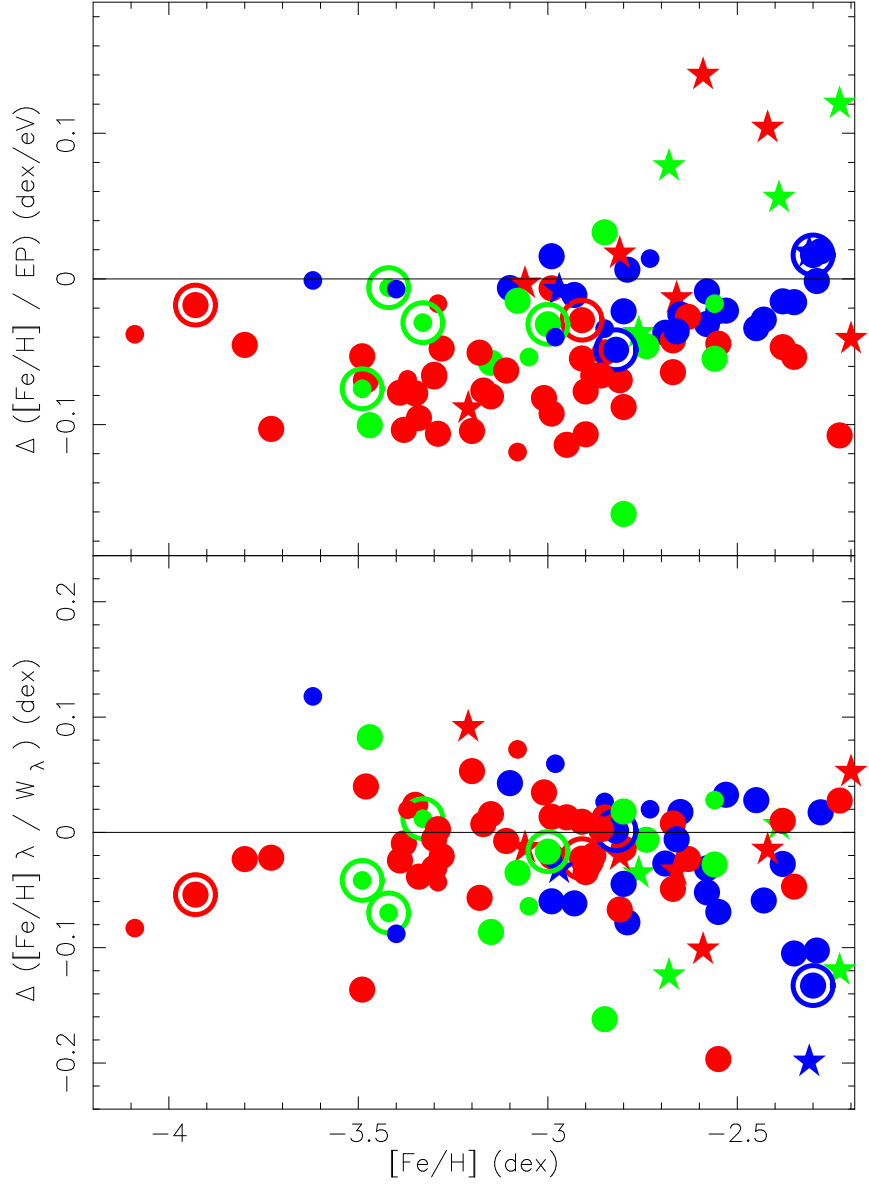}
\caption[]{The slopes of deduced Fe abundance from Fe~I lines
as a function of $\chi$ (typical range: 0 to 4.5~eV) (upper panel) 
or for Log($W_{\lambda}/{\lambda}$) (typical range: $-$5.8 to $-$4.6) 
(lower panel)
are shown for EMP stars with more than 20 detected Fe~I lines.
Symbols are as in  Fig.~\ref{figure_ioneq}.  Smaller symbols
are used for stars with less than 40 detected Fe~I lines.
\label{figure_slopes} }
\end{figure}

The slopes of the linear fit to deduced Fe~I abundances 
as a function of $\chi$, log[$W_{\lambda}/{\lambda}$], and $\lambda$ are
given in Table~\ref{table_feslopes}, and, for the first two,
are shown in Fig.~\ref{figure_slopes}.
The dependence on log[$W_{\lambda}/{\lambda}$] is small (i.e. the slopes are close
to zero), as is also true of the
dependence on $\lambda$.  The quartiles for the absolute value of the 
slope for log[$W_{\lambda}/{\lambda}$] are only
0.017, 0.028, and 0.060~dex/dex, so over the typical range of 
1.8~dex for  log[$W_{\lambda}/{\lambda}$], the maximum change in deduced Fe abundance is only
0.03, 0.05, and 0.11~dex respectively.  
Raising $v_t$ by 0.1~\kms\ decreases this slope by 0.05~dex/dex.
The spread in the lower panel of Fig.~\ref{figure_slopes} of this slope around zero
suggests that the
uncertainty in our adopted values of $v_t$ is about 0.1~\kms,
indicating that our choice of $v_t$ is appropriate.   

The dependence on $\chi$ is small
for the hot EMP dwarfs, but our analyses for the cooler giants tend to 
yield higher Fe abundances from the Fe~I lines with the
lowest $\chi$. A decrease in \teff\ of $\sim$300~K
would be required to make the mean slope
of the linear fit to deduced Fe~I abundances 
as a function of $\chi$ as shown in Fig.~\ref{figure_slopes} 
become zero.  But this introduces problems in the ionization
equilibrium.
No matter how we tried to adjust the various stellar parameters,
the coolest giants tend to have somewhat negative slopes for
the deduced Fe abundance from Fe~I lines as a function of $\chi$;
 the 0~eV Fe~I lines almost always
tend to be too strong in such cool giants.
Use of  MOOG-2010-SCAT instead of MOOG-2002  does not
eliminate this problem.  Presumably this
arises from some deficiency in the model stellar atmospheres
we adopt or in the analysis.  Perhaps this is a symptom of non-LTE,
as suggested by the calculations of \cite{nonlte_fe}.
In any case, since most of the Fe~I lines we use have $\chi > 2$~eV,
the effect of this problem is not large.  But it must be recognized
that for cool giants, something is systematically not correct,
while for the hotter EMP dwarfs, the analyses are completely
satisfactory.  Since Fe~II is the dominant species, many authors,
including for example \cite{thevenin99} and
\cite{nonlte_fe}, recommend using
the abundance derived from Fe~II lines rather than that from
the more numerous and stronger Fe~I lines as the Fe abundance.
Unfortunately with such faint EMP stars, the detectable
Fe~II lines are often very few in number and very weak.

The final derived abundance ratios for our sample of 122 EMP
stars, primarily from the HES, are shown as a function of
[Fe/H] in Figs.~\ref{figure_4plot1} through 
\ref{figure_4plot5}.  
Each of these figures shows the 
abundance ratios for 4 species, including, with the exception of Eu,
upper limits.  We set the vertical scale to be
identical on as many panels in a given one of this set of 5 figures as
possible, but this was not always feasible.

\subsection{Non-LTE \label{section_nonlte} }

The only correction for non-LTE uniformly implemented here 
in the tables and figures is +0.6~dex for Al
for all stars based on the calculations of \cite{al_nonlte}.  Only the 
resonance doublet at 3950~\AA\ can be detected in these
EMP stars; all other
Al~I lines are too weak.  Often
only the 3961~\AA\ line can be used as the 3944~\AA\ line will be blended with
strong CH lines if C is enhanced.

Extensive calculations of non-LTE effects in EMP stars
have recently become available for other species.
Non-LTE corrections for Ca lines have been calculated by several
groups, most recently by \cite{nonlte_ca}.  At the relevant low metallicities,
they are negligible in EMP giants for the resonance line 
of \ion{Ca}{1} at
4226~\AA, but can reach +0.3~dex in EMP giants (they are close
to zero for dwarfs) for typical
subordinate lines of this species. 
The Ca~II line at 3736~\AA\ with $\chi =$ 3.1~eV was picked up in a small number
of stars.  It presumably should be relatively unaffected by non-LTE,
but is a blend in the blue wing of a much stronger
Fe~I line, hence difficult to measure accurately. We cannot
comment on potential differences between abundances inferred
from these two ionization states of Ca with this specific Ca~II line 
from our work.


Other recent calculations include
\cite{nonlte_mgk} for Mg, \cite{nonlte_na} for Na, \cite{nonlte_sr} for Sr, and
\cite{nonlte_ba} for Ba.  For Mg, the non-LTE corrections are 
about +0.1~dex for EMP giants and +0.3~dex for EMP turnoff stars.
\cite{nonlte_sr} find that for the resonance
lines of Sr~II at 4077 and 4215~\AA, the only ones used here,
the non-LTE corrections
are approximately constant at +0.4~dex for EMP dwarfs and are
smaller and negative for EMP giants.
However \cite{hansen13} find much smaller corrections for stars at
[Fe/H] $\sim -3.0$ of Sr abundance from the 4077~\AA\ Sr~II line of
 +0.10~dex and $-0.05$~dex for dwarfs and giants
respectively.
The non-LTE corrections for \ion{Ba}{2} are about $-0.3$~dex
for hot turnoff EMP stars, becoming larger for subgiants,
and reaching +0.3~dex for the coolest EMP red giants.  But,
as we will see later, the
range in Sr and Ba abundance is so large that the differences
in non-LTE corrections with \teff\ for EMP stars are to first
order irrelevant.

Non-LTE  corrections for the
7699~\AA\ line of \ion{K}{1} (the only one detected here) 
have been
 calculated by several groups
\citep[e.g.,][]{ivanova_k,takeda_knonlte,nonlte_mgk}. These non-LTE
corrections are negative, and range from $-0.1$ to $-0.9$~dex (see
Fig.~6 of \citeauthor{ivanova_k} \citeyear{ivanova_k}).  They vary
strongly with \teff\ and with metallicity. But they have
not been applied to our deduced abundances.

There is also the issue of corrections for 3D effects,
which tend to affect lines primarily formed in the outer layers of
RGB giants; see e.g. \cite{asplund05}. 
This becomes important in comparing abundances
for a specific element
deduced from molecules versus those from high excitation atomic lines,
such occurs with O, where the very limited available features include
OH, very high excitation \ion{O}{1} lines (the triplet near 7770~\AA),
and the extremely weak forbidden lines, impossible to detect in these
EMP stars.
Pending the availability of large grids of models which include
such effects, which is still computationally demanding, we ignore
3D effects, but we note that for some atomic lines, and 
in particular for low excitation Fe~I lines in metal-poor stars,
they sometimes work in the opposite
direction so as to cancel at least partially the effects of non-LTE;
see e.g. {\S}5.3 of \cite{stagger2_13}.

Since our primary interest here is outliers from well
characterized abundance ratio trends, these limitations
in our abundance analyses should not be a serious
impediment.  

\subsection{Other Analysis Issues \label{section_other_anal} }

The uncertainties in the deduced abundance ratios are given in
Table~\ref{table_unc_ratio_giants}
(for giants) and Table~\ref{table_unc_dwarfs} (for dwarfs).
These include terms from the adopted
uncertainty in \teff\ of 100~K, that of \grav\ of 0.25~dex
for giants (a slightly smaller value is adopted for main sequence turnoff stars),
that from a change in [Fe/H] of +0.2~dex for the model stellar
atmosphere used, and that from a change in $v_t$ of 0.2~\kms.
A term for the uncertainty in the measurements of \eqw\ is
also included, set to $0.11/{\sqrt{N({\rm{lines}})}}$~dex for the giants, where
$N({\rm{lines}})$ is the number of detected absorption lines of the
species. A somewhat lower value for a single detected line, 0.08~dex, 
was used for the dwarfs,
where in general due to their higher \teff, the lines
are weaker and less crowded.  Random errors in the $gf$ values also contribute to this term.
These terms are added in quadrature, with a minimum value
for the total uncertainty set to 0.05~dex.  Since in many cases
only a few lines of a given species can be detected in these
EMP stars, the total uncertainty
in the abundance ratios [X/H]  is often dominated by that of the \eqw. 
Values are also given in the tables for
[Fe/H] from the neutral and ionized species, and
for [FeII/FeI].

Another issue we have looked at is whether there is any sign of
differing behavior of abundance ratios between the cool giants and
the hotter turnoff region stars.  While in most cases, no such 
effect can be detected, a clear separation is seen for [Si/Fe]
as a function of [Fe/H]. For the vast majority of the stars,
only the 3905 and/or 4102~\AA\ lines of Si~I could
be detected.  As shown in Fig.~\ref{figure_4plot2}, we
find  that 
the hot dwarfs have [Si/Fe] approximately the Solar value,
which is consistently
lower by $\sim$0.4~dex than the values typical of the cool giants.
This trend was already been noticed in our earlier work
\citep[see, e.g. {\S}7.6 of][]{cohen04},
see also \cite{preston06} and, for the First Stars project,
\cite{first_stars_xii}.
\cite{shi09} presented non-LTE
corrections which they suggested might reduce the discrepancy.
Subsequently \cite{zhang11} demonstrated that with these non-LTE corrections
for the two strong Si~I lines they could largely eliminate the \teff\
dependence, and that the hot turnoff stars have positive non-LTE
corrections of $\sim$ +0.25~dex, making the ratio [Si/Fe]
somewhat above the Solar value in EMP stars.

A smaller such trend appears
to be present for the ratio [Co/Fe] vs [Fe/H], with the dwarfs
having this abundance ratio consistently higher by $\sim$0.2~dex
(see Fig.~\ref{figure_4plot3}). Still smaller trends of
separation in abundance ratios at a given [Fe/H] between
dwarfs and giants,
to be discussed later, appear to be present.

Since some of our HIRES spectra extend far into the UV at reasonable
SNR, another issue of interest is to compare the abundances
deduced from features in the UV versus in the commonly utilized
optical features.  Fe~II has 3 strong lines near 3260~\AA\ which
are stronger than those at optical wavelengths.  We find
that for the 8 C-normal stars in which we could determine
reliable \eqw\ for at least two of the three lines, 
the mean difference in deduced Fe abundance is only 0.09~dex,
with $\sigma$ about the mean of only 0.13~dex.  This is very
good agreement for such a difficult measurement.

The Mn~II triplet near 3470~\AA\ was picked up in 23 C-normal stars in
our sample. These UV lines with $\chi \sim 1.8$~eV give a Mn abundance 
$\sim$0.3~dex 
(with $\sigma = 0.14$~dex)  higher
than the 4030, 4033~\AA\ blue lines, which are the strongest optical lines,
and the only ones usually detected
in most EMP stars. (The third line of this Mn~I triplet at 4034.5~\AA\
is badly blended and was not used.)  Ignoring two major outliers and one
star with a rather noisy spectrum at 3470~\AA\ (HE2339$-$5105) for which
only one of the three Mn~II UV lines has a measured \eqw,
a fit to the 20 remaining C-normal stars finds [Mn~II/Fe~II] 
at [Fe/H] $\sim -3.3$~dex is  $-$0.46~dex,
with a small, negative, and uncertain slope.
This value is
0.30~dex higher than that for the neutral species
as determined from the 4030, 4033~\AA\ resonance lines.
Since the bulk of Mn atoms
are singly ionized at the relevant temperatures,
 we therefore infer
that the Mn~II abundances are probably correct, and
that there are problems (presumably non-LTE effects,
as suggested by a non-LTE calculation for Mn by Bergemann \& Gehren 2008) 
with the analysis for the resonance Mn~I 4030~\AA\ lines.

\cite{cayrel04} suggested that Mn abundances derived from
the 4030~\AA\ Mn~I resonance triplet
require a correction of $\sim$+0.3~dex
compared to those deduced from the redder and much weaker
subordinate Mn~I lines.
In principle our spectra could be used to check this.
However, the subordinate optical Mn~I lines are so weak that we cannot
ensure that their measured \eqw\ are
sufficiently accurate.  This test is best done with a sample of
somewhat higher [Fe/H] and with brighter stars
than that presented here.  If this value 
for the offset holds,
the [Mn/Fe] ratio inferred from the 4030~\AA\ resonance lines of Mn~I
would then be quite close (within 0.1~dex) to that from the Mn~II UV
lines.

Table~\ref{table_lines} presents the list of lines
we have adopted with their atomic parameters, and  the number of
stars in which each line was detected.  Hyperfine structure (HFS) corrections
were derived for Sc~II, Mn~I, Mn~II, Co~I, Cu~I, Ba~II, La~II,
and Eu~II as necessary by comparing for each star the abundance of a species 
derived using a multi-component line list for each detected line 
with that derived from the equivalent widths of each line using the closest
model atmosphere for that specific star; see 
\cite{cohen08} and references therein for the sources of the HFS patterns.
The measured equivalent widths are given in Table~\ref{table_eqw};
negative values denote upper limits. 
Table~\ref{table_solar_abund}
gives our adopted Solar abundances, which are largely from 
\cite{grevesse98} with slight modifications to those of C, O, and Fe.
Our derived abundances are given in Table~\ref{table_abund_merge_1to10}.
Upper limits are indicated by setting $N$(lines) negative.
As indicated
in \S\ref{section_intro}, we have already
published analyses for  74 of these stars previously;
the results presented here supersede the earlier ones.
Note that we assume log[$\epsilon$(Fe)] for the Sun
is 7.45~dex, somewhat lower than the value adopted by many recent studies,
typically 7.52~dex.  Hence  our derived [Fe/H]
 will be higher
by $\sim$0.07~dex  and
our derived  abundance ratios [X/Fe] will be lower by this amount
 than those of many other groups.

The lowest [Fe/H] star found in the present sample of EMP candidates
from the HES is HE1424$-$0241, a giant at $-4.1$~dex. There are
only four reasonably well studied stars known at significantly 
lower metallicity.  They are
HE0107$-$5240 \citep{christlieb02} and HE1347$-$2326 \citep{frebel05}
at [Fe/H] $\leq -5$~dex, HE0557$-$4840
at $-4.8$~dex \citep{norris07},  
and the recently discovered case of SDSSJ102915+172027 at [Fe/H] $-4.8$~dex
\citep{caffau11}.
Each of these four UMP or HMP stars has been the subject
of numerous publications, and we exclude discussion of them
from the present work.

\subsection{Tailored Model Atmospheres \label{section_tailored} }

As we will see later, some of the 0Z sample stars show peculiar
abundance ratios.  We picked a subset of 5 stars  exploring
the range of abnormalities we found,
including the most metal-poor C-normal
star, HE1424$-$0241, with very low Si-abundance,
HE0926$-$0546 (a very low metallicity giant with no
$\alpha$-enhancement)
and several extremely C-enhanced stars
(HE1012$-$1540 and HE2323$-$0256).  For each of these
stars, we computed a plane-parallel model atmosphere
in LTE using ATLAS9 \citep{kurucz93} 
with an elemental composition constructed from
the first iteration of our detailed abundance analysis. 
In the absence of a measured abundance for an element
in a given star, the scaled solar abundance was adopted.
In conformance
with our more standard model atmospheres, convective
overshooting was turned off in the customized models, and
the mixing length parameter ($l/H$) was 1.25.  This treatment
of convection is the same as in the often-used model
atmosphere grid computed by \citet{castelli04}.  We also
adopted the same atomic and molecular line lists for the
calculations of the opacity distribution functions as
\citet{castelli04}.  The opacities were recomputed for
the detailed abundance pattern of each star, then a
model atmosphere tailored  for that specific star was generated.

We compared the results
from scaled Solar composition models in the standard Kurucz grids \citep{kurucz93}
(our normal mode of analysis)
with those when the tailored model atmospheres were used, 
in both cases with our adopted stellar parameters.
It should be noted that our normal abundance analysis
interpolates from the nearest (in \teff, log($g$),
and [Fe/H]) models in the standard Kurucz grid, while the
tailored models were computed for the adopted
stellar parameters, hence no interpolation was required.

The maximum difference
between [Fe/H] as obtained from Fe~I and Fe~II lines
at optical wavelengths was only 0.08~dex.  This corresponds
to the maximum change in the Fe ionization equilibrium between
the two solutions, one with scaled Solar abundances, and one
with tailored abundances. The neutral ions show differences
in [X/H]
within 0.03~dex of that for Fe~I, while the ionized
species tend to follow Fe~II.  The total range of the differences
in [X/H]
for the absolute abundances deduced for each element considering
all species with detected absorption lines between the scaled
solar and the tailored model atmospheres
was  0.07~dex for HE1424$-$0241, 0.09~dex for the very C-rich star HE2323$-$0256,
0.12~dex for HE1012$-$1540, 0.06~dex for HE0926$-$0546, and
0.15~dex for HE0533$-$5340 (among the coolest EMP giants in our sample).

The changes in abundance ratios are smaller, assuming
[X/Fe:Fe~I] for neutral species and [X/Fe:Fe~II] for ionized
species are used.  We find from these tests that
for stars with peculiar chemical
inventories, an additional uncertainty in their abundance ratios
of up to 0.05~dex may exist arising from this effect. Furthermore, again only
for stars with severe abundance peculiarities with respect
to the bulk of the large sample studied here, we may have introduced
small additional errors by trying to slightly adjust the
stellar parameters of the star to improve (at least as we perceive it)
the results of the detailed abundance analyses.

\cite{vandenberg12} have recently presented a grid of stellar 
evolutionary tracks 
and isochrones for stars with enhanced abundances of one of a set of
specific elements.  He points out that
C, N, and O do not contribute significantly
to opacities at low temperatures and pressures,
while Mg and Si do.  So it is not surprising that 
some of the stars checked with tailored model atmospheres
that are very C and/or N enhanced do not show large
differences in their deduced abundances from detailed
abundance analyses using our normal interpolation scheme
with scaled solar composition models.  \cite{vandenberg12} notes
that it is variations for Mg and Si
among the heavy elements that have the most impact on
the opacity, but fortunately the bulk of the stars in
our sample have about the same value of [Mg/Fe] and of [Si/Fe],
consistent with the normal $\alpha$-enhancement
seen among EMP stars.  As we will see later in
\S\ref{section_out},
only a few are highly deviant for Mg
and/or Si, and most of them
are depleted in these elements rather than enhanced,
at which point in such metal poor stars, most of the electrons are from
ionization of H.
These deviant stars were all rejected
as outliers in determining the linear fits of [X/Fe] vs [Fe/H].

\subsection{Completeness of the High Resolution Observations}

There are 12 candidates for which 
[Fe/H](HES) $< -3.5$~dex from either 
the 2005 merge of P200 and Magellan follow up spectra or from that
carried out in the spring of 2007.  All of these
have been observed with MIKE or HIRES.  A comparison of
[Fe/H](HES) vs that obtained from a detailed abundance
analysis is given in Table~\ref{table_below-3.5}.  Two of these twelve turned
out to be M dwarfs with strong emission in the core of the 3933~\AA\  line
and one is a QSO.

22 of the 28 EMP candidate stars with 
$-3.2 <$ [Fe/H](HES) $< -3.5$~dex from
the June 2007 merge of the Magellan and Palomar mod-res spectra
using [Fe/H](HES) derived at that time have been observed
at high spectral resolution (see Table~\ref{table_3.2to3.5}).
Four of the six without a high resolution spectrum
are extremely C-enhanced; these four probably have true values of
[Fe/H] higher than $-3.2$~dex.

A perusal of Tables~\ref{table_below-3.5} and Table~\ref{table_3.2to3.5}
demonstrates the prevalence of the underestimation of [Fe/H](HES) in
extremely C-rich stars discussed in \cite{feh_cstars}.  It is also
interesting to note that some of the higher [Fe/H] interlopers in these
two groups have abnormally low [Ca/Fe] ratios, which is probably
why their [Fe/H](HES) values are significantly lower than their
[Fe/H](HIRES) values.

A large number of stars with $-3.2 <$ [Fe/H](HES) $< -2.8$~dex from
the June 2007 P200+Magellan merge have also been
observed with either MIKE or HIRES.

\section{Comparison with Previous Abundance Analyses \label{section_comp} }

The abundance analyses we present here for EMP candidates from the
HES were, as indicated above, carried out using the same method
of determination of the stellar parameters, master line list,
transition probabilities as well as other atomic parameters,
and abundance analysis code for all the stars with the exception
of 9 of the stars from the Keck Pilot Project for which subsequent
repeat observations were not obtained.
Here we compare our results from detailed abundance analyses
to the preliminary [Fe/H] determined from our earlier moderate
resolution followup spectra for each candidate and to
detailed abundance analyses carried out by other major groups
for the small number of stars previously analyzed.  The goal  
from these comparisons is  to establish that the work
of the 0Z project is sufficiently robust that we are able to
determine the trends among abundance ratios, and that stars
we call outliers really are outliers from these mean trends.
Note again that we assume log[$\epsilon$(Fe)] for the Sun
is 7.45~dex, somewhat lower than the value adopted by many recent studies,
typically 7.52~dex.  Hence our derived [Fe/H] 
 will be higher
by $\sim$0.07~dex while our abundance ratios [X/Fe] will
be  $\sim$0.07~dex lower than those of many other groups.

\subsection{Comparison of the Fe-Metallicities from the HES vs 
High Resolution Spectra}

We compare the Fe-metallicities derived from our moderate dispersion
follow up spectra using a simple algorithm based
on the strength of the 3933~\AA\ Ca~II line plus 
H$\delta$ with those from our high resolution spectra and a detailed
abundance analysis.  The former was the initial basis for similar
algorithms used by many recent large surveys of stellar spectra,
see e.g. \cite{segue1} describing the pipeline developed for the SEGUE SDSS-II.
Such a comparison is therefore quite illuminating.  One must
however remember that these  codes
have been extensively improved over the past decade
and their current versions are much more
sophisticated, yielding better performance than the early versions
available to us in the initial phases of this project.

A comparison of the metallicities from the 2007 HES algorithm with those
found from our entire sample of high resolution spectroscopy 
(\nstars~stars), ignoring the extreme C-stars for
which the algorithm fails  \citep{feh_cstars}
and the 2007 version does not return a value, 
shows that for the 35 dwarfs
near the main sequence turnoff (i.e. stars with $T_{eff} > 6000$~K),
the HES algorithm underestimates the metallicity by 0.27~dex
in the mean (with $\sigma = 0.36$~dex).  The EMP main sequence turnoff
region dwarfs have very weak lines, so it is perhaps not surprising
that moderate resolution spectra underestimate their metal abundances.
Presumably recent updates in the calibration of the metallicity algorithm
improve this situation.  
For the 87 cooler subgiants and giants,
we find the mean difference is small (0.07~dex), but the dispersion is
even larger, 0.44~dex.  These differences for the full sample
are shown in Fig.~\ref{figure_feh_compare}.


\subsection{Comparison with the First Stars Project}

We have compared our results for stars in common with 
a number of other major projects focusing on EMP Galactic halo
field stars, in particular the First Stars large project
at the VLT \citep{cayrel04}, in some of our earlier publications, see
e.g. \S{7} and Appendix B of \cite{cohen08}.
First we note that the sample for their project, by design, 
is deliberately strongly biased against C-rich stars. There we found
that there is a $\sim$0.2~dex offset in the absolute abundance
scale between our 0Z  project and the First Stars project.
Some of the factors giving rise to this are discussed in detail
in \cite{cohen04}.  
We have now found one more term contributing to this systematic difference
in derived [Fe/H].  With the use of the 2010 version of MOOG with
a better treatment of isotropic scattering \citep{sobeck_moog},
we find that for the most metal-poor stars in our sample,
[Fe/H] is decreased by 0.1 to 0.15~dex as compared to the 2002 version
of MOOG \citep{moog}. The abundance analysis code used by the
First Stars project already contained a suitable treatment of scattering
in 2004.  
This improvement, together with the factors discussed
in our earlier papers, helps to explain the difference in [Fe/H] between
the First Stars results and our earlier results for these two stars.

There are only two stars in common.
HE2323$-$0256 (aka CS 22949$-$037) was 
discovered by \cite{mcwilliam95a}.  This star was 
subsequently reobserved at higher SNR and analyzed by
\cite{first_stars_2} and also by \cite{cayrel04}.
Comparing our current analysis of the spectrum of  HE2323$-$0256 
with that of \cite{cayrel04}, we first note that our stellar
parameter assignment is essentially identical as that of the First
Stars project,
4900~$\pm$125~K for \teff\ vs 4915~$\pm{15}$~K, 
\grav 1.5 vs 1.7~dex, and $v_t$ 1.8 vs 1.9 \kms\ (First Stars vs 0Z).
Our 0Z analysis which now uses
the updated version of MOOG \citep{sobeck_moog} with a tailored
model atmosphere yields log($\epsilon$) for Fe~I (Fe~II) of 3.52
(3.59)~dex, while that of the  First Stars project is
3.51 (3.56)~dex, which is remarkably good agreement between
the two completely independent observations and analyses.
The abundance ratios
[X/Fe] for the 15 species in common also agree well, with 10 of these
having differences $| \Delta | ~ \leq ~ 0.15$~dex.
The agreement for the second star in common, BS~16467--062, is not as good.
Our \teff\ is 164~K higher than that of the First Stars project
\citep{cayrel04}, in part because of the difference in the
adopted interstellar absorption along the line-of-sight to this star.
This difference in stellar parameters leads to a 0.3~dex difference
in [Fe/H], with our value being higher.  If we had adopted the \teff\
used by the First Stars project, the disagreement in [Fe/H] would
become much smaller.
Comparing derived abundance ratios [X/Fe], the agreement even for
BS~16467--062
 is fairly good; 9 of  the 13
species in common have $| \Delta | ~ \leq ~ 0.15$~dex.


\subsection{Comparison with Other Groups}

HE1300+0157 was the only star found by the HERES project
\citep{barklem05} to have [Fe/H] $< -3.5$~dex.  We observed this
star with HIRES in April 2006.  A detailed
abundance analysis based on high SNR HDS/Subaru spectra was carried out by 
\cite{frebel07}.
Our adopted \teff\ is 100~K hotter than that of \cite{frebel07}
(a difference no larger
than the uncertainties), and thus
our [Fe/H] is higher as well.  The agreement between our results
and those of \cite{frebel07} for
the abundance ratios [X/Fe] for this star is excellent, with the
difference in abundance ratios exceeding 0.12~dex for only three of
the 15 species in common.

\cite{johnson02} published abundances for the rare earths
and a few other elements in the carbon star HE0058$-$0244
(aka CS22183--015).  The \teff\ are very different, that
of the 0Z project is 420~K higher than that of \cite{johnson02}.
Given that they focused on the rare earths, each of which
only has a few relatively weak lines, and that this is a 
rather crowded spectrum since the star is a carbon star, 
the agreement between the two independent analyses 
for the derived abundance ratios is reasonable.

There are two stars in common with the recent study of
Norris and collaborators \citep{norris12a,yong13},
HE1346$-$0427 and BS16545$-$089.  The agreement
for all the stellar parameters and for the deduced
log[$\epsilon$(Fe~I)] is very good, well within the
uncertainties, except for \teff\ for the second star,
where our value is somewhat lower than that adopted
by them; this difference appears to arise, at least in part, 
from differences
in the adopted photometry.  They adopt the Solar Fe abundance
of \cite{asplund09}, which is 0.07~dex higher than the one
we use.  Their Fig.~3c  compares their set of $W_{\lambda}$ from 
a 2008 HIRES/Keck spectrum with ours from a 2002 spectrum;
there is a small offset of $\sim$3~m\AA\ with very small scatter.

Some of the brighter well-studied metal-poor stars observed
as part of the Keck Pilot Project
(e.g. BD 3--740, G139-8) have detailed abundance 
analyses in the literature. 
As was discussed in \cite{keck_pilot_1}
and in \cite{keck_pilot_2}, our results are in good overall agreement in
these cases as well.

\section{Abundance Ratio Outliers \label{section_out} }

After all the abundance analyses were completed, we constructed
plots of [X/Fe] vs [Fe/H] (Figs.~\ref{figure_4plot1}
to \ref{figure_4plot5}) 
based on
the data in Table~\ref{table_abund_merge_1to10}.  
We then selected from these plots the strong (i.e. very discrepant)
outliers, either
high or low, for verification of of their outlier status.
Each of these was checked in detail, looking for mistakes, 
inspecting the original spectra,
remeasuring the equivalent widths of selected lines
again if necessary, eliminating specific lines known to be blends,
etc.  This resulted in one of several outcomes: a correction to [X/Fe]
such that the star is no longer an outlier\footnote{To avoid confusion,
such stars are not specifically identified in the figures.},
a verification of the outlier
status (indicated by ``V'' in the relevant figures for that species) 
or a decision that the data in hand
were not adequate to determine whether or not the outlier status
is correct; such cases are indicated by ``?'' in these figures.  When possible,
additional spectra were obtained to resolve the uncertain cases,
either covering a wider spectral range or of higher SNR.

In the early stages of the 0Z project, we believed that several C-stars
were strong outliers for [Ca/Fe]. 
Our determination of the  Ca abundance  
turned out to be problematic
in those  very C-rich stars whose spectra were obtained
prior to HIRES detector upgrade in mid-2004  and thus
included only a limited wavelength range.  In
an effort to derive Ca abundances from these early HIRES spectra, we ended up using
lines which were crowded/blended, presumably by molecular
features. This was only realized
fairly recently \citep{cohen12} when we obtained additional C-star HIRES spectra extending
out to 8000~\AA\ which covered key
 Ca~I lines in the 6160~\AA\ region.  The 6160~\AA\ region is quite clean; 
 it is not afflicted
by CH, CN, or C$_2$ bands.  We found much lower 
(up to 0.75~dex lower) Ca abundances from the additional
Ca lines in these C-stars than those we previously published in \cite{cohen06}.
Fortunately our earlier paper does not focus at all on the Ca abundances
in the EMP C-stars.

\subsection{Linear Fits to Abundance Ratios \label{section_fits} }

Based on the extensive previous work in this field,
see, e.g. the discussion of toy models in \S6 of \cite{cohen_draco},
we chose to  approximate
the behavior of abundance ratios vs [Fe/H] as a linear relationship
in the extremely metal-poor regime of our HES sample. 
We therefore fit lines to the abundance ratios [X/Fe] vs [Fe/H] 
where there is adequate data for species X.  
In order to demonstrate the slope of
the line, Table~\ref{table_fits} gives the evaluations of the line equation
for each species X at [Fe/H] $= -3.5$ and $-3.0$~dex.  For titanium, the line
was fit to [Ti12/Fe12], which denotes the mean of [TiI/FeI] and [TiII/FeII]. 
The fit to [BaII/FeII] is not as well determined as some of the others
because of the many upper limits for the Ba abundance in the 0Z sample,
which were ignored in constructing this fit.
To at least partially take this into account, a second fit for
[Ba/Fe] vs [Fe/H] was carried out assuming that the upper limits
for [Ba/Fe] in stars with \teff $< 6000$~K are actually detections.  This
increases the sample size and slightly lowers the fit value at low
Fe-metallicities.
 The table also gives the mean and rms dispersion
for [Fe/H](FeI) $-$  [Fe/H](FeII), and the same for the more temperature dependent
Ti.  As noted earlier, the means of 0.0 independent of Fe-metallicity
and the low $\sigma$, especially for Fe, where $\sigma = 0.10$~dex, support the validity
of our procedures and abundance analyses.

These fits 
exclude C-stars and also a small number of  strong outliers 
(either high or low)
as indicated in the last column of Table~\ref{table_fits}.  Only stars with
[Fe/H] $< -2.5$~dex are included.
The histogram of deviations 
from the linear fit is shown for the set of stars from which each
fit was constructed for  Mg, Ca, Cr, and Sr in Fig.~\ref{figure_fit_4plot}.
The fit for each of [Mg/Fe] and  [Ca/Fe] vs [Fe/H] is 
approximately a constant ratio, [Mg/Fe] $\sim$ +0.40~dex,
[Ca/Fe] $\sim$  +0.27~dex.


Our linear fits to [X/Fe] vs [Fe/H] given in Table~\ref{table_fits}
have been compared with those of \cite{cayrel04}, both evaluated at 
[Fe/H] $= -3.0$~dex.  Overall the agreement
is very good.  With the correction of +0.15~dex for [Mg/Fe] advocated by
\cite{first_stars_xii} to the results of \cite{cayrel04}, 
7 of the 10 fits for species in common agree to within 0.05~dex.
The fits for [Si/Fe] evaluated at [Fe/H] $-3.0$~dex
differ by 0.08~dex, but this fit shows a strong dependence
on \teff, discussed in \S\ref{section_other_anal};
see also Fig.~\ref{figure_4plot2}. The fits for [Ca/Fe] differ by only 0.06 dex at the
nominal value of [Fe/H], which is within the uncertainty of
the comparison. The
largest difference is only 0.09~dex (and that is the fit
for [Zn/Fe], an element with at most two detectable weak lines
for EMP star spectra in the optical wavelength regime).
The dispersions around the fits
from the First Stars project of \cite{cayrel04} are somewhat smaller
than ours in most cases, perhaps because their sample is more
homogeneous (in terms of \teff\ and \grav), and is
considerably brighter in the mean, thus making it easier to get
high SNR high spectral resolution spectra.

The strong outliers from the
mean relations of [X/Fe] vs [Fe/H], defined as those stars deviant by
$\geq$5$\sigma$,
are listed in Table~\ref{table_out_fit}.  Given the very large spread
in the abundance ratios of the heavy neutron capture elements Sr and Ba,
we require deviations of 10$\sigma$ for outlier status for these two species.
Ignoring the C-stars, this leads to 
approximately 15\% of the sample being strong
outliers in one or more elements between Mg and Ni.  This rises to
$\sim${19\%} if very strong (${\geq}10{\sigma}$) outliers for Sr and Ba are included.
There
is no question that they are statistically significantly
different from the bulk of the population, and are not in the
far wings of some smooth Gaussian distribution of errors.  For our sample
size, there should only be 1 star more deviant than 2.5$\sigma$, where $\sigma$
is the total observational uncertainty given in Table~\ref{table_unc_dwarfs}
(for dwarfs) or \ref{table_unc_ratio_giants} (for giants), off the
mean distribution in such relations for any given species.

As was first shown by \cite{mcwilliam95b}, several of 
these linear fits have a statistically significant slope
(e.g. [Cr/Fe], [Mn/Fe], [Sr,Fe], [Ba/Fe] etc.).  They
begin with a large positive or negative value at the lowest [Fe/H], then approach
[X/Fe] = 0.0 at some [Fe/H] generally near $-2.5$~dex.  We assume
that once the linear fit reaches the value of 0.0~dex, it remains
at 0.0 for all higher [Fe/H].   This modification only affects the evaluation
the deviation from the linear fits for strong outliers in [Sr/Fe] and in [Ba/Fe] at
metallicities near the high end of the 0Z sample, i.e. those higher than  $-2.5$~dex.

There are several stars which are very deviant low outliers in [Ba/Fe].  One of these
(HE0305$-$5442) has an upper limit to [Ba/H] which is 
0.3~dex lower than that of the previous record holder for the lowest known [Ba/H],
Draco~119 \citep{draco119}, which also has only an upper limit.

We compare the dispersion we measured around the linear fits
for the restricted sample of stars used to
construct each fit  
for various elements with those based on the predicted 
uncertainties in Table~\ref{table_fits_sigma}.
In doing so,  we have assumed that the uncertainty in the [Fe/H] values
does not contribute significantly to that of
the abundance ratios [X/Fe], which is not always valid.  Hence there
is another term contributing to the dispersion around the linear
fits, which can be approximated as
$\sigma(x) { {dy} \over {dx} }$, where $x$ is [Fe/H] and
$y$ is [X/Fe].  For those species with significant slopes
for the linear fits given in  Table~\ref{table_fits}, this term
will increase the observed dispersion about the linear fits
compared to the one predicted based solely the analysis
and measurement errors in [X/Fe].  This extra term only
affects those elements with the largest slopes, which
 are, as shown in  Table~\ref{table_fits},  Sr~II and Ba~II;
it raises the expected dispersions of 
[Sr/Fe] and of [Ba/Fe] from 0.12 and 0.10~dex respectively
to 0.16 and 0.12~dex respectively.  We may have slightly underestimated
the increase in the predicted dispersion for [Ba/Fe] due to the
many upper limits in the Ba abundance among our sample stars.

Such a comparison of the predicted and actual deviations around
the linear fits to [X/Fe] vs [Fe/H] for species X, shown
in Table~\ref{table_fits_sigma}, suggests that, ignoring
major outliers,
there is no detectable range in [X/Fe] vs [Fe/H] that is larger
than the uncertainties for Si, Sc, and Ni.  The ratio of that
measured (see Table~\ref{table_fits}) to that predicted
(see Table~\ref{table_unc_dwarfs} for dwarfs and Table~\ref{table_unc_ratio_giants}
for giants, ranges from 1.1 to 1.3 for these three elements
(see Table~\ref{table_fits_sigma}).
The ratios for Mg, Ti, Cr, and Mn,
are only slightly higher, 1.6 to 1.9.  They are still
larger for Al (2.7), and Ca (2.1), while the heavy
neutron capture elements Sr and Ba have the largest ratios ($\sim$4), hence
the highest intrinsic dispersion, of all the species tested.



\subsection{Low [Ca/Fe] Stars \label{section_lowca} }

A substantial number of stars with [Ca/Fe] below the Solar ratio
(the ``low-Ca'' stars)
are found in the present sample,  a phenomenon that is
virtually non-existent among solar neighborhood stars and
Galactic disk stars, which are considerably closer and with higher
metallicities than the 0Z sample.  Low -Ca stars are also never seen
among Galactic globular cluster stars, which are a population with characteristics
closer to our 0Z sample. 
As is shown in Fig.~\ref{figure_4plot2},
there are 10 C-normal stars, as well as one $s$-process enhanced C-star
with a high [Fe/H] (HE2353$-$1758) which we ignore, with [Ca/Fe] below the Solar ratio.
Table~\ref{table_fits} gives the mean value of [Ca/Fe] for 
C-normal stars in our sample as +0.27~dex independent of [Fe/H].
Note that there are no C-normal stars with [Ca/Fe] $> +0.54$~dex,
which would be the comparable requirement for ``high-Ca'' stars.

The   minimum value for [Ca/Fe] in our sample is
$-0.6$~dex for HE1424$-$0241, an extremely peculiar star
described in detail in \cite{cohen_1424}.  
However, only one of the low-Ca stars
has [Mg/Fe] below the Solar ratio  (see Fig.~\ref{figure_4plot1}), while HE1424$-$0241 has the 
normal EMP halo star
value of +0.44~dex, and only HE1424$-$0241 has [Si/Fe] $< -0.07$~dex
(with an extremely low value of $-1.0$~dex).

Figure~\ref{figure_family_lowca} displays the chemical inventory
of the family of 10 stars (all giants\footnote{One star of these 10 low-Ca
stars has \teff\ 2~K higher
than the upper limit we allow for giants.}  from our sample with [Ca/Fe] $< 0$~dex.
The vertical axis is $\delta$[X/Fe]/$\sigma$,
where the numerator of this expression is the deviation from the linear
fit to [X/Fe] vs [Fe/H] at the Fe-metallicity of the star.  The denominator
is the  predicted observational uncertainty given in
Tables~\ref{table_unc_dwarfs} and \ref{table_unc_ratio_giants}, with
$\sigma$ is set to 0.2~dex for C (from the G band of CH) and N from the 3360~\AA\
band of NH. (The $\sigma$ values used for Mg, Si, and Ca for giants are
0.09, 0.16, and 0.11~dex respectively.)
Low [Ca/Fe] is associated with low [Mg,Si/Fe], but the
relative depletion of Mg and of Si (with the exception of one very anomalous
star with very deficient Si) measured in units of $\sigma$ is 
less than half the size of the Ca deficiency.   If we attempt to compare
the number of low outliers for Mg with those for Ca, 
there are three stars with [Mg/Fe] $< 0$~dex, of which
the lowest  [Mg/Fe] is $-0.24$~dex (HE0533$-$5340),
while the lowest [Ca/Fe] is $-0.57$~dex (HE1424$-$0241). However, 
since the mean value for
[Mg/Fe] is 0.10~dex higher than that of [Ca/Fe], perhaps the relevant number for
comparison is that there are 12 stars with [Mg/Fe] $< +0.10$~dex, one of
which is a C-star, but only four of them are among those with sub-solar  [Ca/Fe].

Sub-solar  ratios of [Ca/Fe] are also clearly
associated with abnormally low [Sr/Fe] and [Ba/Fe] 
(see Figure~\ref{figure_family_lowca} and also Fig.~\ref{figure_familylowca_sr}).
The low-Ca stars tend to lie along the lower envelope
of the [Sr/Fe] and [Ba/Fe] vs [Fe/H] distributions shown in the 2 upper panels
of Fig.~\ref{figure_4plot5}, but not all the anomalously
low neutron-capture element stars have low [Ca/Fe].  A check of
the two stars with the lowest [SrII/FeII] in our sample (see 
Fig.~\ref{figure_4plot5}),
neither of which has low-Ca, does not show any obvious anomalies among the
lighter elements. A few of the low-Ca stars
have high [Mn/Fe] for their Fe-metallicity.  \cite{honda11}
have found a star (BS16920$-$016) with 
unusually high [Zn/Fe], close to normal [Mg/Fe],
slightly low [Si/Fe] and [Ca/Fe],
although not subsolar [Ca/Fe], and  high
[Mn,N/Fe], accompanied by a deficiency of the heavy neutron-capture
elements, which may be related to this group of 10 low-Ca EMP stars
we have isolated from our sample.

Differential analyses of large samples of
solar-neighborhood dwarfs within a small range 
of \teff\ in the thin disk of the Galaxy as compared
to stars in the thick disk such as those of \cite{edvard93},
\cite{bensby05}, \cite{reddy06}, and \cite{nissen1} have been able to achieve
very high precision.  These surveys have
demonstrated that the trends of [X/Fe] versus [Fe/H] are 
close but not identical
between the various disk stellar populations of the Galaxy; the trends
of the various disk components  are separated by not more than 0.15~dex,
a small difference not easy  to detect in the present sample.  However,
in all such studies to date,
sub-solar values of the $\alpha$-element abundance ratios (i.e.
of [Mg, Si, or Ca/Fe]) are {\it{never}} seen.  This also holds for the
First Stars survey of EMP giants \citep{cayrel04}, whose sample
is in the mean $\sim$2.5~mag brighter than ours; there are no
stars with subsolar [Mg, Si, or Ca/Fe] in the First Stars survey;
there are also no stars with 
sub-solar [Ca/Fe] in the pioneering
study of \cite{mcwilliam95b}.
%

A few moderately metal-poor halo field stars have
been found that appear to be $\alpha$-poor. \cite{fulbright02}
suggested that  lower [$\alpha$/Fe] stars are found among those with high
space velocities with respect to the local standard of rest, while
\cite{stephens02} suggest such stars are associated with the outer halo.
The most extreme $\alpha$-poor stars, including that found by
\cite{carney97}, were reviewed by \cite{ivans}.  These
stars show depletions of Na, Mg, Al, Si, Ca and Sc with respect to Fe,
with Sr, Y, and Ba also anomalously low.
The explanation offered by \cite{ivans} is that
their
chemical inventory has a composite origin, with SNIa contributing
a substantial fraction ($> 1/3$) of their chemical inventory, and SNII 
the rest. We calculate a rough guess of this value for the Sun,
assuming that the SNII contribution has the mean [Mg/Fe] of our
0Z sample, +0.37~dex, and that SNIa contribute Fe, but no Mg;
the result is that SNIa contribute ${\sim}30$\% of the
Sun's total chemical inventory.  Given that,
how such a substantial enhancement
of the relative contribution of SNIa at such low metallicities as
are being considered among the 0Z stars could
occur is quite unclear.  

%

Only the prompt SNIa contribution is relevant for the chemical inventory
of EMP stars.
What distinguishes the prompt SNIa from the more numerous delayed
SNIa is not clear \citep{greggio08}. 
Initial calculations of 
SNIa nucleosynthesis yields were carried out by \cite{nomoto84}.  The
set of calculations often used today are those of \cite{iwamoto99},
with more recent multi-dimensional calculations by \cite{travaglio04b}.
These strongly suggest that Mg, given that Mg is produced
almost entirely by SNII, should be even more depleted than Ca.  However,
this is not the case among the EMP stars studied here, although
that does hold, at least roughly, for the sample collected by \cite{ivans}.

If one tries to use the difference between hydrostatic and
explosive $\alpha$-burning nucleosynthesis invoked in \cite{cohen_1424}, this
too fails, as while Mg is produced in hydrostatic $\alpha$-burning,
both Si and Ca are produced by explosive $\alpha$-burning, but
[Si/Fe] is closer to normal in all but one of these stars.
Any proposed explanation for the low-Ca EMP stars must also address the 
accompanying depletion of the
heavy neutron capture elements.

Slightly sub-solar [Ca/Fe] is found in 
the stellar populations of several of the dSph satellites of the Milky Way,
but only 
at the high Fe-metallicity end of their abundance distributions, 
and also in those Galactic globular clusters Pal~12 \citep{cohen_pal12}  
and Rup~106 \citep{brown97}
believed to have originally  been part of
the Sgr dSph galaxy, again with [Fe/H] considerably above that characteristic 
of our 0Z sample.
Examples of this are shown in \cite{tolstoy09} and in \cite{kirby11}.
These stars, however, show the typical SNIa pattern of depleted
$\alpha$-elements, reaching down to $-0.3$~dex for [Ca/Fe] and $-$0.2~dex for
[Mg/Fe] in Fornax \citep[see, e.g. fig.~11 of][]{tolstoy09},  and have [Fe/H] $> ~ -1.1$~dex.  
They also have
highly enhanced [Ba/Fe], presumably arising from the $s$-process
in AGB stars, through which the SNIa progenitors must
pass prior to reaching the white dwarf regime. 

\cite{tolstoy09} suggest that the last
stage of chemical evolution in Fornax 
accompanies a dwindling star formation rate (SFR) before
star formation ceases altogether.  When the SFR is very low, the ratio
of Type Ia to Type II SNe is large, producing 
sub-solar ratios for [$\alpha$/Fe].
AGB stars within the aging population will continuously
produce  $s$-process elements,
leading to
highly-enhanced [Ba/Fe],  a characteristic
not manifest in the EMP low-Ca population.

Our view of these various anomalies, and in particular of the 
Ca-poor nature of $\sim$10\% of our 0Z sample of stars, is that
early SNII must have been more diverse in their ejecta abundances than
current models such as those of \cite{woosley95} or
more recently those of \cite{kobayashi06}, based on the
yields of \cite{nomoto06}, or of \cite{tominaga07} or \cite{heger10}
suggest to date if they are the ultimate cause
of what we have observed.  As the above studies of SNII behavior
indicate, the progenitor mass, the 
mass cut of ejection, the fallback,
and other details of the SN can profoundly affect the chemical
composition of the ejected material, and are difficult to model.

Several of the model SNII yield grids suggest that heavier
progenitors produce more O and Mg (i.e. hydrostatic $\alpha$-elements)
with respect to Ca and Ti (explosive $\alpha$-elements) than do
lower mass progenitors.  The models of \cite{nomoto06}
predict a sizable increase (a factor of $\sim$18) in the Mg/Ca ratio in the ejecta
for progenitor masses
increasing from 13 to 30~$M_{\odot}$.
This suggests that a bias of the IMF toward higher mass stars on
the upper main sequence might also be relevant in producing the
low-Ca EMP stars.



\cite{ngc2419} recently demonstrated that a group of stars with
a very peculiar chemical inventory exists in the extreme outer halo
Galactic globular cluster NGC~2419.  The most obvious symptom is a very
large deficiency of Mg (a factor of 5 compared to the bulk of
the cluster stars).  This is accompanied by very high K, somewhat high Si and Ca,
and high Sc.  The Fe-peak elements are constant in abundance for all the
stars within this globular cluster.  This too, while very peculiar,
is not related to the low-Ca population we have isolated within the
0Z EMP halo field stars.


%

\subsection{Other Smaller Outlier Families \label{section_other_out} }

There are two other families of outliers we recognize within our sample.  The first
is a group of two extremely metal-poor stars, both with [Fe/H] $< -3.5$~dex,
and both highly C-enhanced (HE1012$-$1540 and HE2323$-$0256),
although neither shows C$_2$ bands.
Multiple measurements of $v_r$ from high-dispersion spectra exist for
each of these two stars.  Unlike the higher-Fe-metallicity C-stars with
C$_2$ bands and in most cases $s$-process enhancements,
Table~\ref{table_mult_vr} offers no evidence that either  
HE1012$-$1540 or HE2323$-$0256 is a binary.
Their chemical inventory is shown in 
Fig.~\ref{figure_family_highmg}.  It is characterized by 
high C and N, carrying on into high Na, Mg, and/or Al,
with more or less normal abundances for Ca and for the Fe-peak elements.
[Sr/Fe] and [Ba/Fe] are close to the solar value, which is quite
high for such metal-poor stars. 
Although the higher metallicity C-rich stars clearly have C $>$ O,
a tentative detection of O based on two UV OH lines in HE1012$-$1540
suggests that O is also highly enhanced in HE1012$-$1540;
the upper limits from the 6300~\AA\ and 7770~\AA\ lines
are consistent with this value.  Assuming the UV line
equivalent widths are valid, then
$\epsilon$(C)/$\epsilon$(O) $\sim$ 1/2.  
The issue of the oxygen abundance in the extremely low
metallicity C-rich stars requires additional attention
and future observations.
The origin of these  stars is discussed in more detail in
\S\ref{section_cstars}, where it is suggested that they
are the continuation of the classical C-stars with C$_2$ bands
to still lower [Fe/H]; see \cite{norris12b} for a 
different interpretation.

There is also is a small group of stars
with peculiarities in only a small number of Fe-peak elements
exemplified by HE2344$-$2800, shown in Fig.~\ref{figure_family_highcrmn}.
This subgiant  
has high [Cr/Fe] and [Mn/Fe]
(measured from both Mn~I and Mn~II lines), but everything else
is more or less normal.  There are also a number of other stars
which tend to have low heavy-neutron capture abundance ratios
without having any of the peculiarities discussed above.

The families of outliers discussed above include 
all of the major outliers we have found.

\subsection{Mixed Stars}

Among the C-normal stars in our 0Z sample,
we have found only limited signs of internal mixing of material that has
modified the surface composition of the giants in our sample.
\cite{spite_fs9} detected in the First Stars project sample 
the expected variation of the $^{12}$C/$^{13}$C
ratio, which is a very sensitive diagnostic of such mixing,
showing a correlation with [C/Fe] and an anti-correlation with [N/Fe].
The SNR of our spectra in the region of the G band of CH is not 
generally adequate
to reliably measure the weak $^{13}$CH features.

Although not believed to arise from internal mixing,
we see no sign of widespread anti-correlations between Mg and Al or Mg and Na
such as are common among globular cluster stars, as described in the review
by \cite{gratton04}, where various theories for their origin are
discussed. 
Since these
are common in globular cluster stars, they are
presumably present in at least some of those halo field stars which were formerly in
globular clusters.
The cool giant HE2217$-$4053  has high Na 
([Na/Fe] +0.81~dex) and Al relative to Fe, with Mg relatively
low ([Mg/Fe] +0.06~dex), which is the classical pattern seen in such cases.
Two other cool giants, HE2200$-$1946
and HE2339$-$5105, also show unusually high [Na/Fe] but normal [Mg/Fe].
These all have very low E$(B-V)$, less than 0.02~mag, and are quite cool,
so even though the $v_r$ for two of these are close to 0~\kms,
the NaD lines cannot be 
substantially enhanced by interstellar contributions.
Ignoring C-rich stars, only cool giants have [Na/Fe] $> +0.3$~dex.






\section{Carbon  Enhanced EMP Stars \label{section_cstars} }

There are a substantial number of C-rich stars in our sample.
Several subclasses of these stars have been defined by
\cite{beers05}, with further refinements by \cite{aoki10}.
The key discriminants are whether or not the $s$-process elements
(typically La and/or Ba)
are enhanced as well as having [C/Fe] $> 1$~dex (denoted CEMP-s),
whether both the $s$ and $r$-process (i.e. Eu) are enhanced
(CEMP-rs), or whether there is no enhancement of the heavy
elements beyond the Fe peak.  Here we make an additional distinction
between those stars with detectable C$_2$ bands and those
C-rich stars which do not show C$_2$ bands.

\subsection{C-Stars}

The C-rich stars that show bands of C$_2$ (denoted 
in this paper as  ``C-stars'' as an abbreviation for
``classical carbon stars'')
are indicated here
in the set of figures that display abundance ratios 
by a star symbol; the color of the star symbol indicates \teff\, as is true
for each of the symbols used.  There are 27 C-stars in our sample.
Most of them are CEMP-s stars;
as shown in Fig.~\ref{figure_4plot5}, only four of these C-stars do not have
highly enhanced [Ba/Fe], hence are classified as CEMP-no.  These four
stars span the full range in [Fe/H] over which we have found CEMP-s stars, so
within the very limited statistics of our sample there is no difference
in their Fe-metallicity distribution.  Two of these four stars are
binaries with measured periods (see Table~\ref{table_mult_vr}).
One of these four CEMP-no stars
is a dwarf with \teff $\sim 6500$~K.  

There are two stars with very high
[C/Fe] and [Ba/Fe] which do not show C$_2$ bands.  Both of these are
quite hot with \teff $> 6350$~K; the C$_2$ bands presumably are not
detected due to the high \teff.  Here we classify these two stars
as C-rich, but perhaps they should really be considered C-stars.
One of these two is the
CEMP-rs star, HE2148$-$1247, discussed in detail in \cite{cohen03}.

There are more C-stars in our 0Z sample than one might expect if the
fraction of CEMP stars at low metallicity in the Galactic halo
 is about 15\% (estimates in the
literature range from about 10 to 20\% for the HES sample
below $-2$~dex (14\% Cohen et al 2005,
9\% Frebel et al 2006, 21\% Lucatello et al 2006).
As noted in {\S}\ref{section_sample}, C-stars are over-represented
in our sample of candidate EMP stars from the HES as a consequence
of a problem in the initial sample selection
for our high resolution work; see \cite{feh_cstars}.

The CEMP-s stars are by now well understood as 
low-metallicity analogs of the CH and Ba stars
whose characteristics are reviewed by \cite{araa_chstars}.
  We believe
they are binaries where
we see the secondary star, while the  (initially)
more massive primary star has already become a low luminosity white dwarf.  
The third dredge up 
during the thermal pulses of the AGB phase of stellar evolution brings
material to the
surface of the primary which is 
enhanced in C and in the $s$-process elements.  It is
subsequently transferred to the secondary
via a strong stellar wind.  Our radial velocity monitoring for
these CEMP-s stars (see Table~\ref{table_mult_vr} and the discussion
in \S\ref{section_eqw_vr}) confirms that essentially all
of these stars are binaries.

The theoretical basis for this,
and its application to CEMP-s stars, has been explored extensively
in the literature, see e.g. \cite{busso_araa}.  Recent models
of nucleosynthesis in AGB stars including those of 
\cite{bisterzo11} and of \cite{lugaro12}
apply the latest theoretical stellar evolution models varying
the mass and metallicity of the AGB star, the details of the
$^{13}$C pocket within which neutrons can be produced, and the dilution
factor (the ratio of the mass accreted by the secondary star to that of its
outer convective envelope) to reproduce
in detail the chemical inventory of a large number of such stars
with data from the literature.  Subsequent to our
early suggestion of accretion induced collapse
as a source of
$r$-process elements to produce CEMP-rs stars
 outlined in  \cite{cohen03},
there have been a number of additional 
suggestions of how to accomplish this;
see \cite{bisterzo11} for a recent discussion.

One other point of interest is that the [Sr/Fe] ratio rises only
modestly in CEMP-s stars as compared to the [Ba/Fe] ratio, which
sometimes rises to very high levels, enhancements  approaching
a factor of 100 above the solar value.  This suggests that the
$s$-process production of Ba is much more efficient than that
of Sr.  However, there is no enhancement of either of these
elements above the Solar value at [Fe/H] $< -3.3$~dex.

\subsection{The Most Metal-Poor C-rich Stars}

\begin{figure}
\epsscale{1.05}
\plotone{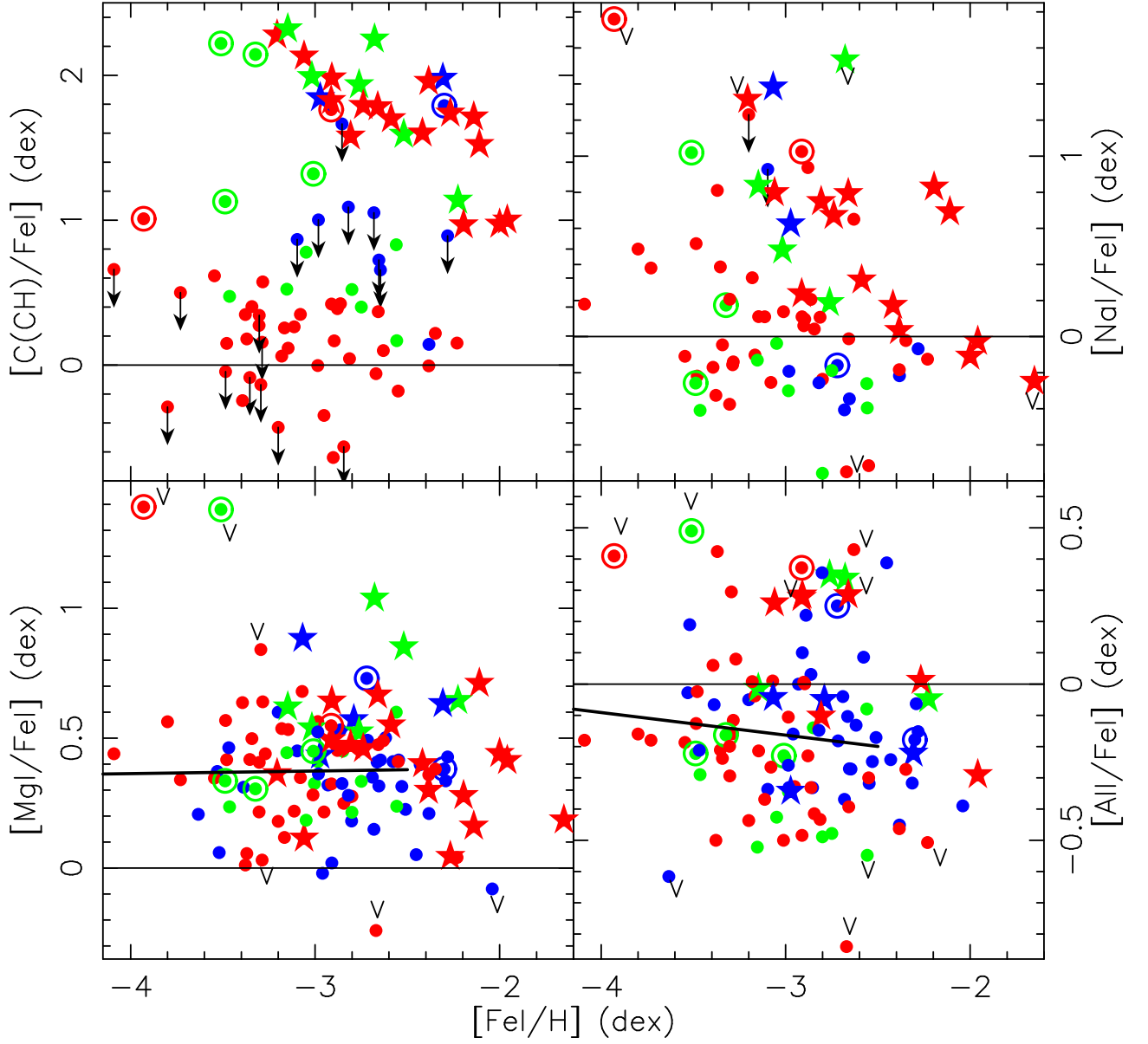}
\caption[]{[C/Fe] (upper left), [Na/Fe] (upper right),
[Mg/Fe] (lower left) and [Al/Fe] vs [Fe/H] for our 0Z sample. 
Symbols are as in  Fig.~\ref{figure_ioneq}.  ``V'' denote an
abundance ratio that was carefully checked, ``?'' denotes one
which when checked could not be verified.  The linear fits from
Table~\ref{table_fits} are shown when available.  Three C-rich
stars from our
sample have [C/Fe] beyond the maximum range shown in the upper left panel.
\label{figure_4plot1} }
\end{figure}

\begin{figure}
\epsscale{1.05}
\plotone{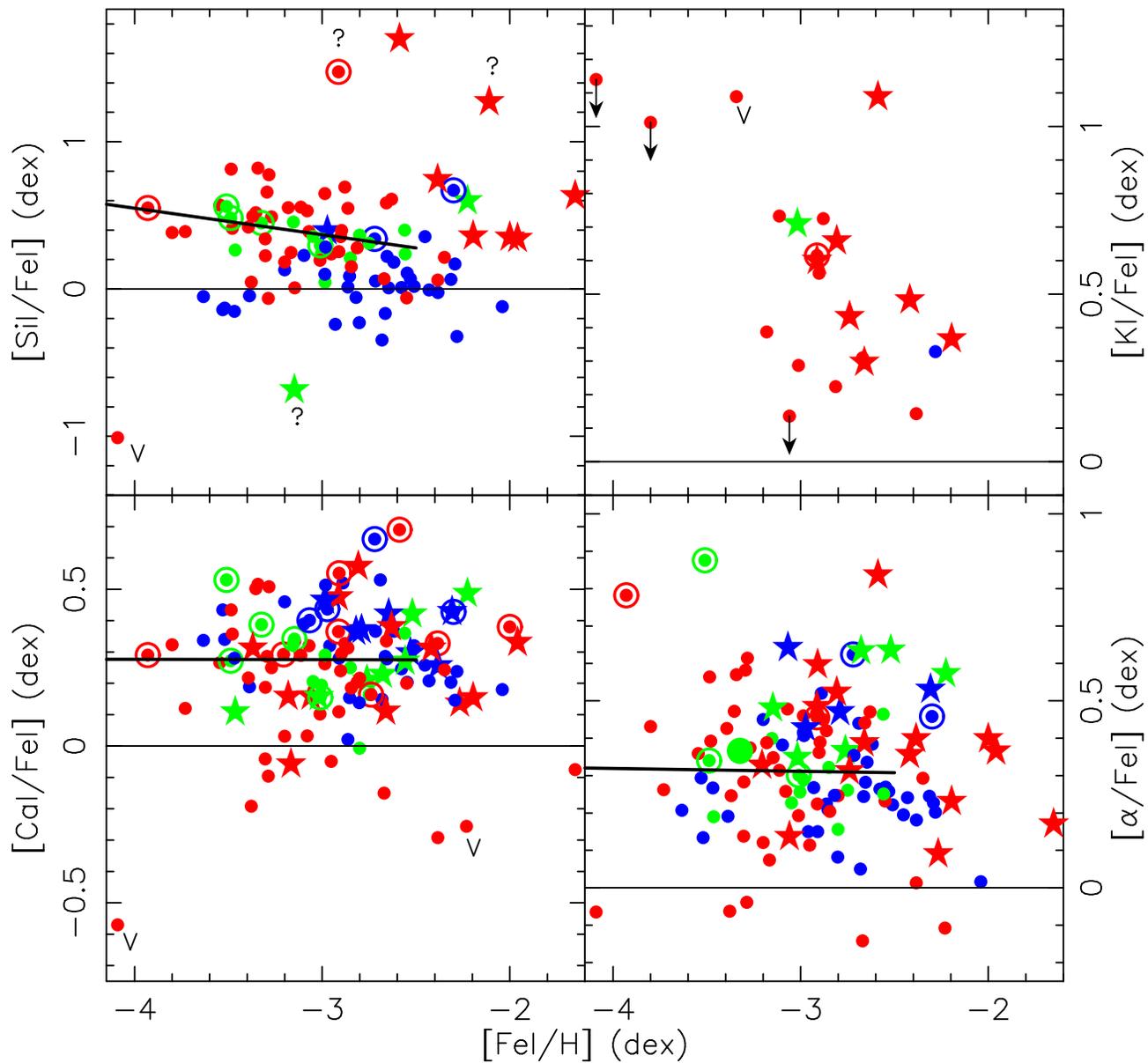}
\caption[]{[Si/Fe], with the linear fit for giants, (upper right), [K/Fe] (upper left),
  [Ca/Fe] (lower left) and [$\alpha$/Fe] (lower right) vs [Fe/H]
 for the 0Z sample. 
 Symbols are as in 
Fig.~\ref{figure_ioneq}.  Notes regarding checks of outliers are as in
Fig~\ref{figure_4plot1}. 
The linear fits from
Table~\ref{table_fits} are shown when available. 
\label{figure_4plot2} }
\end{figure}

\begin{figure}
\epsscale{1.05}
\plotone{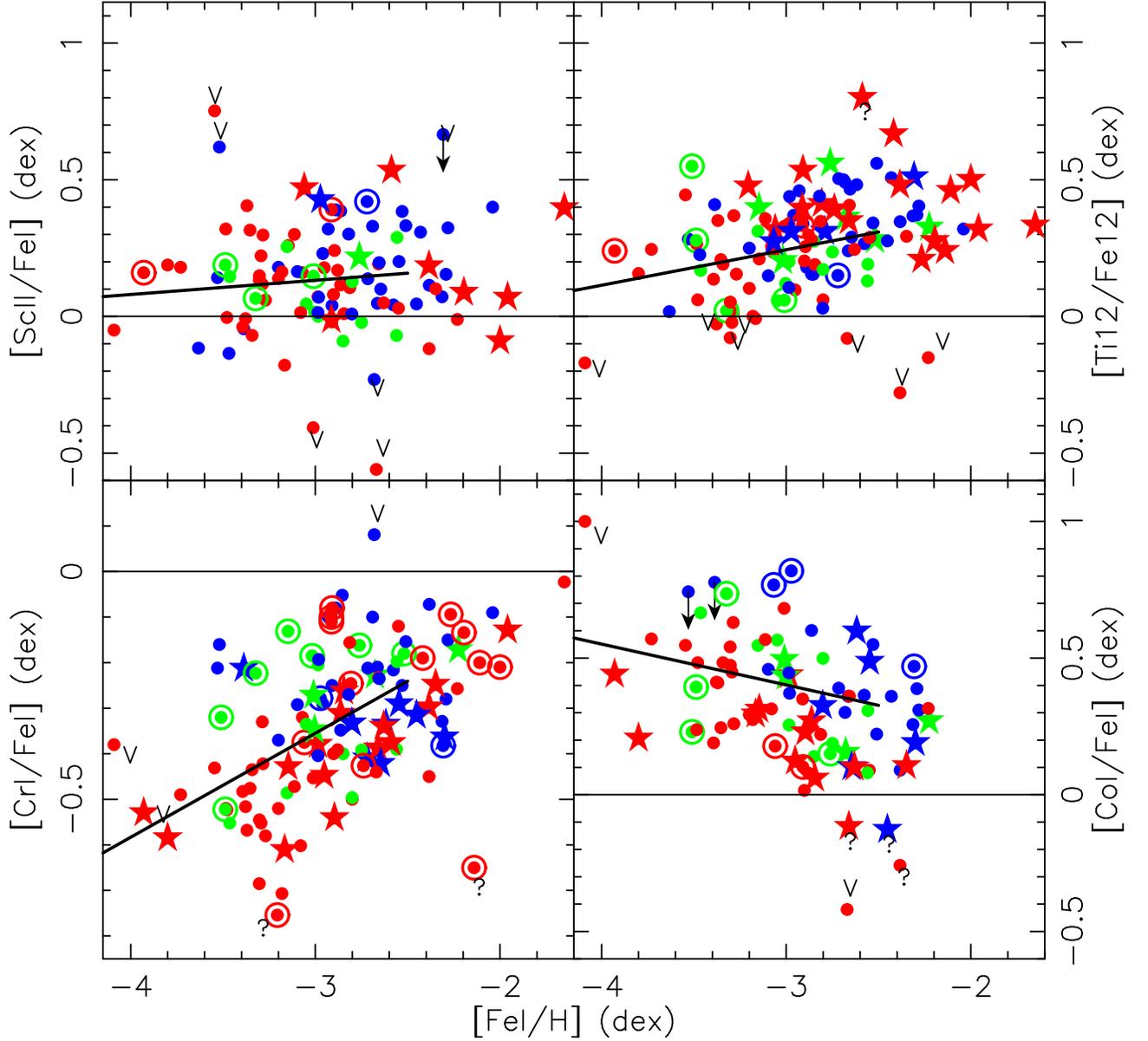}
\caption[]{[ScII/Fe] (upper left), [Ti12/Fe12], which is
the mean of [TiI/FeI] and [TiII/FeII]  (upper right),
[Cr/Fe] (lower left), and [Co/Fe] (lower right) vs [Fe/H]
for the 0Z sample.   Symbols are as in 
Fig.~\ref{figure_ioneq}.  Notes regarding checks of outliers are as in
Fig~\ref{figure_4plot1}.
The linear fits from
Table~\ref{table_fits} are shown when available. 
 \label{figure_4plot3} }
\end{figure}

\begin{figure}
\epsscale{1.05}
\plotone{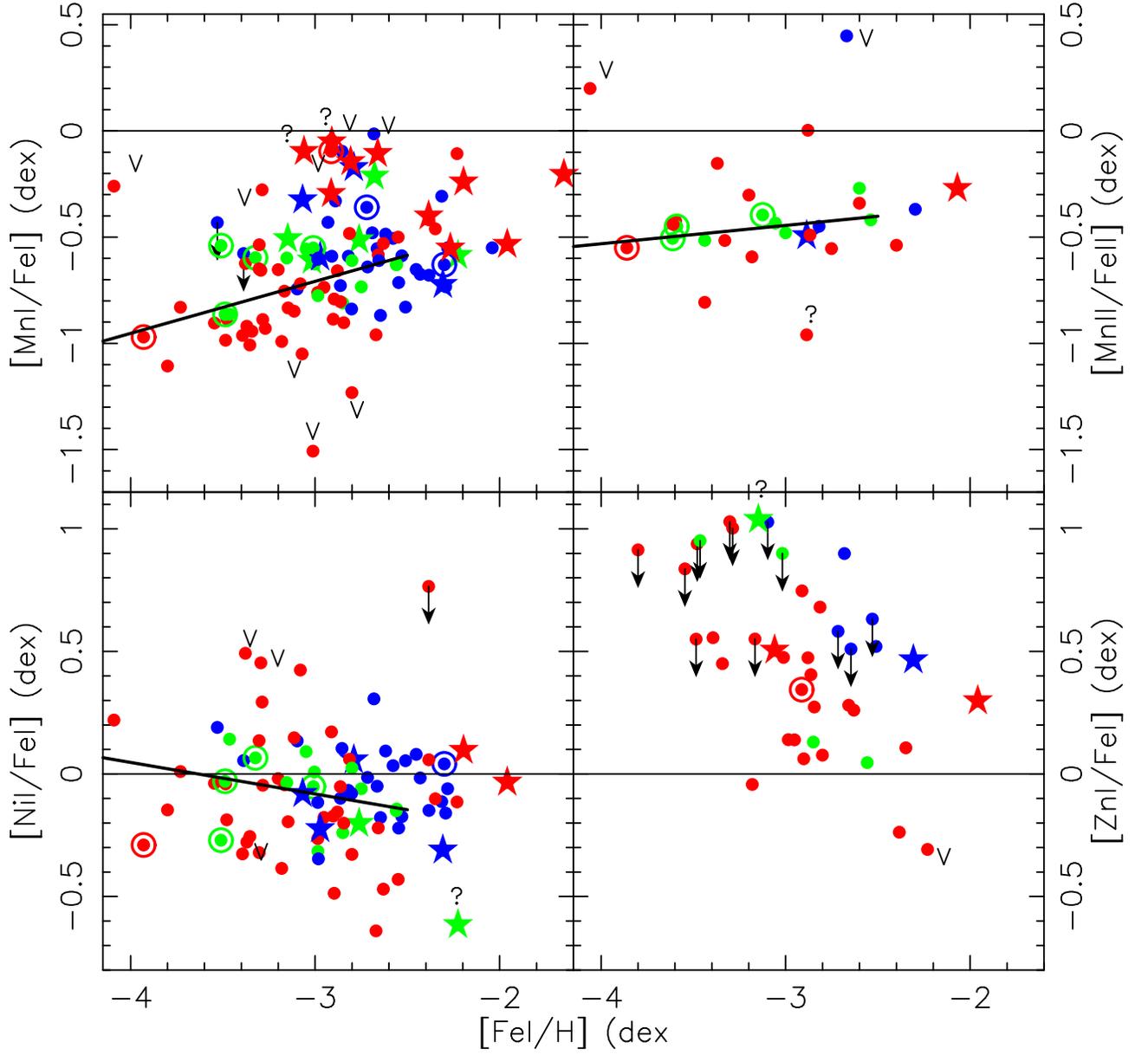}
\caption[]{[MnII/Fe] (upper right), [MnII/Fe] (upper left),
[Ni/Fe] (lower left), and [Zn/Fe] (lower right) vs [Fe/H]
for the 0Z sample.  Symbols are as in 
Fig.~\ref{figure_ioneq}.  Notes regarding checks of outliers are as in
Fig~\ref{figure_4plot1}.
The linear fits from
Table~\ref{table_fits} are shown when available. 
 \label{figure_4plot4} }
\end{figure}

HE1150-0428 is
the most metal-poor genuine C-star in our sample, with [Fe/H] $-$3.2~dex.
Below this metallicity,  
there are no stars with detected C$_2$ bands, even though there
are still a few C-rich stars.  Of the 23 stars in our sample
with lower [Fe/H], four (17\%) have [C/Fe] $> 0.7$~dex, but
none of these show C$_2$ bands. 
We believe that the apparent deficit of C-stars with 
C$_2$ bands at the lowest metallicities is,
at least in part, an artifact of our nomenclature.       
Three of these four C-rich stars
have $T_{eff} \geq 5550$~K.  The fourth has [Fe/H] a factor of 5
lower than the most Fe-poor C-star.  Since
the strength of bands of C$_2$ 
is roughly proportional to $N(C)^2$, while those of CH depend only
on the first power of $N(C)$, we ascribe the disappearance
of genuine C-stars with detectable bands of C$_2$
(see Fig.~\ref{figure_4plot1}) to the
lower Fe abundance (taking Fe as a proxy for C) and to the
(by chance)
tendency towards somewhat higher \teff\ 
to explain the abrupt disappearance of C-stars at the lowest
metallicities.  There may also be a shift towards higher O abundance, i.e.
lower C/O ratios, but evidence to support this is weak at present.

In terms of other abundance ratios, roughly half the C-rich stars
below $-3.3$~dex have high [Na/Fe] and [Al/Fe], but comparably
high values can be found in somewhat higher Fe-metallicity C-stars.
The only genuine outliers beyond any expectation are the very high
[Mg/Fe] ratios found for HE1012$-$1540 and HE2323$-$0256 (aka CS22949$-$037)
(see Fig.~\ref{figure_4plot1}); CS22948$-$034 \citep{aoki02} is similar to these
two CEMP outliers.   This small family of C-rich stars is discussed in
\ref{section_other_out}; see Fig.~\ref{figure_family_highmg}.
All their other abundance ratios are within or close to the range spanned
by the somewhat higher Fe-metallicity C-stars for the species Mg to Ni.

Three of these four stars have
sub-solar [Ba/Fe], while the fourth has [Ba/Fe] $\sim$0.3~dex,
so none of these would be considered CEMP-s stars using a conventional
definition of highly super-solar [Ba/Fe].  However, at these
low metallicities, the Ba/Fe ratio is falling rapidly as
[Fe/H] decreases, and these four stars do lie at the
upper end of the sample in [Ba/Fe] for their low Fe-metallicities.
Unfortunately none of them has a detection of Eu, so we cannot
distinguish between $r$ and $s$-process production of the Ba.

We suggest  that the C-rich stars  at the lowest
Fe-metallicities we have found in our sample from the HES
are the remnants of binary
mass transfer systems, but that the evolution and nucleosynthesis involved
are somewhat different from the picture at higher (but still very
low) Fe-metallicity.  We make this suggestion in part because
these stars comprise a fraction of the total in the sample
at such low metallicity which is similar to that of the
known binaries (i.e. the CEMP-s stars) at only slightly higher
metallicity.  Presumably there are binary systems among the most extreme
EMP stars in our sample, and some of them would engage in mass transfer,
and perhaps be seen today as the C-rich EMP stars below [Fe/H] $-3.2$~dex.
If this scenario is valid,
then the total amount of C transfered in these extreme cases
must be lower, otherwise
we would see a detectable rise in [C/Fe] as [Fe/H] decreases
below $-3$~dex, which we do not. This would
require either less 
C-enriched material being transferred from the donor star in very low metallicity
binary systems or less C-enhancement of that material.

Another possibility that we have offered earlier is that
these stars are the EMP equivalent
of the R-type carbon stars, which are not binaries.
However, these stars are somewhat hotter than the conventionl
R-type carbon stars.

A number of theoretical papers simulating low metallicity AGB models
\citep{lau07,ventura09}
comment that AGB stars of lower metallicity
are expected to reach a higher maximum temperature
at the base of the convective envelope. Thus the lowest
$Z$ models are expected to undergo a more advanced nuclear processing.
Perhaps the reaction $^{12}$C $+ \alpha ~ \rightarrow ^{16}$O
might keep $\epsilon$(C)/$\epsilon$(O) $< 1$ in the accreted
material, thus avoiding the formation
of C$_2$.  With regard to the absence of $s$-process enhanced C-rich stars
below [Fe/H] $-3.2$~dex, we first note that [Ba/Fe]
is dropping rapidly at the lowest metallicities in 
the entire stellar population; see Fig.~\ref{figure_4plot5}.  
It is difficult to make a lot
of Ba when there are fewer and fewer Fe seeds to which neutrons
can be added; one tends to make lead instead.  
This has been known for quite a while;
Fig.~12 of \cite{busso_araa} indicates the shift from the production
of Ba to Pb among the most metal-poor AGB stars they 
modeled, all of which are more Fe-rich than the stars we are considering here.
Furthermore,
\cite{lau07} point out that zero metallicity
stars avoid the third dredge up and hence do not produce
any $s$-process elements at all.  So the absence of $s$-process
material in the most extremely metal-poor C-rich stars
should not be a surprise.

The strongest evidence against our hypothesis is that the radial
monitoring programs, fragmentary as they are, since most of these
stars have been discovered quite recently, do not support
the suggestion that these stars are binaries,
as is shown in Table~\ref{table_mult_vr}.  
However,
the models of \cite{lau07} suggest that zero metallicity massive AGB stars,
even with mass as low as 5~M$_{\odot}$,
end their evolution as SN1.5 after/during
the AGB stage as their degenerate cores are able to grow
to the Chandrasekhar limit  before the star loses its
envelope.  The envelope loss is delayed due to the weaker stellar
winds expected in the lowest metallicity stars.
Hence the primary explodes before it can become a white dwarf.
In this case, the secondary which we now see might get a velocity
kick and be ejected from the system, and that may be why the
apparent fraction of binaries among these C-rich stars as viewed
at the present time is not as high as in the
somewhat higher metallicity CEMP-s stars.  

\cite{norris12b} on the other hand suggest that the
C-rich stars with Fe-metallicity below $-3.1$~dex represent
a different mode of star formation in a very low metallicity
environment than do the C-normal stars of similar Fe-metallicity.
They ascribe the star formation for the C-rich population to
cooling via fine structure transitions of C~II and O~I, while a
second unspecified but different cooling mechanism led to the formation
of the C-normal stars.

A careful study of the crucial elements, including CNO, in the most metal-poor 
C-rich stars, as well as further radial velocity monitoring
and theoretical modeling, and of course increasing still further
the sample size of stars at these extremely low metallicities,
will help to unravel this puzzle.

\section{$r$-process Enhanced Stars \label{section_rstars} }

A small number of stars with highly enhanced 
europium (i.e. with [Eu/Fe] $> 1$~dex)
have been discovered among the very metal-poor stars, as
well as a larger number with more modest $r$-process enhancements.
Among the best studied of these is CS22892$-$052, which was analyzed in
great detail by \cite{sneden03}.  Even initially it was apparent
that the heavy elements beyond the Fe-peak in this and similar highly
Eu-enhanced stars follow the canonical
$r$-process distribution (scaled by a constant value for a given star).
\cite{sneden09} refine their abundances
for five $r$-process rich stars with improved atomic data
and spectra to demonstrate how extremely closely the 
distributions of these heavy neutron capture elements
in these stars follow the Solar $r$-process
distribution with $\sim$15 rare earth elements detected, while
BD+17~3248 now has 32 heavy neutron capture elements with spectroscopic
abundances \citep{roederer10}.  \cite{aoki10} recently discovered the most
extreme such star known to date, with [Eu/Fe] +1.9~dex, also unusual
as it is a main-sequence star rather than a giant; this star is not C-enhanced.

\begin{figure}
\epsscale{1.05}
\plotone{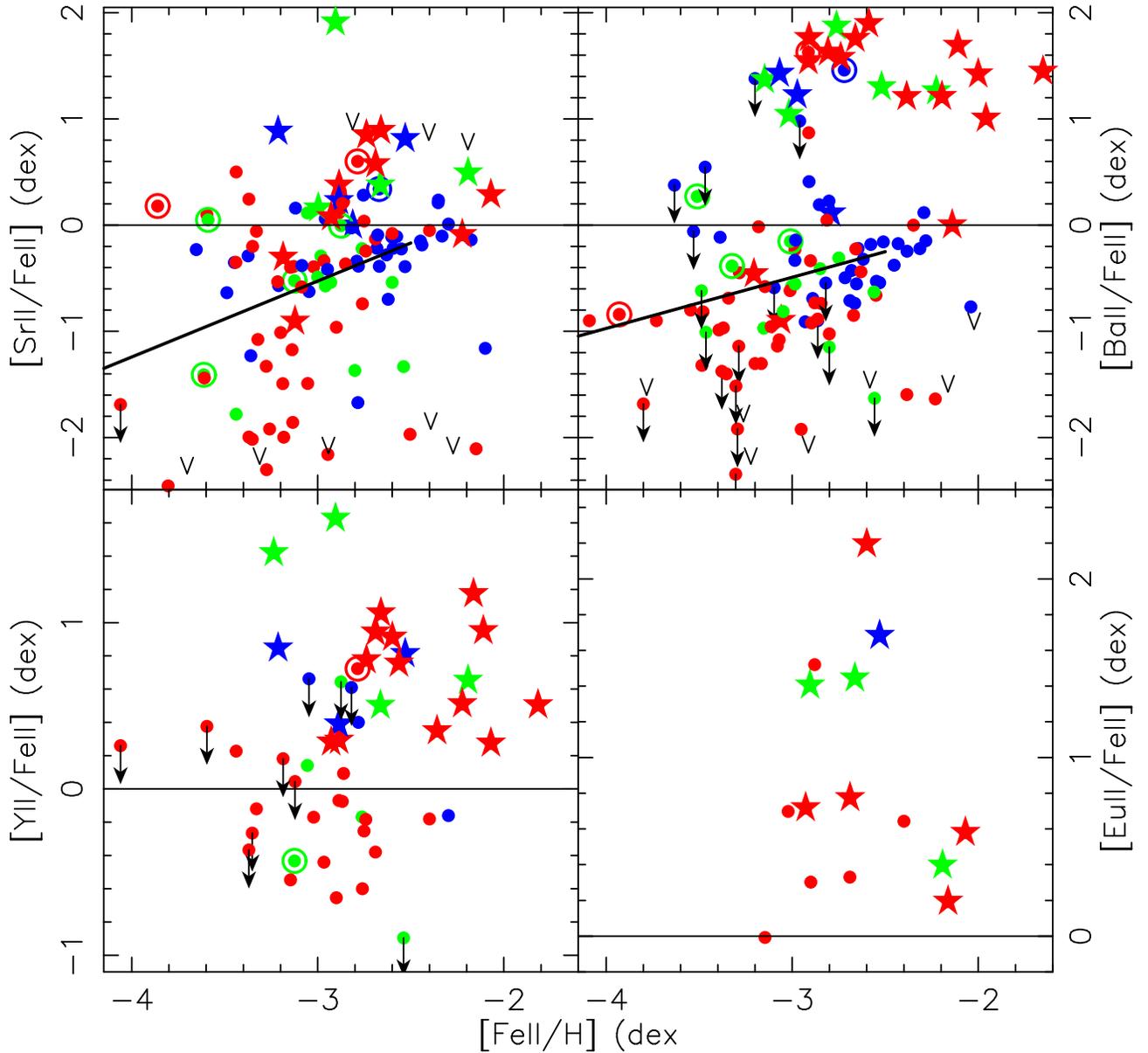}
\caption[]{[SrII/Fe] (upper left), [BaII/Fe] (upper right),
[YII/Fe] (lower left), and [EuII/Fe] (lower right) vs [Fe/H]
for the 0Z sample.  Upper limits are ignored for EuII.
Symbols are as in 
Fig.~\ref{figure_ioneq}.  Notes regarding checks of outliers are as in
Fig~\ref{figure_4plot1}.
The linear fits from
Table~\ref{table_fits} are shown when available. 
 \label{figure_4plot5} }
\end{figure}

\begin{figure}
\epsscale{1.05}
\plotone{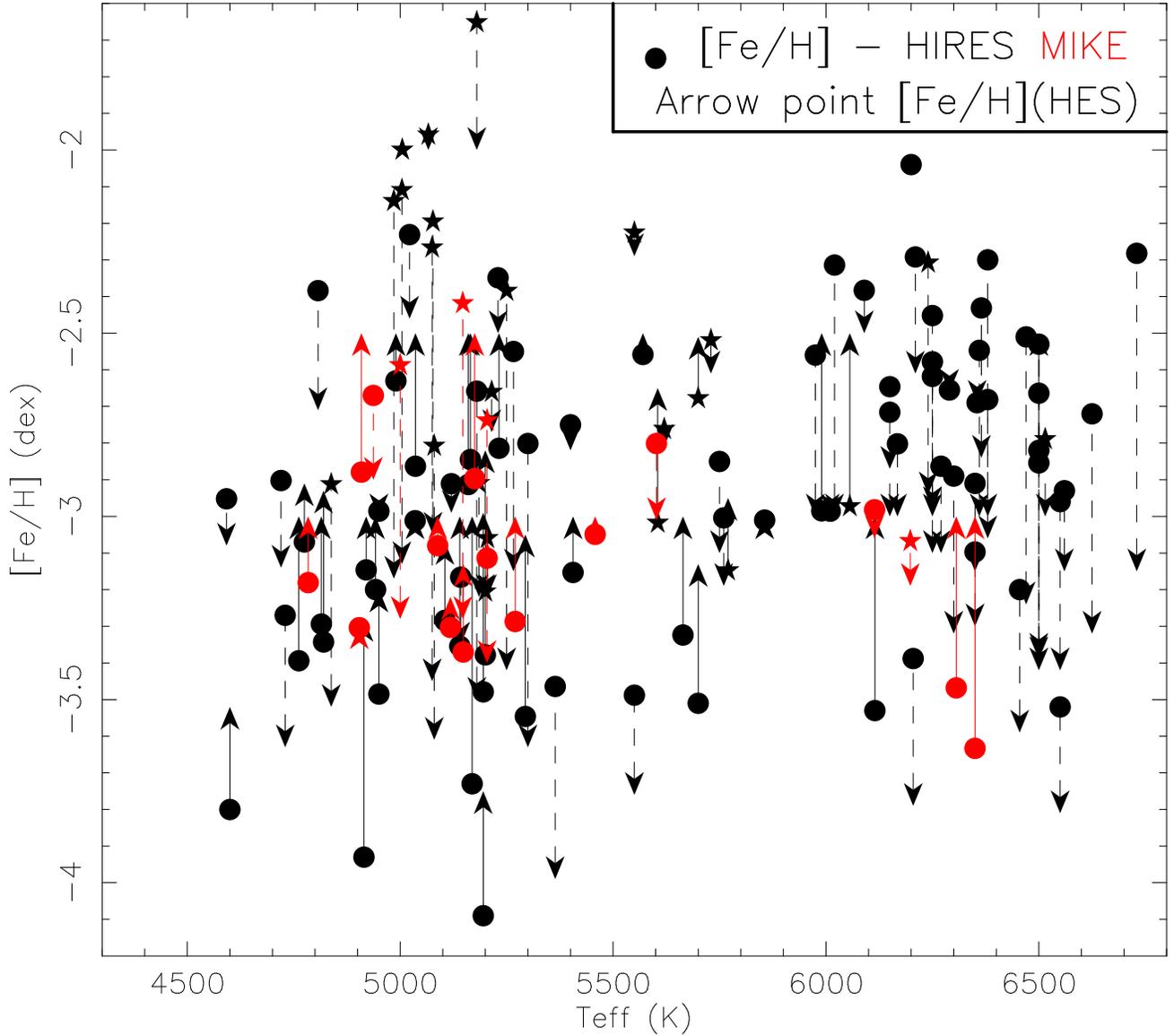}
\caption[]{The difference between the [Fe/H] deduced from the high
resolution Keck and Magellan spectra, indicated by the filled circles 
(C-stars are denoted by small stars), and the moderate resolution
follow up spectra analyzed with a modified version of the
\cite{beers98} algorithm, indicated by the points of the arrows.
If [Fe/H](HIRES) $>$ [Fe/H](HES), the arrow is shown as a solid line;
otherwise it is drawn as a dashed line.
For the hot dwarfs, which are on the right side of this figure,
the arrows mostly point downward indicating that
[Fe/H](HIRES) $>$ [Fe/H](HES).
\label{figure_feh_compare} }
\end{figure}

\begin{figure}
\epsscale{0.70}
\plotone{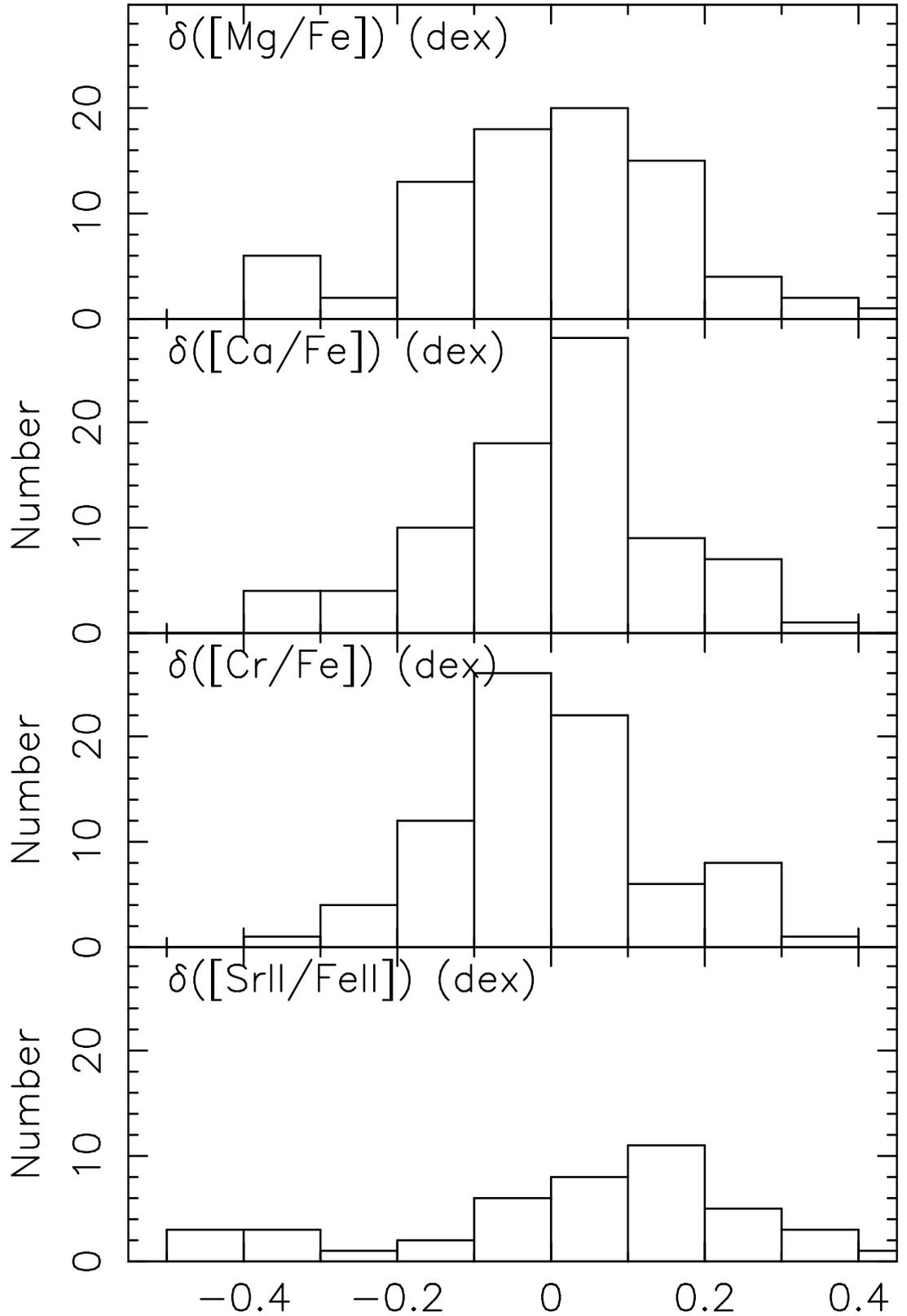}
\caption[]{Deviations from the linear fits for [Mg/Fe],
[Ca/Fe], [Cr/Fe], and [SrII/FeII] (from top to bottom) are shown as histograms.
C-stars, outliers, and stars with  [Fe/H] $> -2.5$~dex 
were not used in the construction of the linear fits.  
\label{figure_fit_4plot} }
\end{figure}

\begin{figure}
\epsscale{1.05}
\plotone{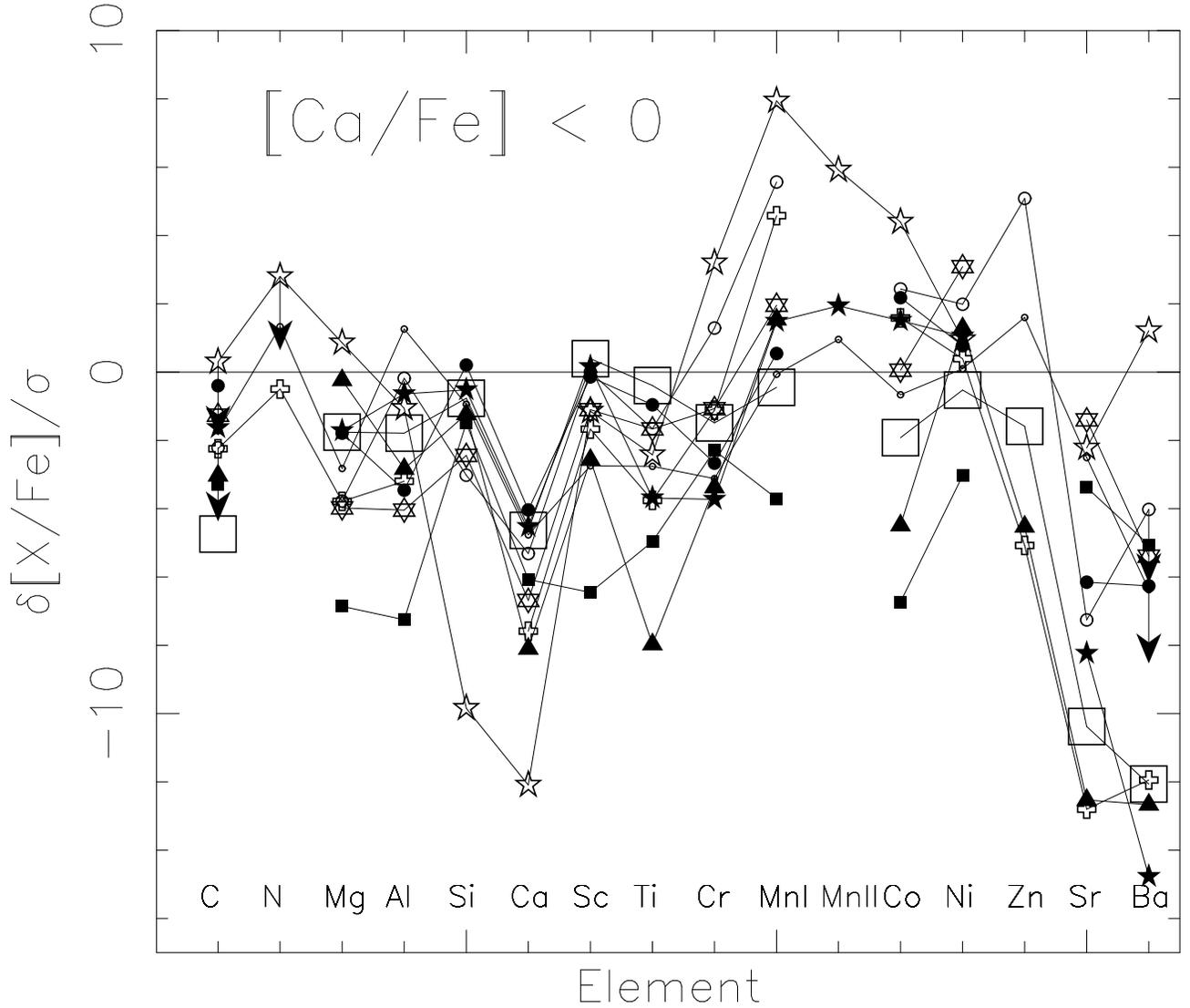}
\caption[]{$\delta$[X/Fe]/$\sigma$ for 10 EMP stars in our sample
with subsolar [Ca/Fe], where $\delta$[X/Fe]/$\sigma$ is the difference
between the derived value of [X/Fe] and that from the linear fit to [X/Fe] vs [Fe/H]
using  the  derived Fe-metallicity of the star divided by a generous
estimate of the uncertainty in the abundance ratio.
A different symbol is used for each of the 10 stars.
\label{figure_family_lowca} }
\end{figure}

\begin{figure}
\epsscale{1.05}
\plotone{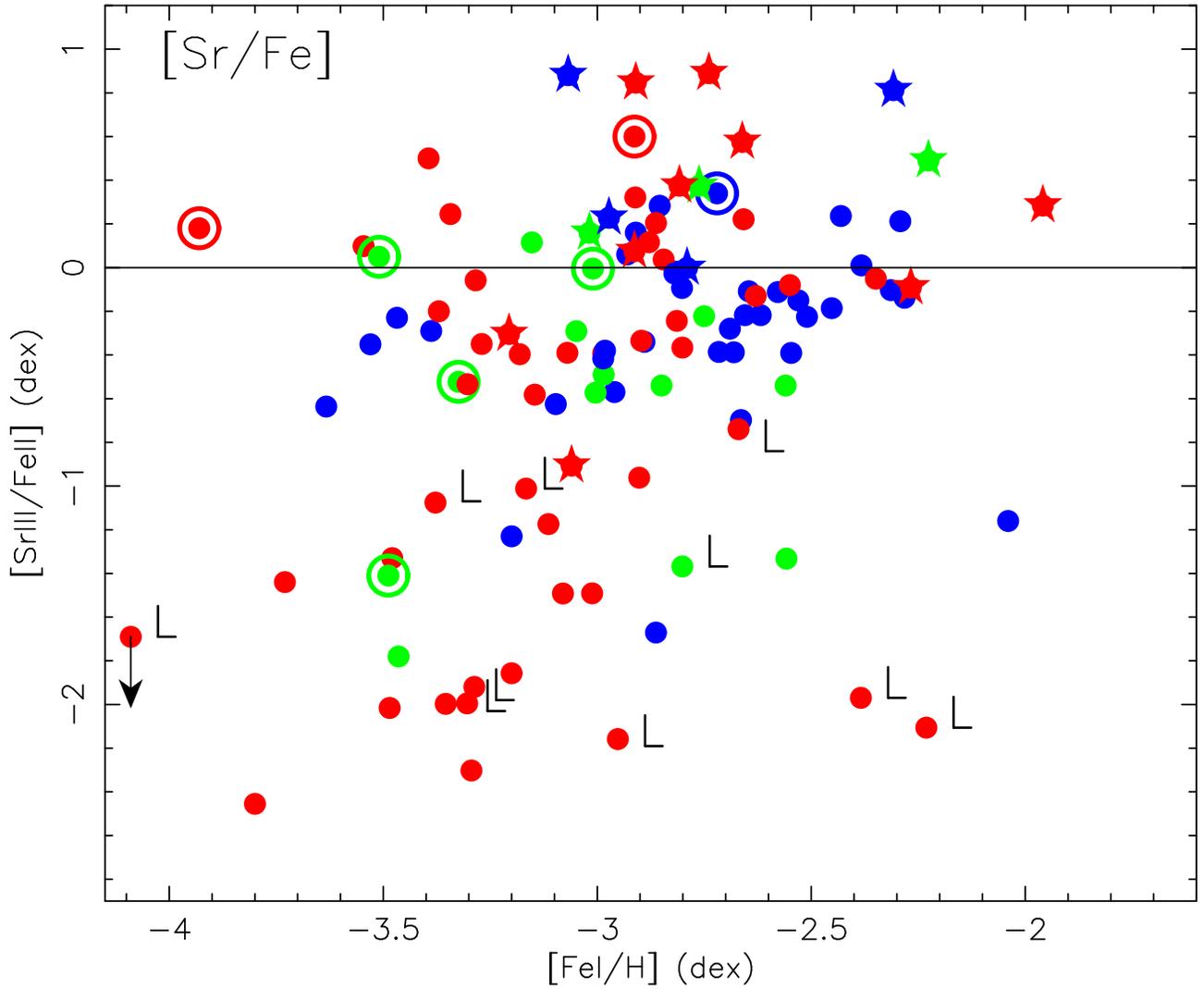}
\caption[]{[SrII/FeII] vs [Fe/H] with the stars with 
sub-solar [Ca/Fe] marked by the letter ``L''.
Note that the low-Ca family also has low heavy-neutron capture elements,
well below the mean of our 0Z sample.
Symbols are as in 
Fig.~\ref{figure_ioneq}. A typical error bar is shown for a single star.  
\label{figure_familylowca_sr} }
\end{figure}

\begin{figure}
\epsscale{1.05}
\plotone{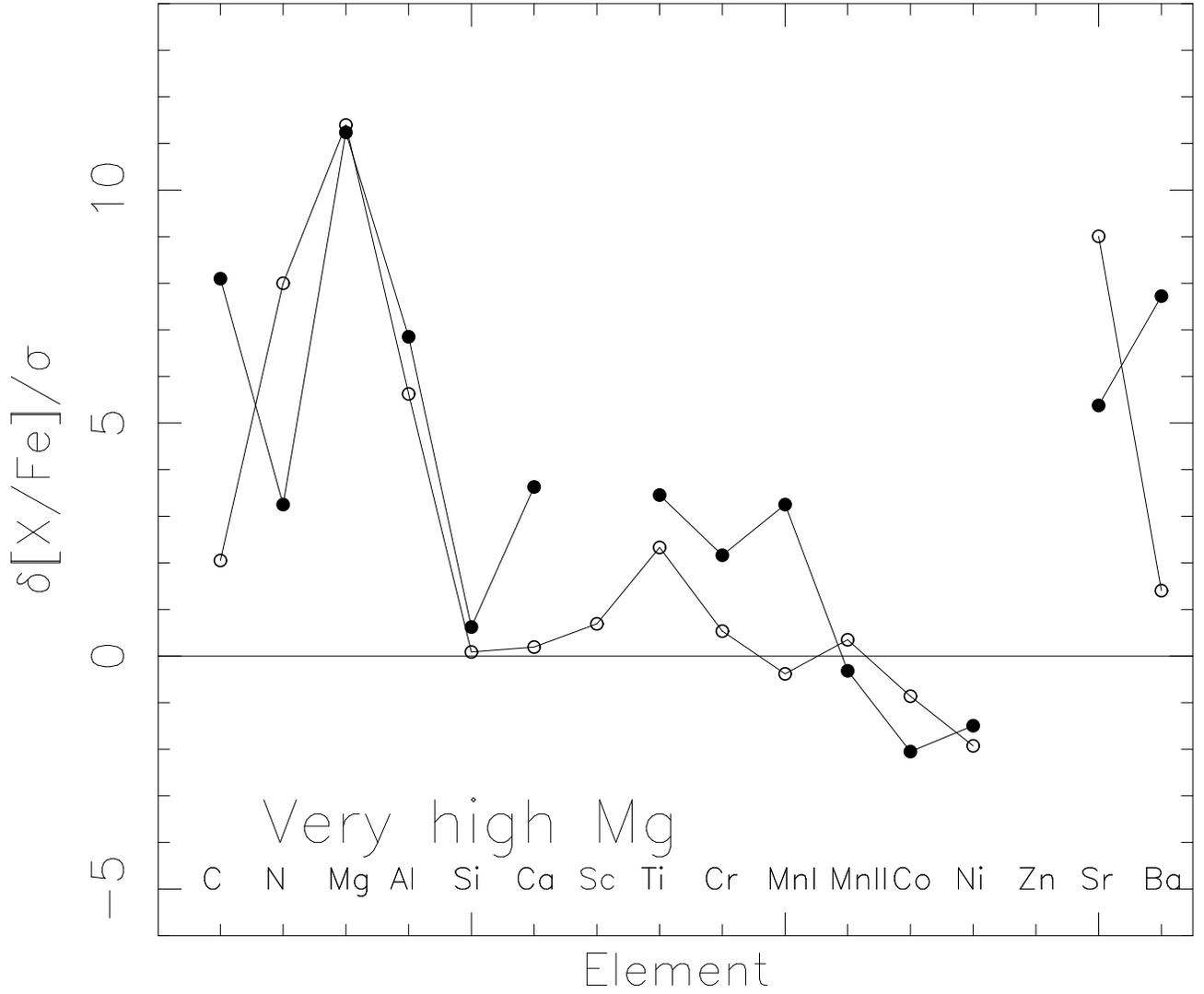}
\caption[]{The same as Fig.~\ref{figure_family_lowca} for 2 EMP stars in our sample
with very high [Mg/Fe]. 
\label{figure_family_highmg} }
\end{figure}

\begin{figure}
\epsscale{1.05}
\plotone{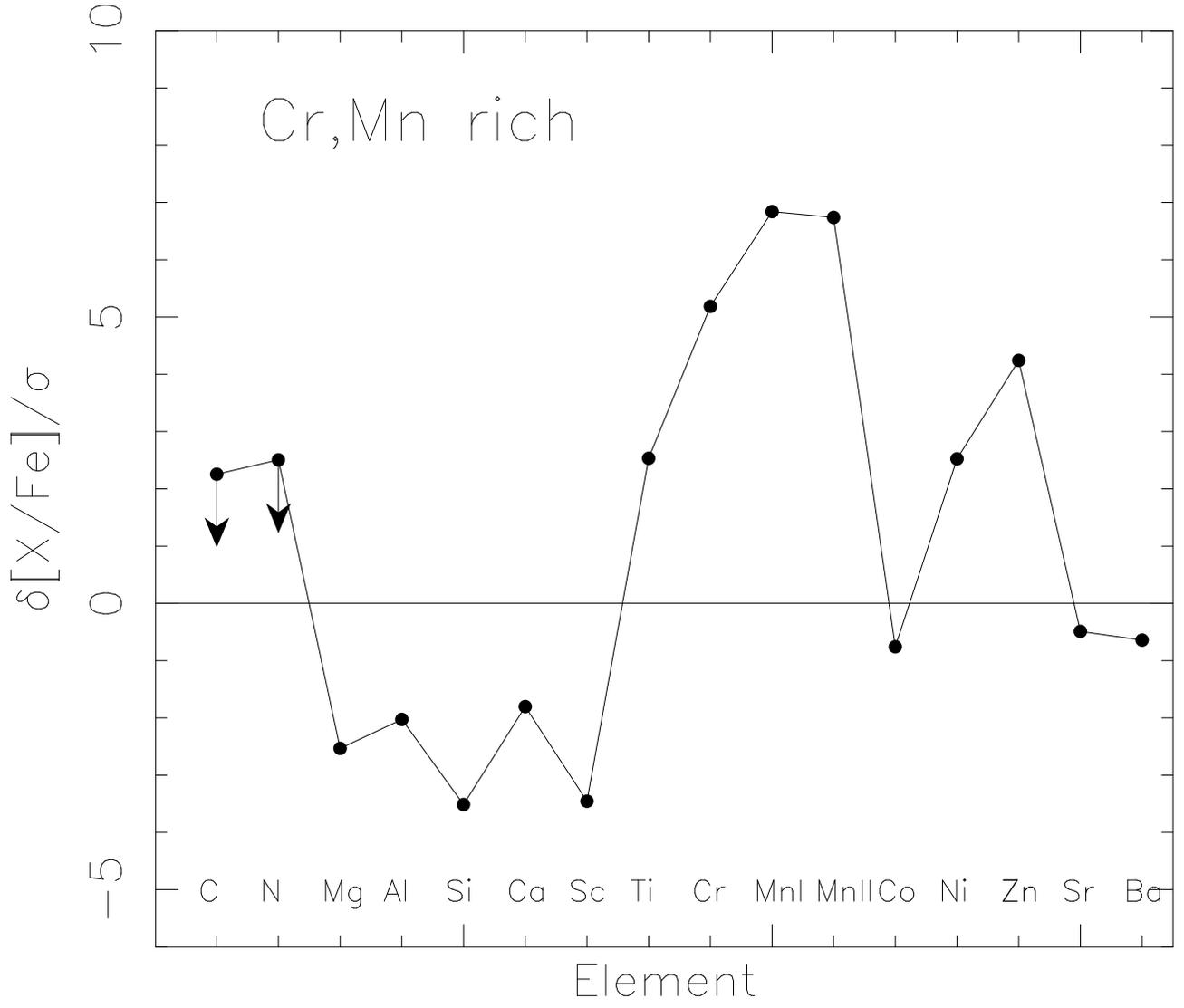}
\caption[]{The same for HE2344$-$2800, a main sequence turnoff region star, 
which has abnormally high Cr and Mn.
\label{figure_family_highcrmn} }
\end{figure}

\begin{figure}
\epsscale{0.90}
\plotone{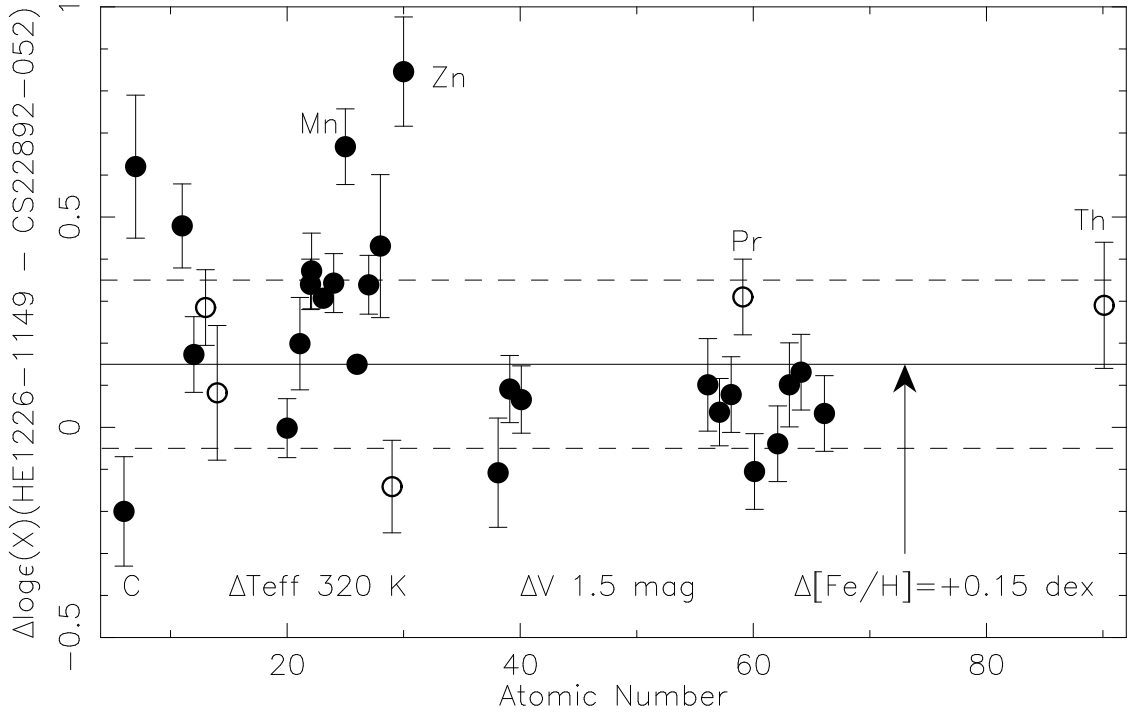}
\caption[]{[X/Fe] vs atomic number for the extreme $r$-process star HE1226$-$1149
found in our sample with respect to that of CS22892$-$052, an extreme $r$-process
star with a very comprehensive analysis by \cite{{sneden03}}.  Open circles denote
species with only a single detected absorption line in HE1226$-$1149.  Typical uncertainties
for EMP giants in the abundance ratios are shown.
The solid horizontal line is the difference for [Fe/H] between these two stars, with
the dashed lines representing differences $\pm$0.2~dex above or below this value.
\label{figure_he1226} }
\end{figure}

We have found one new extreme $r$-process enhanced star in our
sample,  HE1226$-$1149, with [Eu/Fe] +1.6~dex
and [Eu/Ba] +0.65~dex; see Fig.~\ref{figure_4plot5}. This star is
not C-rich.  As is typical
for such stars, it has [Fe/H] $-2.9$~dex and and is a giant with
\teff ~ = 5120~K.   The abundance ratios [X/Fe] for this star 
 for 13 heavy-neutron capture elements 
are compared
with those of CS22892$-$052 taken from \cite{sneden03} in 
Fig.~\ref{figure_he1226}.  HE1226$-$1149,
has approximately the same absolute
abundances of these elements as does CS22892$-$052, 
but a slightly higher (by only 0.15~dex) Fe abundance.
To within the observational
errors the abundance ratios of the heavy neutron-capture
elements follow the standard $r$-process abundance
distribution.  HE0011$-$0035 and
HE2244$-$2116, both with [Eu/Fe] $\sim$ +0.65~dex, are
less extreme $r$-process enhanced stars.

It is interesting that extremely $r$-process enhanced stars
seem to occur only within a small range in [Fe/H]
near $-3.0$~dex, see, e.g. \cite{barklem05}.  Presumably
they result from an input from a single source (i.e. a single
SN of the appropriate type), which adds  a fixed (to first order)
$\Delta\epsilon$(Eu).
At higher [Fe/H], this will not be enough to significantly
raise the [Eu/Fe] ratio.  Their absence at lower [Fe/H] 
may be related to the rapid decline of [Ba/Fe] at the lowest
metallicities shown in Fig.~\ref{figure_4plot5}.

The detections of Eu are quite uncertain among the extreme C-stars
as there are no strong lines of Eu~II redder than 4130~\AA, and the
spectra of such C-rich stars are very crowded with molecular features of CH and CN
at the relevant wavelengths. We  rechecked the detections of
Eu in all such stars again in Dec. 2012, and modified the most uncertain of them
to upper limits, but there still may be some cases where molecular
features are being mistaken for Eu lines.   However, in most cases,
even in the 9 C-stars for which we still claim a detection of Eu, 
multiple lines of Eu have been detected which each give
consistent abundances for this element.  For these 9 C-stars, the
median [Eu/Ba] is $-$0.8~dex, a typical $s$-process ratio, showing the significant difference
in this key abundance ratio between the $r$-process seen in HE1226$-$1149, where
[Eu/Ba] is +0.6~dex,
and the $s$-process of neutron capture formation of the heavy elements
seen in the C-stars.

At the lowest metallicities probed here, 
even in stars which are not C-rich,
it is quite difficult to detect Eu
unless it is highly enhanced.  The predicted line strengths
for [Eu/Fe] +0.5~dex at $T_{eff} \sim 5000$~K and
[Fe/H] $-3.2$~dex are below 10~m\AA\ for even the strongest
Eu~II lines in the optical wavelength regime reaching down to
3800~\AA, making it difficult to reliably detect
any Eu lines.    Since  our sample stars are faint, there are many
upper limits for their europium abundances, and, 
as shown in Fig.~\ref{figure_4plot5}, there are no
solid detections of Eu for any star in our sample
below [Fe/H] $= -3.3$~dex.  Compounding the difficulty
of detecting Eu in EMP stars
is the general trend of an increasing deficiency in the
heavy neutron capture elements towards lower Fe-metallicity
best shown in Fig.~\ref{figure_4plot5} for [Ba/Fe].

With these caveats in mind, we compare the frequency of 
extreme $r$-process enhanced stars found by the HERES project
\citep{barklem05}
with that found here.  In both cases, hot dwarfs are excluded as
almost all extreme $r$-process enhanced stars known have \teff\ near
5100~K, and stars with strong CH are also excluded.  
The agreement for the fraction of extremely $r$-process
enhanced stars, given the very small numbers involved,
is reasonable.  
Adopting the definition of \cite{barklem05} 
for moderately $r$-process enhanced stars of $0.3 < [Eu/Fe] < 1.0$~dex,
with the same exclusions,
we find that 8\% of our 0Z sample 
is  moderately $r$-process enhanced.
This is somewhat smaller than the fraction found by \cite{barklem05} of
14\%.  The lower mean [Fe/H] values in our 0Z sample lead to a higher fraction
of stars with only upper limits for their Eu abundance,
and this probably gives rise to  the lower fraction of moderately
enhanced $r$-process stars in our sample as compared with that of the
HERES project. 



\subsection{Heavy Neutron Capture Elements In Normal EMP Stars}

With the exception of the C-stars, the ratios among the heavy neutron capture elements 
among EMP stars
appear to follow the $r$-process rather than the $s$-process ratios, when
such a distinction is possible.  However, given the many
upper limits to Eu, this is not easy to verify for most of our
sample of EMP stars from the HES for the reasons discussed in \S\ref{section_rstars}.
The relative abundances [X/Fe] of the heavy neutron capture elements show both 
a tremendous variation in their overall depletion with
respect to Fe, as is illustrated in Fig.~\ref{figure_4plot5},
and also variations in the ratios of the first peak (i.e. Sr)
vs those in the second peak (i.e. Ba).  The need for an additional
channel to reach the
unexpectedly  high [Sr/Ba] ratios seen in some EMP
stars has been widely discussed in the literature \citep[see, e.g.][]{mcwilliam95b,travaglio04a,bisterzo11}.

\begin{figure}
\epsscale{1.05}
\plotone{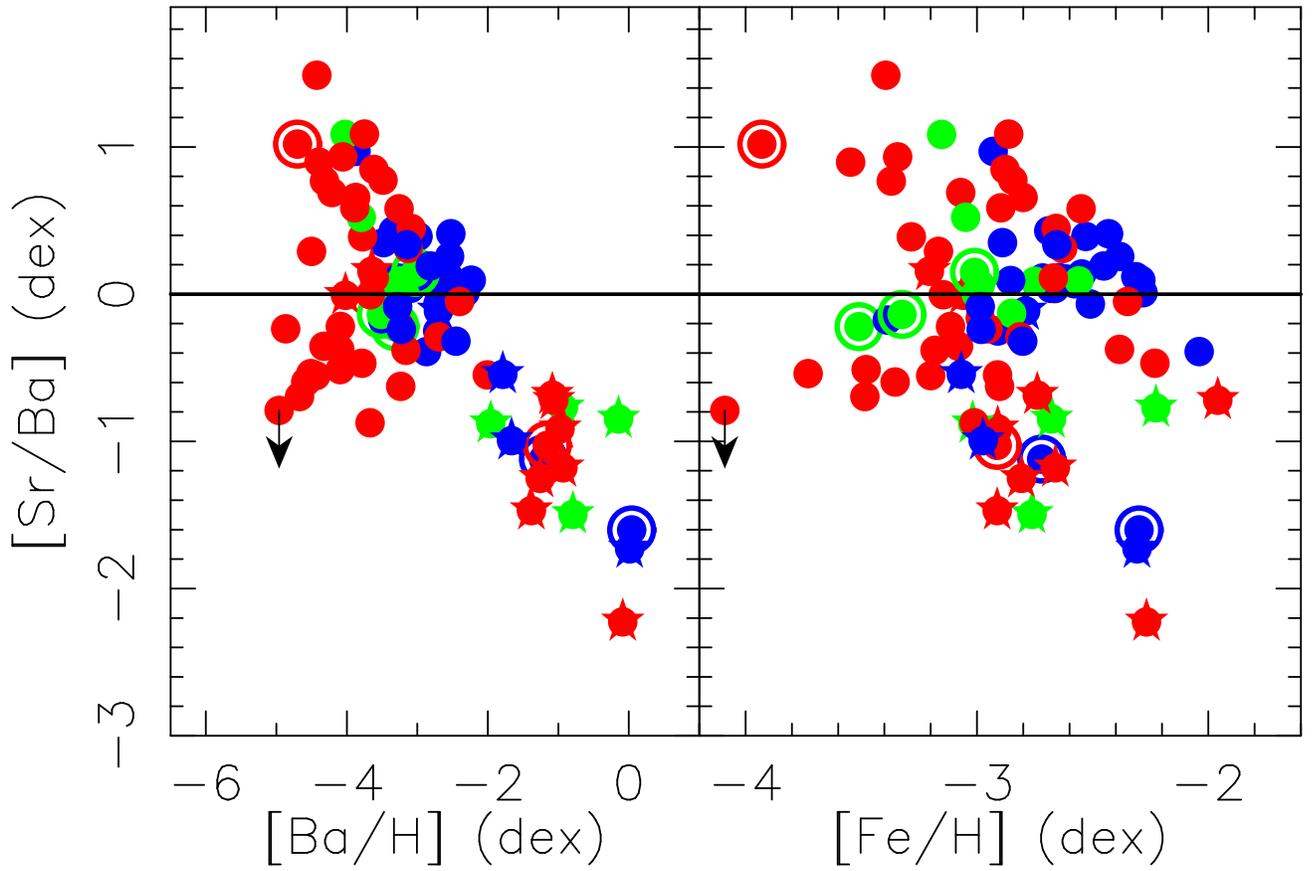}
\caption[]{[Sr/Ba] is shown as a function of [Ba/H] (left panel)
and of [Fe/H] (right panel).  Upper limits
are not shown unless [Ba/H] $< -4.5$~dex.  
Symbols are as in Fig.~\ref{figure_ioneq}.  
The error bars in each panel of this figure are comparable in size to the points.
\label{figure_srba_2} }
\end{figure}

Fig.~\ref{figure_srba_2} shows
[Sr/Ba] both as a function of [Ba/H] and also of [Fe/H].
Stars with only upper limits are not shown except for the
most metal-poor star in our sample, with an upper limit for Sr
and a very marginal detection of Ba.  
As contrasted with the large range
and chaos seen in plots of [Sr/Fe] and [Ba/Fe] vs [Fe/H]  
(see Fig.~\ref{figure_4plot5}) or the somewhat less
chaotic relation of [Sr/Ba] vs [Fe/H],
the left panel of our
Fig.~\ref{figure_srba_2}\footnote{A figure similar to
the left panel of our Fig.~\ref{figure_srba_2} 
based on results from the First Stars project can
be found in \cite{first_stars_viii} (their Fig.~15).}
shows a much clearer relation such that the lowest Ba stars seem to have
the largest enhancement of Sr relative to Ba.  This implies
that the production of the heavy neutron capture elements is occurring
with well defined relative ratios which are a function of the
Ba abundance, but these processes are decoupled in their efficiency
from those that produce the Fe-peak elements.  
One can also view this
in terms of mixing rather than production, i.e. mixing 
into the almost pristine gas of the proto-Galactic halo
of the 
ejected neutron-capture process material.  Perhaps this
is very localized  compared to that for the Fe-peak elements.
Perhaps the neutron-capture processes require a special 
rare type of SNII,
so that stochastic local mixing becomes important.

\begin{figure}
\epsscale{1.05}
\plotone{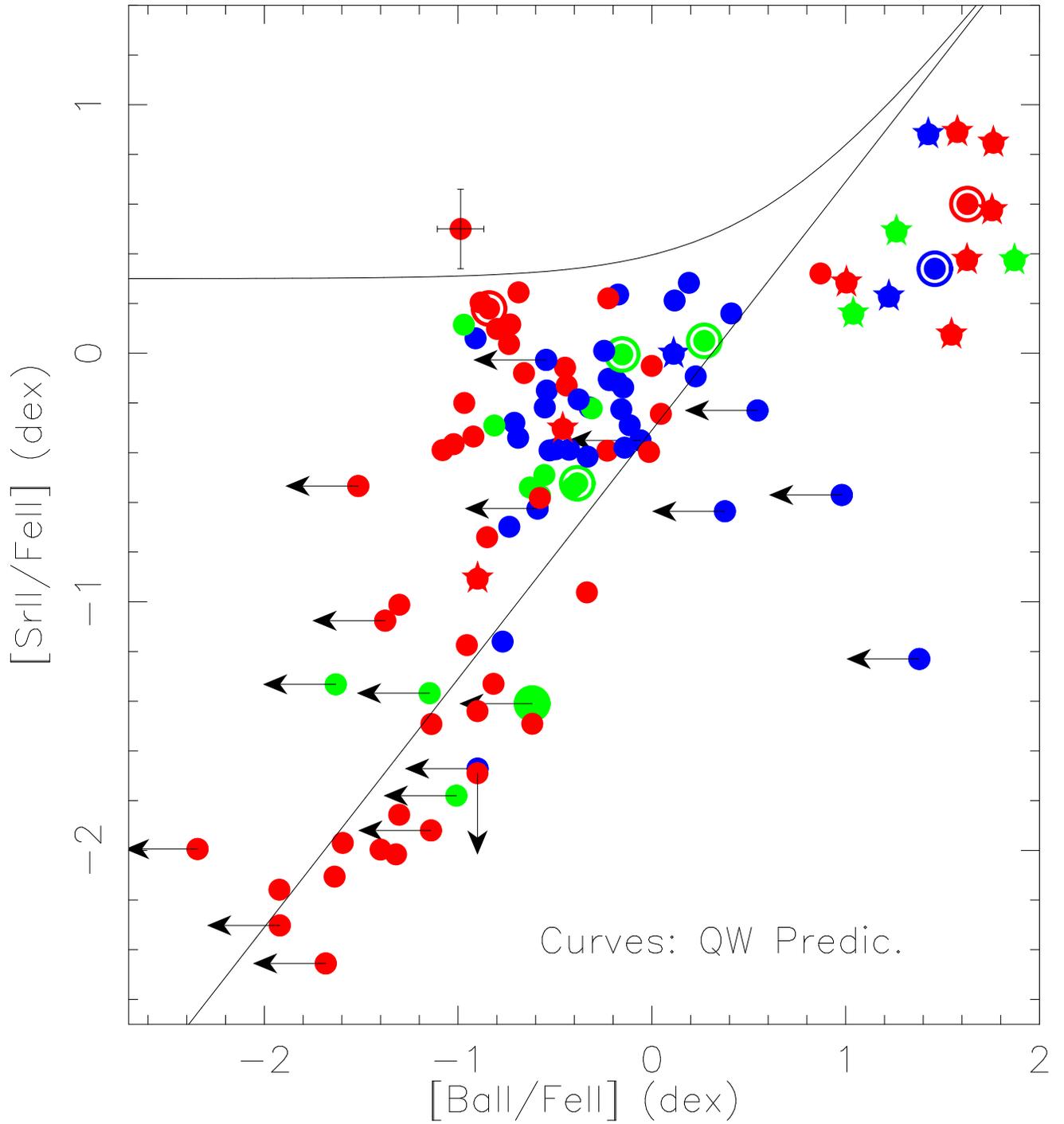}
\caption[]{The prediction by \cite{qian08} of the allowed region 
(the area between the two curves) of
[Sr/Fe] as a function of [Ba/Fe]  is compared
 with our derived  abundance ratios.
Symbols are as in 
Fig.~\ref{figure_ioneq}.  A typical error bar is shown for a single star.
\label{figure_qian_srba} }
\end{figure}

\cite{qian08} have presented a phenomenological theory to explain the abundances
of the heavy-neutron capture elements by mixing  ejecta of several 
different types of SNII that might occur in the early formation of the
Galaxy.  They predict  allowed regions in the plane [Sr/Fe] as a function
of [Ba/Fe], which are illustrated as lines in Fig.~\ref{figure_qian_srba}, with our
data superposed.  The agreement is quite good in that essentially
all of the stars in our sample (ignoring the C-stars) lie
within the area designated as allowed.

\section{Distance Effects \label{section_distance} }

We have looked for a dependence of chemical inventory on distance
within our in-situ HES EMP sample.
We first calculate the absolute luminosity for each star from its
observed $V$ mag and adopted reddening.    We then interpolate
between the $\alpha$-enhanced isochrones of \cite{yi03}  for [Fe/H] $-1.2,~-2.2$
and $-3.2$~dex at a fixed age of 12~Gyr
to find the predicted $M_V$ at the \teff\ we adopt given the [Fe/H] 
we have deduced for the star.  We assume the star is at or  more luminous than
the main sequence turnoff.  The distances thus derived
for EMP stars from the HES range from 
0.5~kpc for the nearest dwarfs\footnote{Our sample includes two bright previously known
very metal-poor dwarfs which are both closer than 0.5~kpc.} to 34~kpc,
with 10 EMP giants lying beyond 15~kpc.  We divide our 0Z sample into three groups,
close ($D < 4$~kpc, 54 stars), inner halo ($4 < D < 15$~kpc, 59 stars),
and outer halo ($D > 15$~kpc, 10 stars).  We looked for a relationship 
between [Fe/H] and $D$, but, as indicated in
Table~\ref{table_ratio_distance}, did not find anything.  The mean and dispersion
for the 10 outer halo giants observed in situ is [Fe/H] = $-3.02$~dex
with $\sigma~0.39$~dex; the median is $-2.90$~dex.  For the
54 inner halo stars we find a median of $-2.88$~dex, essentially identical.
Note that neither of the two peculiar
CEMP-no extreme EMP stars with highly enhanced Na and Mg discussed in 
\S\ref{section_cstars} belongs to the outer halo group.

Recently the huge and homogeneous data archive of SDSS and SEGUE spectra
has been used to search for metallicity gradients in the halo, but the results
of \cite{carollo07} \citep[see also][]{dejong10}
are in conflict with those of \cite{sesar11}, who used the CFHT Legacy Survey.
\cite{sesar11} find the median metallicity to be independent of distance out to
at least 25~kpc.  The most recent SDSS results from \cite{schlaufman12}
find no gradient in the mean [Fe/H] in the halo out to at least 17~kpc.
Although ours is an in-situ sample, our data
is not ideal for this purpose, as
we have not taken into account the potential effect of the
selection function for EMP stars within the HES, which
strongly suppresses the higher metallicity part of the distribution
\citep{hes_mdf}, and thus tends to make the medians for the two distance
ranges approach each other.  Furthermore, due to lack of proper motions,
we do not know the 3D orbits of most of our sample stars, so there
may be considerable blurring of any distinction between the inner and
outer halo.

\begin{figure}
\epsscale{1.05}
\plotone{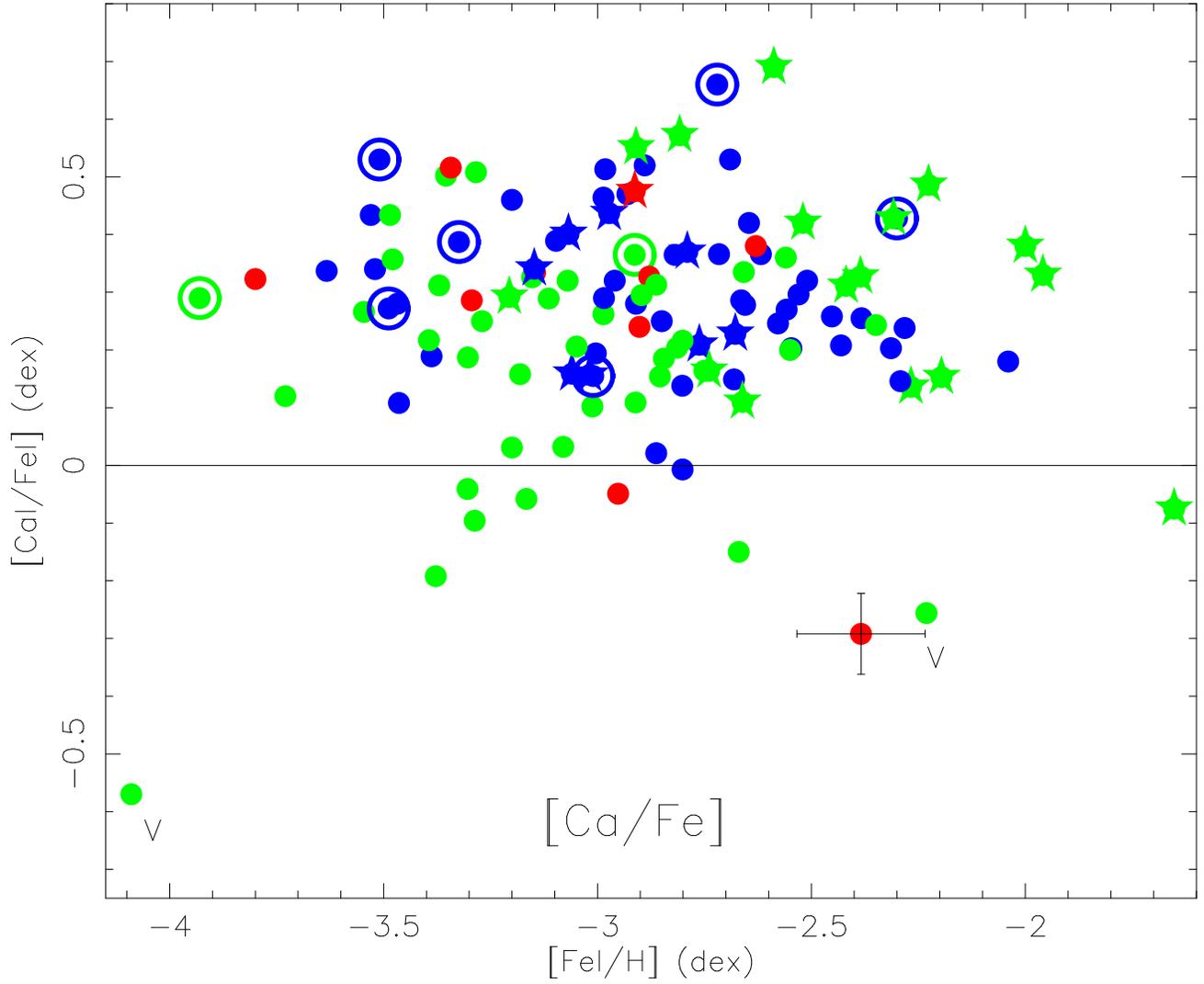}
\caption[]{As in Fig.~\ref{figure_4plot2}, but with colors denoting distance;
blue: close, green: inner halo, red: outer halo.  A typical error bar
is shown for a single star.
\label{figure_ca_distance} }
\end{figure}

We have also looked for distance effects on the abundance ratios.
The ratios [Mg/Fe] and [Ca/Fe] are of particular interest as there
have been suggestions in the literature  of a decline
in $\alpha$-enhancement between the inner and outer halo, which is
shown most clearly in the SDSS-based recent survey of the smooth
halo by \cite{schlaufman12} (see their Fig.~4).  However, we have
not found any
 systematic difference with distance for most of the abundance
ratios with suitable data from our in situ halo field star sample. 
In particular, as shown in
Fig.~\ref{figure_ca_distance}, the mean [Ca/Fe] ratio appears 
to be independent of distance.  However,
the lack of a strong gradient in $\alpha$/Fe with distance in
our sample may result at least in part from sampling issues;  our data are confined
to the most metal-poor population at all distances.

\begin{figure}
\epsscale{1.05}
\plotone{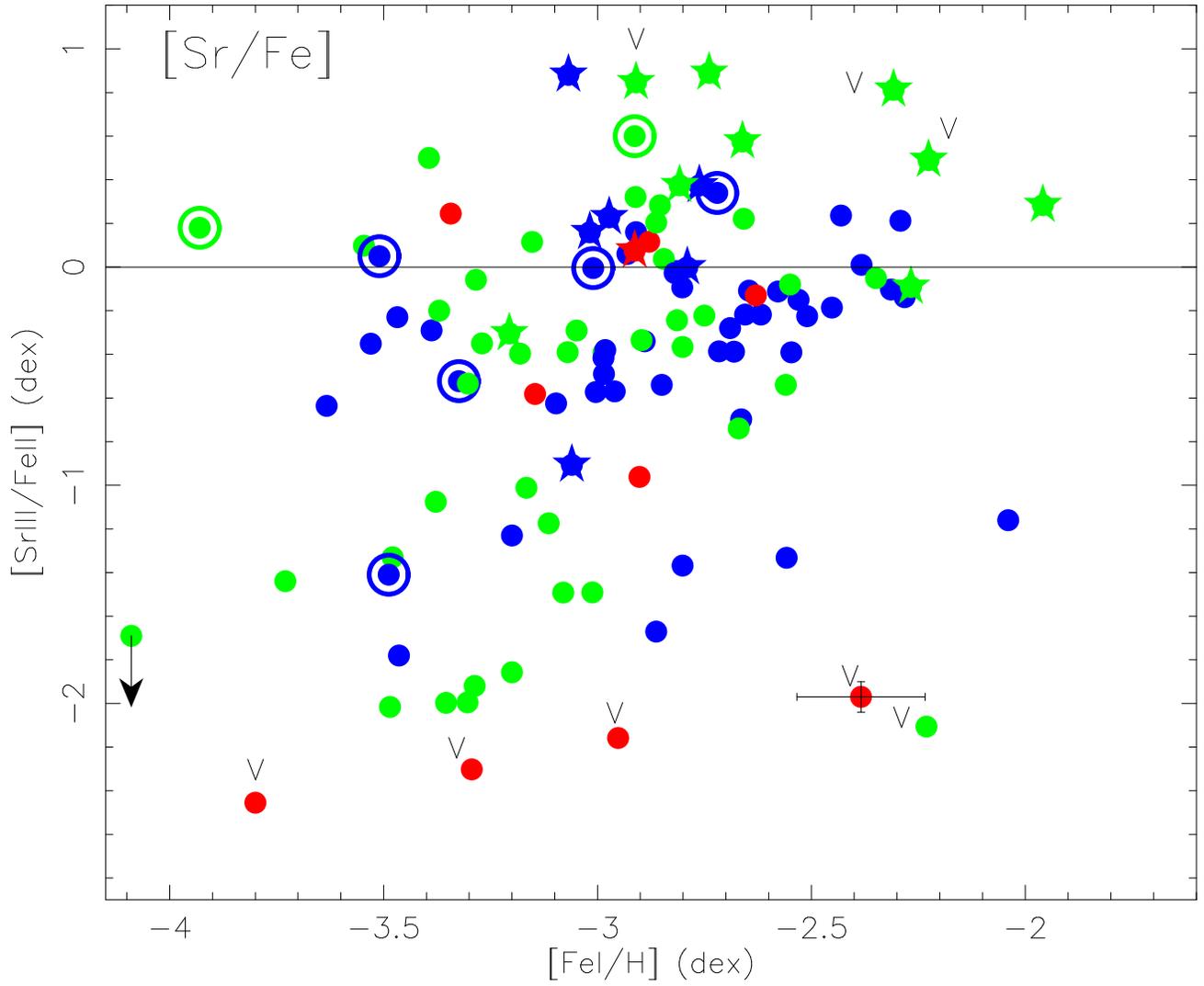}
\caption[]{The same for the ratio [Sr/Fe].   The ratio [Sr/Fe]
appears to be lower among the most distant stars than among the nearer
stars of the same [Fe/H]. 
\label{figure_sr_distance} }
\end{figure}

The only potential distance effect that we noticed in checking all
the abundance ratios for those species for which we had enough
data was for [Sr/Fe] and for [Ba/Fe].  As shown in Fig.~\ref{figure_sr_distance}
and in Table~\ref{table_ratio_distance},
there is a hint that the outer halo stars have systematically
ratios of [Sr/Fe] systematically lower by $\sim$0.4~dex
at a given [Fe/H] than do inner halo stars; 
[Ba/Fe] appears to display the same trend with distance, but is
complicated by the many upper limits
for  Ba abundances among our sample stars.
Statistical measures of abundance ratios as a function of distance 
are given in Table~\ref{table_ratio_distance}.  Ignoring [Si/Fe], where
there is an obvious analysis problem between dwarfs and giants 
(see \S\ref{section_abund_proc}),
the total range in the mean from $D < 4$~kpc to $D > 15$~kpc with
in situ measurements is large only for [Sr~II/Fe~II]
and for [Ba~II/Fe~II]; most of the other abundance ratios show maximum
differences in the mean abundance ratio between the 
three distance bins of 0.15~dex or less.

There have been a large number of studies of halo stars passing
through the solar neighborhood, see, e.g. \cite{stephens02},
\cite{fulbright02},
 \cite{gratton03}, \cite{roederer09}, \cite{nissen1}, \cite{nissen2}.
 These have the advantage that the stars, being much closer
 than our in situ halo sample, are brighter, and often
 have parallaxes, so that a Galactic orbit can be computed,
allowing a clean kinematic separation between inner and outer halo stars.
Overall, these studies find small systematic differences
in abundance ratios between  the mean [Mg/Fe] at a level
of 0.1 to 0.2~dex, with the outer halo stars having lower
$\alpha$/Fe.  Smaller such trends are sometimes claimed for Si, Ca, Ti, and Ni,
with even more subtle and smaller differences claimed by 
Nissen \& Schuster \citep{nissen1,nissen2} in the behavior of 
[Zn,Y, Ba/Fe].   Many of
the low $\alpha$ stars have retrograde orbits,
have ages 2~Gyr younger than those
of the normal-$\alpha$ stars \citep{nissen3}, and have $R_{apo} > 20$~kpc.
As indicated above, our sample, which lacks kinematic data, is not ideal
for this purpose.  Furthermore, we cannot hope to achieve the precision
abundances that can be obtained for the much brighter solar neighborhood stars selected
to have very similar stellar parameters;
a 0.1~dex difference in abundance ratio is
is too small to be detectable
in our sample of EMP stars given the uncertainties in our data.

\section{Comments on Nucleosynthesis and Related Issues \label{section_nucleosyn} }

Models of nucleosynthesis in SNIa and of SNII physics including
ejection mechanisms, fallback, and mixing within the ejecta
\citep{kobayashi06,tominaga07,heger10}
 have been developed that can
successfully explain the chemical inventory of (most) EMP
Galactic halo stars.  Here we highlight the outliers from
the general trends established by the many recent studies
of EMP halo field stars, the major one of which (in addition
to our own 0Z project), is the First Stars project \citep{cayrel04}.

\begin{figure}
\epsscale{1.05}
\plotone{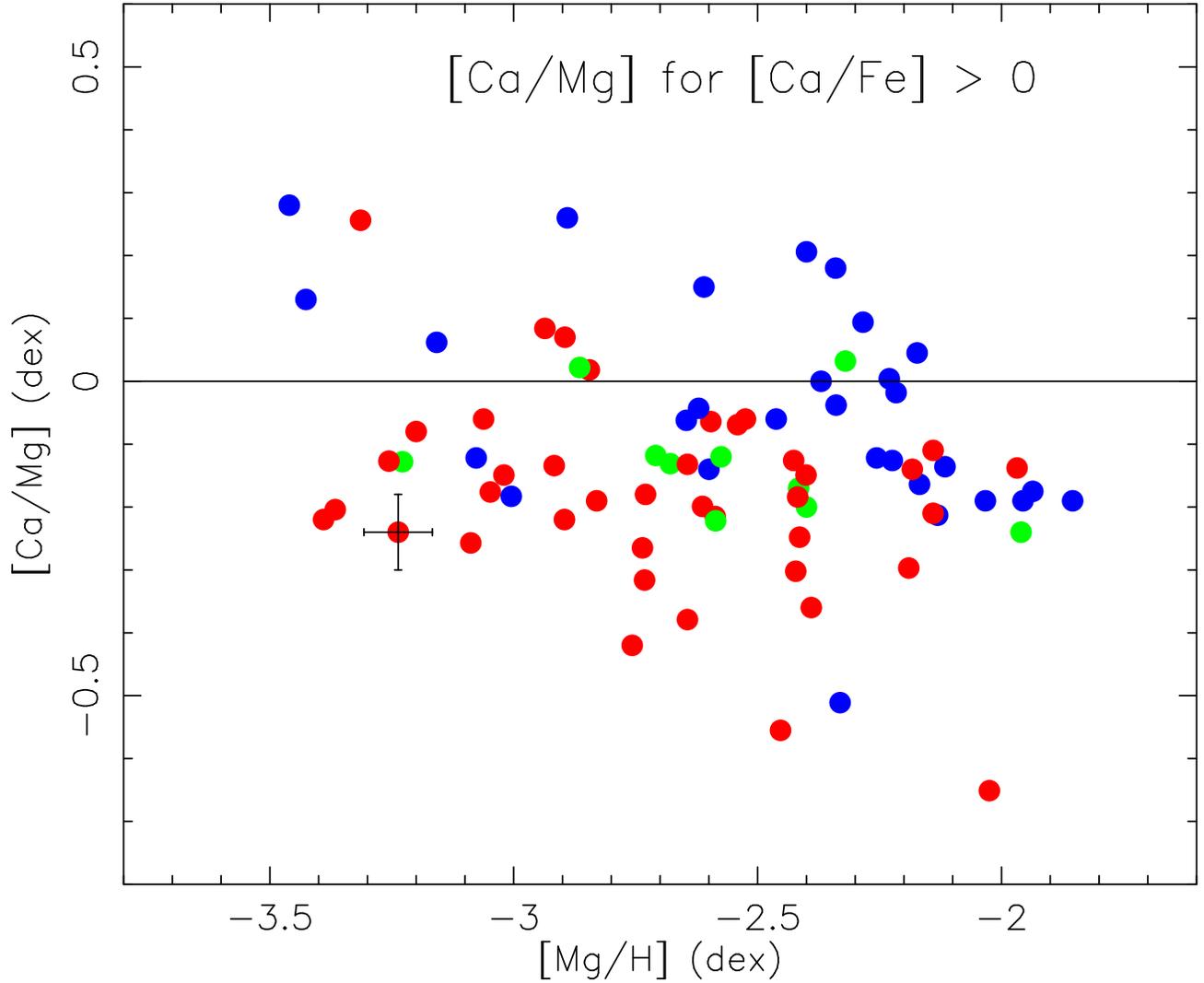}
\caption[]{[Ca/Mg] is shown as a function of [Mg/H] for the
C-normal stars with [Ca/Fe] $> 0$ in our 0Z sample.
Symbols are as in 
Fig.~\ref{figure_ioneq}.  Notes regarding checks of outliers are as in
Fig~\ref{figure_4plot1}.  A typical error bar is shown for a single star.
\label{figure_mgca} }
\end{figure}

Before doing so, we consider evidence from the 0Z project regarding
the possible variations in SNII progenitor mass. The ratio
Ca/Mg varies significantly with the progenitor mass of SNII,
see, e.g.,  the grid of SNII yields by \cite{heger10}.
Fig.~\ref{figure_mgca} displays that ratio for our 0Z sample as a function
of [Mg/H], ignoring the 
C-rich stars and the largest group of outliers, the low-Ca family.
It is apparent that there are small systematic offsets
between the mean relation for the giants, subgiants, and main
sequence turnoff region stars, with the giants having higher 
[Mg/Ca] than the subgiants, and those higher than the dwarfs.
This is presumably a result of the small dependence of non-LTE
corrections for Mg and for Ca on stellar parameters
discussed in \S\ref{section_nonlte}. The nature of these offsets and their
magnitude are close to those predicted by 
combining the calculations
of \cite{nonlte_mgk} and \cite{nonlte_ca} for 
non-LTE in giants vs dwarfs for Mg and for Ca respectively.

The most interesting point
of Fig.~\ref{figure_mgca} is that there is only a small change
in the ratio [Ca/Mg] with increasing Mg abundance, with a slope of approximately
$-$0.1~dex/dex for giants.  A similar figure was constructed by \
\cite{mcwilliam95b} (their Fig.~24), who found a very similar slope
for their sample of giants.
This strongly suggests that the mean SNII progenitor mass, and hence
the IMF (at least for massive stars capable of becoming SNII) varied
somewhat but not significantly
over the period of time during which the relevant SNII occurred.
Note that this only applies for the C-normal stars with [Ca/Fe] $> 0$~dex.

The many C-stars found among the EMP halo field stars can be explained
by binary mass transfer from an AGB companion as discussed
in \S\ref{section_cstars}.
Although theoretical population syntheses of C-rich star production via
binary mass transfer have trouble reproducing the observed high fraction
of C-rich stars quantitatively, as well as the apparent relative
absence of N-rich stars,
\citep[see, e.g.][]{izzard09}, our radial velocity surveys (as well as those
of others, see \S\ref{section_eqw_vr}) and corresponding
binary fractions strongly support the binary mass transfer scenario.

Ignoring the C-rich stars, the major class of outliers
is the low-Ca stars.  We believe that these are
the result of  variations of SNII nucleosynthesis
yields with progenitor mass and perhaps other factors.
Heavier progenitors produce more O and Mg with respect to Ca
and Ti, with the yields of \cite{nomoto06} showing
changes in the ratio of Mg to Ca of a factor of 18 
in the ejecta with progenitor mass increasing from 13 to 30~$M_{\odot}$.
This change could perhaps explain the low [Ca/Fe] family of stars
if one assumes a local effect (i.e. non-global, inhomogeneous mixing) from
a rare higher mass SN progenitor.  Presumably stochastic effects
from individual early SN and limited mixing in the proto-Galactic halo
led to stars with unusually low Ca/Fe
being formed at a low frequency ($\sim$10\%) in the young halo, and
not at all (or at a much lower frequency)
once the cumulative number of SNII contributing to the
chemical inventory of a star becomes much larger and the halo
became better mixed, as is the case  for
the higher-metallicity disk.  An explanation in terms of contributions
from prompt SNIa
may also be feasible, but has trouble reproducing the detailed behavior
of the $\alpha$-elements in the low-Ca EMP population.

%

A very peculiar halo red giant, this time with normal C,
a huge excess of Ca and a small deficit of Mg relative to its Fe,
was found by \cite{lai09}.  If the lighter $\alpha$-elements
such as Mg were also highly enhanced, one might view
this star as having a strong contribution from a PISN,
but Mg is not enhanced at all. This star is faint and quite cool
with very uncertain stellar parameters \citep[see][for details]{lai09},
and their results should be checked.
 If their result holds, it could have arisen from a very
low-mass progenitor Type II SN.  Alternatively, it could be heavily
contaminated by Type Ia SN ejecta.  We note that no C-normal 
star in our sample has [Ca/Fe] $> 0.55$~dex.

The recent surge in detailed abundance studies of the stellar population
in the dwarf satellites of the Milky Way 
\citep[see, e.g.][]{tolstoy09,kirby11}
has led to new insights on
how the star formation history and the influence of galactic winds
affects the chemical inventory of a low luminosity galaxy.     
In part motivated by such results, as well
as by the rapid improvement in the mass resolution of cosmological dark matter 
simulations, simulations
of the stellar component of galaxy halos have become detailed
enough to follow the assembly of the halo of a galaxy
such as the Milky Way.  These take into account both  in situ star formation as well
as accretion of stars from subhalos, and follow the impact
of the accretion on the chemical inventory within the Galaxy's
halo.  Such techniques, strongly focused on the $\alpha$/Fe ratios, 
were initially developed by \cite{bullock05};
\cite{zolotov10} and \cite{tissera12} 
have considerably advanced the art and are much
more successful in reproducing the observed mean trends in our survey,
although not the outliers.

Another converging thread for the study of EMP Milky Way halo stars
is the increasing quantity and quality of data on damped Lyman absorbers found
in the spectra of high redshift QSOs.  The latest study by
\cite{rafelski12} finds that the metallicity distribution
and $\alpha$/Fe ratios of $z > 2$ DLAs are 
consistent with being drawn from the same parent population
as those of Milky Way halo stars.  Furthermore \cite{cooke12} have found
a modestly C-enhanced DLA with [Fe/H] $\sim  -2.8$~dex.  \cite{fumagalli11}
recently broke the long-standing lower metallicity limit for 
high redshift gas clouds (Lyman limit systems seen in QSO spectra)
by finding two with upper limits of $Z < 10^{-3.8}~Z_{\odot}$,
which is well within the EMP Galactic halo star range we are exploring here.






\section{Summary}

Our 0Z survey had the goals of finding more EMP stars with
[Fe/H] $< -3.0$~dex and using them (and in particular
their chemical inventories) to improve our understanding
of the young Milky Way, focusing on the evolution
of its heavy element content, how well the ISM in the young
Galaxy was mixed, and what nucleosynthesis processes contributed.
This work began in 2000 with the Keck Pilot Project 
\citep{keck_pilot_1,keck_pilot_2},
and with this paper it concludes.
Although we have published many papers along the way describing
stars of particular interest, here we have attempted an overview
of the entire ensemble of 122 stars for which we have
obtained high dispersion spectra at either Keck or Magellan
and subsequently have carried out detailed abundance analysis
for each sample star.  Our sample is an in-situ probe of the
Milky Way halo, and 8\% of the sample is more distant than
15~kpc. 

The most interesting thing we have found is that, ignoring Carbon stars, 10\%  of
the sample has sub-solar [Ca/Fe], a phenomenon that is
virtually non-existent among solar neighborhood stars and
Galactic disk stars, which are considerably closer and with higher
metallicities than the 0Z sample, as well as among Galactic globular cluster stars. 
These low-Ca stars also have
slightly low abundances for the other $\alpha$-elements
we detect (Mg and Si), as well as 
low abundance ratios for the heavy neutron-capture elements.
There are two potential explanations for producing such
an unusual chemical inventory.  The first involves
a more top-heavy IMF in the young Milky Way,
biasing the mass of the SNII progenitors higher so that the
Mg/Ca ratio in the SN ejecta becomes higher
\citep{nomoto06}, thus producing a ``deficiency'' of Ca.
The second is a substantial contribution from 
prompt SNIa elevating the Fe-peak but not producing
much of anything else, but this hypothesis has some problems reproducing
the detailed $\alpha$-element distributions.

There are smaller groups of outliers within our sample,
of which the most unusual is a group of two C-rich stars
where the high C-enhancement is carried further through
to high N, Na, Mg, and/or Al.
Both of the two members of this class 
have [Fe/H] $< -3.5$~dex and presumably the lack of detectable C$_2$ bands
is a consequence of their very low Fe-metallicity.
These stars appear
``normal'' for the elements we can probe between
Ca and the Fe-peak, but then show relatively high heavy-neutron capture
element abundances (e.g. [Ba/Fe] +0.27 and $-$0.84~dex) 
given their very low [Fe/H].

Multiple epochs of high-dispersion
spectroscopy are available for a substantial fraction
of the sample stars.   While all the higher Fe-metallicity
C-stars that show $s$-process enhancement appear to be
binaries, and most of the stars with C$_2$ bands but no
$s$-process enrichment also appear to be binaries, there is
no evidence to support a binary origin for these
two anomalous C-rich stars.  We speculate that these
two stars may be remnants from mass transfer binary
systems involving a very low metallicity primary
which ended  its evolution as SN1.5 during or after the AGB
phase, such the primary was able to grow to the Chandrasekar
limit before it lost its envelope.  The primary thus would
explode and end its stellar evolution before becoming a white dwarf.

We have also found a new extreme $r$-process enhanced star
with [Eu/Fe] +1.6~dex.  As expected, for the 13 heavy neutron-capture
elements we could detect, its chemical inventory follows the
standard $r$-process abundance ratios.

Another surprise is the very low abundances of Sr and Ba found
in a substantial number of the EMP giants; 13 stars have
[Sr/Fe] $< -1.5$~dex and 8 lie below this value for Ba, with
a few more having upper limits slightly higher than 
[Ba/Fe] $=  -1.5$~dex.  The Sr/Ba ratio is well behaved as
a function of [Ba/H], but not so well behaved as a function of
[Fe/H], illustrating again the decoupling of the production
of the heavy neutron-capture elements from that of Fe.

The behavior of C-rich stars changes below [Fe/H] $-3.2$~dex;
no stars showing bands of C$_2$ were seen, whereas this
is common among higher Fe-metallicity stars.  Some
potential contributing factors were discussed, including
the possibility of substantial O enhancement keeping
C/O $< 1$.

Overall, ignoring C-stars, approximately 15\% of the sample are strong
outliers in one or more elements between Mg and Ni.  This rises to
$\sim$19\% if very strong outliers
for Sr and Ba are included.
Ignoring the strong outliers, the dispersions around linear fits
of [X/Fe] vs [Fe/H] for the C-normal stars are comparable to those predicted from
the observational and analysis uncertainties for Si, Sc, and Ni.
They rise to $\sim$1.7 times the expected values for Ti, Cr, and Mn,
and to $\sim$2 for Mg and Ca, rising to $\sim4$ for Sr and Ba.
The dispersion around a similar linear fit for
[Ca/Mg] vs [Mg/H] shows a small scatter once a dependence
on \teff, presumably from non-LTE effects, is removed.

The most metal-poor star in our sample has [Fe/H] $-4.1$~dex.
Given how arduous was the path to find it, future searches for
UMP or HMP stars  
will require a survey where extensive screening to eliminate
higher metallicity interlopers is not necessary. The initial photometry 
(presumably including a narrow band filter centered on the
3933~\AA\ Ca~II line) or spectra must discriminate well
enough to cleanly isolate such stars for detailed study,
which was not the case for the HES.  SDSS spectra are much more
suitable, but UMP and HMP stars are so rare that it may require
an ``all star'' survey 
to find them in the halo of the Milky Way.  While 
selected dSph satellites of
the Milky Way have much less advanced chemical evolution, and
mean Fe/H values for their member stars more than a factor of 10 
below that of the halo of the Milky Way, their low luminosity
and in many cases large distances may render them less productive
than a Milky Way halo field star survey.

\acknowledgements

We are very grateful
to the Palomar, Las Campanas, and Keck time allocation committees for
their long-term support of
this  campaign during the initial phase of moderate resolution 
spectroscopy which began
began in 2000 and ended in 2006 as well as the subsequent high resolution
spectroscopy.
J.~Cohen acknowledges partial support from NSF grants
AST--0507219 and AST--0908139. I.~Thompson acknowledges partial
support from NSF AST-0507325. ENK acknowledges support
from the Southern California Center for Galaxy Evolution, a multicampus
research program funded by the University of California Office of Research,
and partial support from NSF grant AST-1009973.
This work was partially supported by Sonderforschungsbereich SFB 881 ``The Milky
Way System'' (subproject A4) of the German Research Foundation (DFG).
We are grateful to the many people  
who have worked to make the Keck Telescopes and their instruments,
and the Magellan Telescopes and their instruments,  
a reality and to operate and maintain these observatories.
The authors wish to extend special thanks to
those of Hawaiian ancestry on whose sacred mountain we are privileged
to be guests.  Without their generous hospitality, none of the
observations presented herein would have been possible.

\clearpage

Appendix A

Notes on Individual Stars

The hot turnoff region dwarf HE0102$-$0633 has very strong Na D lines and a very
strong resonance line of K~I at 7699~\AA.   This star has a high reddening for
a EMP star from the HES, E($B-V$) = 0.14~mag, and $v_r~ = ~ +2$~km s$^{-1}$.  Hence
we only derive an upper limit for [Na/Fe] for this star.
We believe these features are
mostly  interstellar in origin.

A similar situation appears to hold for much cooler star HE1416$-$1032.
The detected features include strong NaD lines and an upper limit
for the 5688~\AA\ Na~I line, but $v_r$ for this star is $\sim$ 0 \kms, and
E($B-V$) is also rather high for the present sample, 0.08 mag.
One ISM component can be seen about 0.4~\AA\ to blue of each component
of the NaD lines, and there may be other components within the
feature we consider the stellar NaD lines.  We consider the equivalent
widths of all three Na~I features as upper limits, and set
the stellar [Na/Fe] upper limit to the most stringent of these, which
comes from the 5688~\AA\ line.

{}

\clearpage


   




\begin{deluxetable}{l r r r r r   rrrr}
\tabletypesize{\footnotesize}
\tablewidth{0pt}
\tablecaption{Sample of Candidate EMP Stars, Mostly from the HES
\label{table_sample} }
\tablehead{
\colhead{ID} & \colhead{RA} & \colhead{Dec} & \colhead{V} & 
   \colhead{Date} & \colhead{t(exp)} & \colhead{S/N\tablenotemark{a}} &
   \colhead{Spec Mode\tablenotemark{b}} & \colhead{$v_r$\tablenotemark{c}} \\
\colhead{ } & \colhead{(2000)} &  \colhead{(2000)} & \colhead{(mag)} &  \colhead{} & 
  \colhead{(sec)} & \colhead{} & \colhead{} & \colhead{(km~s$^{-1}$)}
}
\startdata

HE0007$-$1832 & 00 09 52.8 & $-$18 16 12 & 15.46 & 9/2001 & 8400 & 90 & O & 26.4 \\
HE0009$-$6039 & 00 11 31.0 & $-$60 22 21 & 16.54 & 9/2004 & 1800 & 35 &   M & 25.0 \\
HE0011$-$0035 & 00 14 15.8 & $-$00 18 59 & 15.04 & 9/2001 & 9600 & 100 & O & $-170.6$ \\
HE0012$-$1441 & 00 15 27.1 & $-$14 24 37 & 16.40 & 5/2002 & 10,800 & 80 & O & 11.2 \\
HE0017$-$4346 & 00 20 16.2 & $-$43 30 18 & 15.88 & 9/2004 & 1800 & 57  & M & 58.9 \\
HE0024$-$2523 & 00 27 27.6 & $-$25 06 26 & 14.91 & 9/2000 & 6000 & 115 & O & $-181.6$ \\
HE0025$-$2334 & 00 28 29.2 & $-$23 17 57 & 14.95 & 9/2001 & 7200 & 100 & O & 12.7 \\
HE0028$-$2152 & 00 31 07.7 & $-$21 36 07 & 15.80 & 9/2002 & 9600 & 130 & O & 162.5 \\
\enddata
\tablenotetext{a}{SNR per spectral resolution element in the continuum near 5300~\AA.
(HIRES: spectral resolution 36,000.  MIKE 42,000 blue camera, 32,000 red camera).}
\tablenotetext{a}{Spectrograph mode: O = HIRES prior to detector upgrade,
     B = HIRES-B after upgrade, R = HIRES-R after upgrade, M = Mike/Magellan}
\tablenotetext{c}{Heliocentric radial velocity.}
\tablenotetext{d}{V mag adopted from the HES in the absence of a better source.}
\end{deluxetable}


\clearpage

\begin{deluxetable}{l r r r r r}
\tablewidth{0pt}
\tablecaption{$v_r$ for EMP Stars with Multiple Spectra From HIRES
\label{table_mult_vr} }
\tablehead{
\colhead{ID} & \colhead{[Fe/H](Hires)} & \colhead{[C/Fe])} & 
     \colhead{$T_{eff}$} & 
   \colhead{$s$-rich\tablenotemark{a} } & \colhead{Notes} \\
\colhead{ } & \colhead{(dex)} &  \colhead{(dex)} & \colhead{(K)} &  \colhead{(+ or 0)} \\ 
\colhead{ } & \colhead{Date} & \colhead{$v_r$} & \colhead{($\sigma v_r$\tablenotemark{b})} \\ 
\colhead{ } &  \colhead{ } &  \colhead{(km~s$^{-1}$)} & \colhead{(km~s$^{-1}$)}
}
\startdata
C-Stars & with C$_2$ \\
HE0007$-$1832 & $-2.79$ & 2.7 & 6515 & 0 \\
~ & 9/2001 & 27.7 & 0.2 \\
~ & 9/2002 & 26.9 & 0.5 \\
~ & 9/2004 & 27.7 & 0.5 \\
HE0012$-$1441 & $-2.52$ & 1.6 & 5730 & + & c \\
HE0024$-$2523 & $-2.72$ & 2.6 & 6625 & + & d \\
HE0058$-$0244\tablenotemark{e} & $-2.76$ & 1.7 & 5620 & +  \\
\enddata
\tablenotetext{a}{High [Ba/Fe] (+) vs normal [Ba/Fe] (0).}
\tablenotetext{b}{Uncertainty in the mean $v_r$ computed from the rms dispersion.}
\tablenotetext{c}{Spectrum shows double lines.}
\tablenotetext{d}{\cite{keck_pilot_3} gives an orbit for this
close binary with period 3.14 days and with full amplitude of $v_r$ of 104~km~s$^{-1}$.}
\tablenotetext{e}{HE0058$-$0244 is a rediscovery of CS22183-015, see
        \cite{johnson02} and \cite{tsangarides04}. }
\tablenotetext{f}{MIKE/Magellan spectrum}
\tablenotetext{g}{Andersen, Hansen \& Nordstr\"om (2013)
   found this star to be a long period binary.}
\tablenotetext{h}{Andersen, Hansen \& Nordstr\"om (2013)
   have determined a period of 289 days for this binary.}  
\tablenotetext{i}{A spectrum with low SNR.}
\tablenotetext{j}{HE2356$-$0410 is a rediscovery of CS22957-27, previously known
    binary with $v_r$ range $-62$ to $-78$~km~s$^{-1}$  \citep{preston01}.}
\tablenotetext{k}{HE2344$-$2800 is a rediscovery of CS22966-048.}
\tablenotetext{l}{HE2323$-$0256 is a rediscovery of CS22949-037.}
\tablenotetext{m}{Double lined \citep{keck_pilot_1}.}
\end{deluxetable} 

\clearpage

\begin{deluxetable}{l rrr | llr  r}
\tablewidth{0pt}
\tablecaption{Nominal and Adopted Stellar Parameters
\label{table_param} }
\tablehead{
\colhead{ID} & \colhead{} & \colhead{Nominal\tablenotemark{a}} & {} & \colhead{} &
 \colhead{Adopted} \\
\colhead{} & \colhead{$T_{eff}$} & \colhead{$\sigma(T_{eff}$)\tablenotemark{b} } 
    & \colhead{log($g$)} &  
    \colhead{$T_{eff}$} & \colhead{log($g$)} & 
   \colhead{$v_t$} & \colhead{[Fe/H]\tablenotemark{c}} \\
\colhead{} & \colhead{(K)} & \colhead{(K)} & \colhead{dex)}  & 
  \colhead{(K)} & \colhead{(dex)} & \colhead{(km~s$^{-1}$)} &
  \colhead{(dex)}
}
\startdata
HE0007$-$1832 &  6500 &   30 &   3.8 &  6500 &   3.8 &   1.7 & $-2.79$  \\
HE0009$-$6039 &  5270 &  150 &   2.7 &  5250 &   2.7 &   1.8 & $-3.29$  \\ 
HE0011$-$0035 &  4955 &   25 &   1.9 &  4950 &   1.8 &   2.0 & $-2.99$  \\ 
HE0012$-$1441 &  5610 &  155 &   3.3 &  5750 &   3.5 &   1.6 & $-2.52$  \\ 
HE0017$-$4346 &  6198 &   53 &   3.8 &  6198 &   3.8 &   1.2 & $-3.07$ \\
HE0024$-$2523\tablenotemark{d}  &  
                 6625 & \nodata & 4.3 & 6625 & 4.3 & 0.6 & $-2.72$ \\
HE0025$-$2334 &  6168 &   75 &   3.8 &  6168 &   3.8 &   1.4 & $-2.80$  \\ 
HE0028$-$2152 &  6150 &   67 &   3.6 &  6150 &   3.6 &   1.6 & $-2.72$  \\
\enddata
\tablenotetext{a}{The nominal $T_{eff}$ and log($g$) are based solely
  on the available photometry, specifically the mean of that deduced
  from each of $V-I$, $V-J$, $V-K$.  There
  are two possible solutions for stars near the main sequence turnoff; the one
  adopted is given.}
\tablenotetext{b}{Dispersion about the mean for the values (maximum of three)
   of $T_{eff}$.}
\tablenotetext{c}{Value from our high-resolution analysis
   adopting [Fe/H](Sun) = 7.45~dex.}   
\tablenotetext{d}{See Keck Pilot Project analyses,   
    \citep{keck_pilot_1,keck_pilot_2,keck_pilot_3}. } 
\tablenotetext{e}{The HIRES spectrum of this star suggests that it is
   on the horizontal branch, not the RGB.}   
\end{deluxetable}

\clearpage

\begin{deluxetable}{l l rrrrr}
\tablewidth{0pt}
\tablecaption{Abundance Ratio Changes for Small Changes in Stellar 
Parameters for Giants\tablenotemark{a}
\label{table_unc_ratio_giants}}
\tablehead{\colhead{Species} & \colhead{$\sigma$[X/Fe]\tablenotemark{b}} &
 \colhead{${\Delta}$log$\epsilon(X)$} &
 \colhead{${\Delta}$log$\epsilon(X)$} &
 \colhead{$\Delta$log$\epsilon(X)$} &
 \colhead{$\Delta$log$\epsilon(X)$} & \colhead{$\Delta$log$\epsilon(X)$} \\
\colhead{} & \colhead{1 Star} & \colhead{$T_{eff}$} & \colhead{\grav} &
   \colhead{[Fe/H]} & \colhead{$v_t$} & \colhead{\eqw} \\
\colhead{} & \colhead{(1$\sigma$)(dex)} & \colhead{(dex)} & 
  \colhead{(dex)} & \colhead{(dex)} & \colhead{(dex)} & \colhead{(dex)}   
}
\startdata
{$[$}FeI/H{$]$}\tablenotemark{c} & 0.17 &  0.11 & $-$0.02 & $-$0.04 & $-0.05$  & 0.01 \\
{$[$}FeII/H{$]$}\tablenotemark{c} & 0.10 & 0.00 & $-0.08$ & $-0.03$ & $-0.03$ & 0.04  \\
{$[$}FeII/FeI{$]$}\tablenotemark{d} &    0.14 &  0.07 &     0.11 &     0.04 &     0.01 &     0.04    \\
{$[$}TiII/TiI{$]$}\tablenotemark{d} & 0.12 & 0.08 & $-0.08$ & $-0.01$ & $-0.02$ & 0.03 \\
~ \\
C(CH) &  0.13 & 0.11 & $-0.04$ & $-0.04$ & 0.00 & 0.05\tablenotemark{e} \\
N(NH) & 0.17 & 0.09 & $-0.12$ & 0.05 & $-0.05$ & 0.05\tablenotemark{e} \\
NaI &    0.10 &     0.01 &    $-$0.07 &    $-$0.01 &     0.03 &     0.06   \\
\enddata
\tablenotetext{a}{Computed from the line list of HE1416$-$1032 with selected
additions as required.}
\tablenotetext{b}{1$\sigma$ uncertainty in [X/Fe] for a single (typical) star.
Fe~II is used for the ionized species. The adopted uncertainties
in stellar parameters ($+100$~K in $T_{eff}$, $+0.25$ dex in \grav,
$+0.2$~dex in [Fe/H] of the adopted model stellar atmosphere, and
$+0.2$~\kms\ in $v_t$) and from the uncertainties in the measured
\eqw\ are discussed in \S\ref{section_abund_proc}.
Potential systematic errors are not included.}
\tablenotetext{c}{Uncertainty in log$\epsilon$(Fe) 
 of a single (typical) star for the adopted uncertainties
 in stellar parameters.  Potential systematic errors are
 not included.}
\tablenotetext{d}{These are of the form [XI/H] -- [XII/H], not [XI/FeI] -- [XII/FeII].}
\tablenotetext{e}{A minimum uncertainty in fitting a synthesis
 to an observed high resolution spectrum of either CH or NH
 bands corresponding to 0.05~dex is assumed.}
\tablenotetext{f}{HFS was used when necessary.  See \S\ref{section_other_anal} for details.} 
\tablenotetext{g}{A larger uncertainty is assigned to the far UV $W_{\lambda}$.}
\end{deluxetable}

\clearpage

\begin{deluxetable}{l l rrrrr}
\tablewidth{0pt}
\small
\tablecaption{Abundance Ratio Changes for Small Changes in Stellar 
   Parameters for Turnoff Region Stars \tablenotemark{a}
\label{table_unc_dwarfs}}
\tablehead{\colhead{Species} & \colhead{$\sigma$[X/Fe]\tablenotemark{b}} &
  \colhead{${\Delta}$log$\epsilon(X)$} &
  \colhead{${\Delta}$log$\epsilon(X)$} &
  \colhead{$\Delta$log$\epsilon(X)$} &
  \colhead{$\Delta$log$\epsilon(X)$} \\
\colhead{} & \colhead{1 Star} & \colhead{$T_{eff}$} & \colhead{\grav} &
   \colhead{[Fe/H]} & \colhead{$v_t$} & \colhead{\eqw} \\
\colhead{} & \colhead{(1$\sigma$)(dex)} & \colhead{(dex)} & 
  \colhead{(dex)} & \colhead{(dex)} & \colhead{(dex)} & \colhead{(dex)}  
}
\startdata
$[$FeI/H$]$\tablenotemark{c}  &  
     0.13  & $-$0.08 & 0.00    & 0.04  &  0.04 & 0.01 \\
$[$FeII/H$]$\tablenotemark{c} & 
     0.09  & $-$0.02 & $-$0.07 & 0.03  &  0.03 & 0.03 \\
$[$FeII/FeI$]$\tablenotemark{d} & 
    0.10 &  0.06 & $-$0.07 & $-$0.01 & $-$0.01 & 0.02  \\
$[$TiII/TiI$]$\tablenotemark{d} &    
    0.12 & 0.06 & $-$0.07 & $-$0.01 & 0.04 & $-$0.06 \\
~ \\
NaI &  0.06 &   0.01 &   0.00 &  $-$0.01 &   0.02 &   0.06  \\ 
MgI &  0.07 &   0.02 &   0.02 &  $-$0.03 &  $-$0.03 &   0.05  \\ 
AlI &  0.06 &   0.00 &   0.00 &   0.00 &  $-$0.01 &   0.06  \\
\enddata
\tablenotetext{a}{Computed from the line list of 
  HE2344$-$2800, except for EuII, which is from HE0458$-$1346.}
\tablenotetext{b}{1$\sigma$ uncertainty in [X/Fe] for a single (typical) star.
  Fe~II is used for the ionized species. The adopted uncertainties
  in stellar parameters ($+100$~K in $T_{eff}$, $+0.2$ dex in \grav,
  $+0.2$~dex in [Fe/H] of the adopted model stellar atmosphere,
  $+0.2$~\kms\ in $V_t$, and from the uncertainties in the measured
  \eqw\ are discussed in \S\ref{section_abund_proc}.
  Potential systematic errors are not included.}
\tablenotetext{c}{Uncertainty in log$\epsilon$(Fe) 
  of a single (typical) star for the adopted uncertainties
  in stellar parameters.  Potential systematic errors are
  not included.}
\tablenotetext{d}{These are of the form [XI/H] -- [XII/H], not 
     [XI/FeI] -- [XII/FeII].}
\tablenotetext{e}{HFS was used when necessary.  See \S\ref{section_other_anal} for details.}
\tablenotetext{f}{A larger uncertainty is assigned to the far UV $W_{\lambda}$.}
\end{deluxetable}

\clearpage

\begin{deluxetable}{l rrrr }
\tablewidth{0pt}
\tablecaption{Slopes of Linear Fits  to Abundances of Individual Fe~I Lines
\label{table_feslopes} }
\tablehead{
\colhead{ID} & \colhead{N lines} & \colhead{} & {Slopes} & \colhead{} \\
\colhead{} & \colhead{} & \colhead{EP\tablenotemark{a}} & 
\colhead{$W_{\lambda}/\lambda$\tablenotemark{b}} & 
\colhead{$\lambda$\tablenotemark{c}}  \\
\colhead{} & \colhead{(Fe~I)} & \colhead{(dex/eV)} &\colhead{(dex)} &
    \colhead{(dex/\AA)}
}
\startdata
HE0007$-$1832 &    18 &  \nodata &  \nodata & \nodata    \\   
HE0009$-$6039 &    31 &   $-$0.017 &   $-$0.043 &   0.4672E$-$04    \\   
HE0011$-$0035 &    63 &   $-$0.092 &   $-$0.021 &   0.1104E$-$04    \\   
HE0012$-$1441 &    14 &  \nodata &  \nodata &  \nodata   \\   
HE0017$-$4346 &    15 &  \nodata & \nodata & \nodata   \\   
HE0025$-$2334 &    50 &   $-$0.022 &   $-$0.045 &  $-$0.2217E$-$04    \\   
HE0028$-$2152 &    55 &    0.006 &   $-$0.078 &   0.1138E$-$03    \\   
HE0037$-$0348 &    72 &   $-$0.081 &    0.016 &  $-$0.1647E$-$04    \\
\enddata
\tablenotetext{a}{ This is $\Delta[log(\epsilon(Fe)]/ \Delta(\chi$),
  where $\chi$ is the excitation potential of the lower level of the line.}
\tablenotetext{b}{ This is $\Delta[log(\epsilon(Fe)]/ \Delta(W_{\lambda}/\lambda)$.}
\tablenotetext{c}{This is $\Delta[log(\epsilon(Fe)]/ \Delta(\lambda$), where $\lambda$
  is the wavelength of the transition.}
\end{deluxetable}

\clearpage

\begin{deluxetable}{l l  rrrrr}
\tablewidth{0pt}
\tablecaption{Atomic Data For Lines Used
\label{table_lines} }
\tablehead{
\colhead{$\lambda$} & \colhead{Elem\tablenotemark{a}} & \colhead{$\chi$} &
    \colhead{log(gf)} & \colhead{No. Stars} \\
\colhead{(\AA)} & \colhead{ } & \colhead{(eV)} & \colhead{(dex)} & 
     \colhead{Detected\tablenotemark{b}} 
}
\startdata

3189.30 &   108.0\tablenotemark{c} &    1.03 &  $-$1.990 &    2   \\
3255.50 &   108.0\tablenotemark{c} &    1.30 &  $-$1.940 &    3   \\
6707.80 &     3.0 &    0.00 &   0.178 &    8   \\
6300.30 &     8.0 &    0.00 &  $-$9.780 &    5  \\
6363.78 &     8.0 &    0.02 & $-$10.300 &    2   \\
7771.94 &     8.0 &    9.15 &   0.369 &    6   \\
7774.17 &     8.0 &    9.15 &   0.223 &    2   \\
7775.39 &     8.0 &    9.15 &   0.001 &    2   \\
\enddata
\tablenotetext{a}{Element code: atomic number + (ionization stage/10),
  i.e. 26.0 = Fe~I, 26.1 = Fe~II.}
\tablenotetext{b}{Upper limits are also included.} 
\tablenotetext{c}{UV OH lines.} 
\end{deluxetable}

\clearpage

\begin{deluxetable}{l l  rrrrrr rrrrrr}
\tabletypesize{\scriptsize}
\tablewidth{0pt}
\tablecaption{Equivalent Widths, First 12 Stars
\label{table_eqw} }
\tablehead{
\colhead{$\lambda$} & \colhead{Elem\tablenotemark{a}} & \colhead{W$_{\lambda}$} &
    \colhead{W$_{\lambda}$} & \colhead{W$_{\lambda}$} & \colhead{W$_{\lambda}$} &
    \colhead{W$_{\lambda}$} & \colhead{W$_{\lambda}$} & \colhead{W$_{\lambda}$} &
    \colhead{W$_{\lambda}$} & \colhead{W$_{\lambda}$} & \colhead{W$_{\lambda}$} &
    \colhead{W$_{\lambda}$} &  \colhead{W$_{\lambda}$} \\
\colhead{(\AA)} & \colhead{ } & \colhead{(m\AA)} & \colhead{(m\AA)} & \colhead{(m\AA)} &
     \colhead{(m\AA)} & \colhead{(m\AA)} & \colhead{(m\AA)} & \colhead{(m\AA)} &
     \colhead{(m\AA)} & \colhead{(m\AA)} & \colhead{(m\AA)} & \colhead{(m\AA)} &
     \colhead{(m\AA)} \\
\colhead{ } & \colhead{HE} & \colhead{0007} & \colhead{0009} & \colhead{0011} & 
   \colhead{0012} & \colhead{0017} & \colhead{0025} &
      \colhead{0028} & \colhead{0037} & \colhead{0039} & 
     \colhead{0048} & \colhead{0054} & \colhead{0055} \\
\colhead{ } & \colhead{ } & \colhead{-1832} & \colhead{-6039} & \colhead{-0035} &
    \colhead{-1441} & \colhead{-4346} &  \colhead{-2334} &
     \colhead{-2152} & \colhead{-0348} &\colhead{-4154} & 
     \colhead{-0611} & \colhead{-2123} & \colhead{-2314} 
}
\startdata

6707.80 &     3.0 &    \nodata &   $-$10.0 &    \nodata &    \nodata &    \nodata &    \nodata &    \nodata &    \nodata &    \nodata &    \nodata &    \nodata &    29.7  \\
5889.95 &    11.0 &    \nodata &    78.0 &    \nodata &    \nodata &   154.0 &    \nodata &    \nodata &   133.1 &   141.0 &   139.8 &    \nodata &    64.9  \\
5895.92 &    11.0 &    \nodata &    71.0 &    \nodata &    \nodata &   125.0 &    \nodata &    \nodata &   111.0 &   122.4 &   124.2 &    \nodata &    51.5  \\
3829.36 &    12.0 &    \nodata &   101.0 &    \nodata &    \nodata &    \nodata &    \nodata &    \nodata &   145.2 &    \nodata &    \nodata &    \nodata &    \nodata  \\
4057.52 &    12.0 &    \nodata &    \nodata &    \nodata &    34.0 &    \nodata &    \nodata &     9.6 &    14.5 &    \nodata &    29.1 &     7.2 &    \nodata  \\
4167.28 &    12.0 &    25.0 &    \nodata &    36.7 &    \nodata &    \nodata &    \nodata &    17.6 &    27.5 &    28.0 &    47.6 &    10.8 &    16.5  \\
4703.00 &    12.0 &    24.3 &    17.1 &    56.9 &   100.7 &    31.5 &    21.9 &    37.5 &    35.2 &    42.7 &    60.4 &    23.4 &    26.0  \\
5172.70 &    12.0 &   105.6 &    85.0 &   167.7 &   258.9 &   125.1 &    94.5 &   124.0 &   151.1 &   155.0 &   176.8 &   124.0 &   120.7  \\
\enddata
\tablenotetext{b}{Element code: atomic number + (ionization stage/10),
  i.e. 26.0 = Fe~I, 26.1 = Fe~II.}
\tablenotetext{b}{Negative W$_{\lambda}$ denote upper limits.}
\end{deluxetable}

\clearpage

\begin{deluxetable}{r r | r r | r r}
\tabletypesize{\normalsize}
\tablewidth{0pt}
\tablecaption{Adopted Solar Abundances For  HES Stars
\label{table_solar_abund} }
\tablehead{
  \colhead{Elem\tablenotemark{a}} &  \colhead{Log[$\epsilon$(X)] } & 
   \colhead{Elem\tablenotemark{a}} &  \colhead{Log[$\epsilon$(X)] } & 
    \colhead{Elem\tablenotemark{a}} &  \colhead{Log[$\epsilon$(X)] }     \\
  \colhead{ } & \colhead{(dex)} & \colhead{ } & \colhead{(dex)} &
     \colhead{ } & \colhead{(dex)} 
}
\startdata
    6.0  &       8.59  &   7.0  &       7.93 &  8.0  &       8.93 \\
   11.0  &       6.32  &  12.0  &       7.54 & 13.0  &       6.47   \\
   14.0  &       7.55  & 19.0  &       5.12 & 20.0  &       6.36 \\
   20.1  &       6.36  & 21.0  &       3.10 & 21.1  &       3.10  \\
   22.0  &       4.99  & 22.1  &       4.99 & 23.0  &       4.00  \\
   23.1  &       4.00  &  24.0  &       5.67 & 24.1  &       5.67 \\
   25.0  &       5.39  &    25.1  &       5.39  &  26.0  &       7.45  \\
   26.1  &       7.45  & 26.2  &       7.45  & 27.0  &       4.92 \\
   28.0  &       6.25  & 29.0  &       4.21 &  30.0  &       4.60 \\
   38.0  &       2.90  & 38.1  &       2.90   & 39.1  &       2.24 \\
   40.1  &       2.60  &   56.1  &       2.13  &  57.1  &       1.14  \\
   58.1  &       1.55  & 59.1  &       0.71  & 60.1  &       1.50   \\
   62.1  &       1.00  & 63.1  &       0.51  & 64.1  &       1.12 \\
   66.1  &       1.10  & 82.0  &       1.95 &  90.1  &       0.12 \\
\enddata
\tablenotetext{a}{Species code: atomic number + (ionization stage/10),
  i.e. 26.0 = Fe~I, 26.1 = Fe~II.  Special case: 26.2 is for three 
  strong Fe~II lines
  in the UV near 3260~\AA.}
\tablenotetext{b}{Adopted log[$\epsilon$(X)] where the value for H is 12.00.}
\end{deluxetable}

\clearpage

\begin{deluxetable}{r rrr rrr rrr rrr rrr}
\tabletypesize{\tiny}
\tablewidth{0pt}
\tablecaption{Relative Abundances, Stars 1 to 5 From The HES
\label{table_abund_merge_1to10} }
\tablehead{
\colhead{} &  \multicolumn{3}{c}{HE0007$-$1832} &  \multicolumn{3}{c}{HE0009$-$6039}  &
     \multicolumn{3}{c}{HE0011$-$0035}  &   \multicolumn{3}{c}{HE0012$-$1441}  & 
     \multicolumn{3}{c}{HE0017$-$4346} \\
  \colhead{Elem\tablenotemark{a}} & 
     \colhead{[X/Fe]\tablenotemark{b}} & \colhead{$\sigma$\tablenotemark{c}} &
        \colhead{N\tablenotemark{d}} &     
  \colhead{[X/Fe]\tablenotemark{b}} & \colhead{$\sigma$\tablenotemark{c}} &
        \colhead{N\tablenotemark{d}} &
  \colhead{[X/Fe]\tablenotemark{b}} & \colhead{$\sigma$\tablenotemark{c}} &
        \colhead{N\tablenotemark{d}} &
  \colhead{[X/Fe]\tablenotemark{b}} & \colhead{$\sigma$\tablenotemark{c}} &
        \colhead{N\tablenotemark{d}} &
  \colhead{[X/Fe]\tablenotemark{b}} & \colhead{$\sigma$\tablenotemark{c}} &
        \colhead{N\tablenotemark{d}} \\       
\colhead{} & \colhead{(dex)} & \colhead{(dex)} & \colhead{} &
     \colhead{(dex)} & \colhead{(dex)} & \colhead{} & 
     \colhead{(dex)} & \colhead{(dex)} & \colhead{} & 
     \colhead{(dex)} & \colhead{(dex)} & \colhead{} &
     \colhead{(dex)} & \colhead{(dex)} & \colhead{} 
}
\startdata
  26.0 &  $-$2.79 &    0.16 &  18 &  $-$3.29 &    0.21 &  31 &  $-$2.99 &    0.20 &  64 &  $-$2.52 &    0.16 &  14 &  $-$3.07 &    0.17 &  15  \\
   6.0\tablenotemark{e} &   2.66 &  \nodata &   1 &   0.16 &  \nodata &  $-$1 &   0.00 &  \nodata &   1 &  
         1.59 &  \nodata &   1 &   2.90 &  \nodata &   1  \\
   7.0 &   2.11 &  \nodata &   1 & \nodata &  \nodata & \nodata & \nodata &  \nodata & \nodata &   0.54 &  \nodata &   1 & \nodata &  \nodata & \nodata  \\
  11.0 & \nodata &  \nodata & \nodata &  $-$0.16 &    0.14 &   2 & \nodata &  \nodata & \nodata & \nodata &  \nodata & \nodata &   1.38 &    0.00 &   2  \\
  12.0 &   0.57 &    0.16 &   4 &   0.03 &    0.27 &   4 &   0.56 &    0.05 &   4 &   0.85 &    0.10 &   4 &   0.88 &    0.13 &   3  \\
  13.0 &  $-$0.05 &  \nodata &   1 &  $-$0.16 &    0.22 &   2 &  $-$0.11 &  \nodata &   1 & \nodata &  \nodata & \nodata &  $-$0.04 &  \nodata &   1  \\
  14.0 & \nodata &  \nodata & \nodata &  $-$0.06 &  \nodata &   1 &   0.65 &  \nodata &   1 & \nodata &  \nodata & \nodata & \nodata &  \nodata & \nodata  \\
  20.0 &   0.37 &  \nodata &   1 &  $-$0.10 &    0.19 &   3 &   0.26 &    0.07 &   6 &   0.42 &    0.15 &   3 &   0.40 &    0.15 &   2  \\
\enddata
\tablenotetext{a}{Species code: atomic number + (ionization stage/10),
  i.e. 26.0 = Fe~I, 26.1 = Fe~II.  Special case: 26.2 is for three 
  strong Fe~II lines
  in the UV near 3260~\AA.}
\tablenotetext{b}{[X/Fe](Fe~I), except for Fe~I, where we give [Fe/H]
   from Fe~I lines only.}
\tablenotetext{c}{rms dispersion for the lines of this species around
  the mean abundance ratio.}
\tablenotetext{d}{Number of lines of this species measured.
   A species with an upper limit for its relative abundance
   is indicated by setting the  number of
   lines to its negative value.  $N < -1$ means several lines were 
   searched for, but none were firmly detected.}
\tablenotetext{e}{C from synthesis of G band of CH.}   
\end{deluxetable}

\clearpage

\begin{deluxetable}{l c c  l }
\tablewidth{0pt}
\tablecaption{Data for Candidates with [Fe/H](HES) $< -3.5$~dex
\label{table_below-3.5} }
\tablehead{
\colhead{ID} & \colhead{[Fe/H](HES)} & \colhead{[Fe/H](high-res)} &
     \colhead{Comments} \\
\colhead{} & \colhead{(dex)} & \colhead{(dex)} &  \colhead{} 
}
\startdata
HE1159$-$0525 &  $-$3.52  &   $-$2.91   &    Extreme C$-$star \\                       
HE1150$-$0428 &  $-$3.57  &   $-$3.21   &  C$-$star \\
HE2209$-$3909 &  $-$3.57  &   $-$3.47   & hot turnoff dwarf \\  
HE1031$-$0020 &  $-$3.61  &   $-$2.81   &   C$-$star \\
HE0122$-$1616 &  $-$3.63  &   $-$2.80   \\   
HE0305$-$5442\tablenotemark{a} &  $-$3.68  &   $-$3.30   & extreme CN \\                                      
HE0313$-$3640 &  $-$3.72  &   $-$3.63  &  hot turnoff dwarf \\
HE0911$-$0512 &  $-$3.84  &  \nodata &       M dwarf \\
HE2323$-$0256\tablenotemark{b} &  $-$3.96  &   $-$3.93   &  extreme CN \\
HE1432$-$1819 &  $-$4.43  &   $-$2.56  &    hot turnoff dwarf \\
HE0208$-$5335 &  $-$4.14  &   \nodata &       QSO  \\
HE1030+0137 &  $-$4.37  &   \nodata &        M dwarf \\
\enddata
\tablenotetext{a}{HE0305$-$5442 is a rediscovery of
CS22968$-$014.  It has  [Fe/H] from \cite{cayrel04} of $-$3.56~dex.} 
\tablenotetext{b}{HE2323$-$0256 is a rediscovery of CS22949$-$037.
It has  [Fe/H] from \cite{cayrel04} of $-$3.79~dex.}
\end{deluxetable} 

\clearpage

\begin{deluxetable}{l c  c l }
\tablewidth{0pt}
\tablecaption{Data for Candidates with $-3.4 <$ [Fe/H](HES) $< -3.2$~dex
\label{table_3.2to3.5} }
\tablehead{
\colhead{ID} & \colhead{[Fe/H](HES)} & \colhead{[Fe/H](high-res)} &
    \colhead{Comments} \\
\colhead{} & \colhead{(dex)} & \colhead{(dex)} & \colhead{} 
}
\startdata
HE0014$-$3607 &  $-$3.21  & \nodata &    Prob. extreme C$-$star \\
HE2356$-$0410 &  $-$3.22  &  $-$3.06 &   Extreme C-star \\
HE1509$-$0806 &  $-$3.22  & $-2.91$ &   Extreme C-star \\
HE0218$-$4136 &  $-3.22$  &   \nodata  \\
HE2203$-$3723 &  $-3.22$  & $-3.30$  \\
HE2128$-$4248 &  $-3.23$  &   \nodata  \\   
HE1500$-$1628 &  $-3.25$  & $-2.38$ &  v. low [Ca/Fe] \\
HE2155$-$3750 &  $-3.28$ &  $-2.59$ &  Extreme C-star \\
HE0349$-$0045 &  $-3.28$ &  $-2.23$ & v. low [Ca/Fe] \\
HE0048$-$0611 &  $-3.28$ &  $-2.67$  \\
HE2129$-$4652 &  $-3.28$ &  \nodata &  Prob. extreme C$-$star \\
HE2122$-$4707 &  $-3.29$ & $-2.42$ &   C-star \\
HE0024$-$2523 &  $-3.32$ &  $-2.72$ &  C-star \\
HE2339$-$5105 &  $-3.32$ & $-2.88$  \\  
HE0216$-$3838 &  $-3.33$ &  \nodata &  Prob. extreme C$-$star \\
HE2251$-$0821 &  $-3.34$ &  $-$2.91 &  Very strong CH, no C$_2$ \\
HE0926$-$0546 &  $-3.35$ &  $-3.73$  \\
HE2357$-$0701 &  $-3.35$ &  $-3.17$  \\
HE0009$-$6039 &  $-3.35$ &  $-3.29$  \\
HE0344$-$0243 &  $-3.36$ &  $-3.35$  \\  
HE0226+0103   &  $-3.36$ &  $-2.81$  \\
HE2215$-$2548 &  $-3.36$ &  $-3.01$  \\  
HE1434$-$1442 &  $-3.42$ &  $-2.39$  & C-star \\
HE1245$-$0215 &  $-3.42$ &  $-2.85$ & hot turnoff star\\
HE1416$-$1032 &  $-3.44$ &  $-3.20$  \\
HE0134$-$2504 &  $-3.45$ &  $-2.74$  &  C-star \\
HE0212$-$0557 &  $-3.45$ &  $-2.27$  & Extreme C-star \\
HE2138$-$4250 &  $-3.46$ &  \nodata  & Prob. extreme C$-$star \\
\enddata
\end{deluxetable}

\clearpage

\begin{deluxetable}{l r r r c r }
\tablewidth{0pt}
\tablecaption{Linear Fits for C-Normal Stars from the HES
\label{table_fits} }
\tablehead{
\colhead{Element} & \colhead{N(stars)} & \colhead{At [Fe/H] $-$3} &
  \colhead{[At Fe/H] $-3.5$ } &  \colhead{$\sigma$\tablenotemark{a}} &
   \colhead{Outliers\tablenotemark{b}} \\
\colhead{} & \colhead{} & \colhead{(dex)} & \colhead{(dex)} &
  \colhead{(dex)}    &   \colhead{High,Low} 
}
\startdata      
All \\    
$[$FeI/FeII$]$ &  120 &  0.00 & 0.00 & 0.10   \\
$[$TiI/TiII$]$  &  86 & 0.04 & 0.04 & 0.18   \\
Exclude & C-Stars  \\
$[$Mg/Fe$]$  &  81 &  +0.37  & +0.37  &    0.17   &    2H,1L \\   
$[$Al/Fe$]$ &  81 & $-0.16$ & $-0.13$  &     0.24  &   1H,1L \\
$[$Si/Fe$]$\tablenotemark{c} &  29  & +0.03 &  $-0.04$   &  0.18  & none   \\     
$[$Si/Fe$]$\tablenotemark{d} &  48  & +0.37 &  +0.46  &   0.20  &  1H,1L \\
\enddata
\tablenotetext{a}{$\sigma$ of [X/Fe] about the linear fit.}  
\tablenotetext{b}{Strong outliers with [Fe/H] $< -2.5$~dex excluded from fit.}
\tablenotetext{c}{For hot dwarfs.} 
\tablenotetext{d}{For cool giants.} 
\tablenotetext{e}{Upper limits are ignored.}
\tablenotetext{f}{Fit is very uncertain.} 
\tablenotetext{g}{Includes upper limits as detections for \teff $< 6000$~K.}
\end{deluxetable}   
 
\clearpage 

\begin{deluxetable}{l r r r }
\tablewidth{0pt}
\tablecaption{Strong Outliers Among the C-Normal Stars from the HES
\label{table_out_fit} }
\tablehead{
\colhead{ID)} & \colhead{[X/Fe]} & \colhead{$\Delta$[X/Fe]\tablenotemark{a}} &
     \colhead{$\Delta$[X/Fe]/$\sigma$\tablenotemark{b}} \\
\colhead{} & \colhead{(dex)} & \colhead{(dex)} 
}
\startdata
Mg  \\
HE0533$-$5340  & $-$0.234 & $-$0.64 & $-$7 \\
HE1012$-$1540\tablenotemark{c} & +1.38 &  +0.98 & +11 \\
HE2323$-$0256\tablenotemark{c} & +1.39 &   +0.99 & +11 \\    
Al \\
HE0533$-$5340 & $-0.84$ & $-0.65$ & $-$6 \\
HE1012$-$1540\tablenotemark{c} & +0.49 & +0.62 & +7 \\
HE2200$-$1946  & +0.43 &  +0.62 & +7 \\
\enddata
\tablenotetext{a}{The difference between [X/Fe] and the linear fit
  for [X/Fe] from Table~\ref{table_fits} evaluated at the [Fe/H] of the star.}
\tablenotetext{b}{The ratio of $\Delta$[X/Fe] to $\sigma$ for a single star given in the
  second column of Table~\ref{table_unc_ratio_giants}.}
\tablenotetext{c}{This star is not ``C-normal''.} 
\tablenotetext{d}{[Fe/H] $> -2.5$~dex, not used in constructing
  linear fits.}
\tablenotetext{e}{The trend of [Co/Fe] vs [Fe/H] is not 
  as well established as for other species; these stars may 
  be at the extremes of the trends rather
  than genuine outliers.} 
\tablenotetext{f}{The trend of [Zn/Fe] vs [Fe/H] is not 
  as well established as for other species and 
  at best only two weak Zn~I lines are detected in our spectra.}
\tablenotetext{g}{Fit value set to 0.0~dex. See \S\ref{section_fits}.} 
\tablenotetext{h}{$\sigma$ includes extra term discussed in \S\ref{section_fits},
and is 0.16~dex for [Sr/Fe] and 0.12~dex for [Ba/Fe].} 
\tablenotetext{i}{Upper limit to [BaII/FeII], not used in constructing fit.} 
\end{deluxetable} 

\clearpage

\begin{deluxetable}{l r r r  }
\tablewidth{0pt}
\tablecaption{Dispersion About Linear Fit vs Observational Uncertainty for C-normal Stars
\label{table_fits_sigma} }
\tablehead{
\colhead{Element} & \colhead{$\sigma$(Lin. Fit)} & \colhead{Pred. $\sigma$} &
  \colhead{Ratio }  \\
\colhead{} & \colhead{(dex)} & \colhead{(dex)} 
}
\startdata
$[$Mg/Fe$]$  &  0.17   & 0.09 & 1.9 \\
$[$Al/Fe$]$ &  0.24 & 0.09 & 2.7 \\
$[$Si/Fe$]$\tablenotemark{a} &  0.18 & 0.16 & 1.2  \\     
$[$Si/Fe$]$\tablenotemark{b} &  0.20 & 0.16 & 1.2 \\ 
$[$Ca/Fe$]$  &  0.15 & 0.07 & 2.1 \\
$[$ScII/FeI$]$ &  0.14 & 0.11 & 1.3 \\
$[$TiII/FeII$]$ &  0.14 & 0.09 & 1.6 \\
$[$Cr/Fe$]$ & 0.13 & 0.07 & 1.9 \\
$[$Mn/Fe$]$  & 0.16 & 0.09 & 1.8 \\
$[$Ni/Fe$]$ & 0.21 & 0.17 & 1.2 \\
$[$SrII/FeII$]$  & 0.66 & 0.16\tablenotemark{c} & 4.1 \\
$[$BaII/FeII$]$  & 0.55 & 0.12\tablenotemark{c} & 4.6 \\
\enddata
\tablenotetext{a}{Fit for dwarfs.}
\tablenotetext{b}{Fit for giants.}
\tablenotetext{c}{Uses augmented $\sigma$ with term from slope, 0.16~dex
  for [SrII/FeII] and 0.12~dex for [BaII/FeII].}
\end{deluxetable}

\clearpage

\begin{deluxetable}{l rll rll rll}
\tablewidth{0pt}
\tablecaption{Abundance Ratios Grouped By Distance
\label{table_ratio_distance} }
\tablehead{
\colhead{Species} & 
   \colhead{} & \colhead{Close\tablenotemark{a}} & \colhead{} & 
   \multicolumn{3}{c}{Inner Halo\tablenotemark{b}} & 
   \multicolumn{3}{c}{Outer Halo\tablenotemark{c}} \\
\colhead{} &   
   \colhead{$<$[X/H]$>$} &  \colhead{N\tablenotemark{d}} &\colhead{$\sigma$} &
    \colhead{$<$[X/H]$>$} & \colhead{N\tablenotemark{d}} & \colhead{$\sigma$} & 
    \colhead{$<$[X/H]$>$} &  \colhead{N\tablenotemark{d}} & \colhead{$\sigma$} \\
\colhead{} &   
   \colhead{(dex)} & \colhead{(dex)} &  \colhead{} &
   \colhead{(dex)} & \colhead{(dex)} &  \colhead{} &
   \colhead{(dex)} & \colhead{(dex)} &  \colhead{} 
}
\startdata
$[$Fe/H$]$
    &  $-$2.86 & 54 & 0.37 &  $-$2.88  & 58 & 0.51  &
       $-$3.02 & 10 & 0.40  \\
$[$Na/Fe$]$ &    0.16 & 21 & 0.65 &   0.18 & 49 & 0.52 &  
       0.28 &  8 & 0.38 \\
$[$Mg/Fe$]$ &    0.39 & 54 & 0.24  &  0.39 & 59 &
     0.25 &  0.49 & 10 & 0.16  \\
$[$Al/Fe$]$ &   $-$0.12 & 53 &  0.25 &  $-$0.18 & 47 &  0.27 &
    $-$0.06  & 10 & 0.30 \\
$[$Si/Fe$]$\tablenotemark{e} &    0.09 & 44 & 0.26 &   0.42 & 47 & 0.41  &
       0.41 &   9 & 0.31 \\   
$[$Ca/Fe$]$ &  0.30 & 54 & 0.13 &  0.22 & 56 & 0.22 &  0.25&  10 & 0.25 \\
$[$ScII/FeII$]$ &   0.20 & 47 & 0.27 &    0.13 & 46 & 0.23  &
    0.09 & 10 & 0.13 \\
$[$TiII/FeII$]$ &   0.28 & 53 & 0.14  &  0.26 & 59 & 0.22 &  0.10 & 10 & 0.18 \\
\enddata
\tablenotetext{a}{$D < 4$~kpc.}
\tablenotetext{b}{Inner halo: $4 < D < 15$~kpc.}
\tablenotetext{c}{Outer halo: $D > 15$~kpc.}
\tablenotetext{d}{Number of stars.}
\tablenotetext{e}{There is a significant difference between
[Si/Fe] for dwarfs and for giants.  See \S\ref{section_abund_proc}. }
\tablenotetext{f}{Upper limits are treated as detections.}
\tablenotetext{g}{C-rich stars are omitted.}
\end{deluxetable}

\clearpage

\clearpage

\clearpage

\clearpage

\clearpage

\clearpage

\clearpage

\clearpage

\clearpage

\clearpage


\clearpage


\clearpage

\clearpage

\clearpage

\clearpage

\clearpage

\clearpage

\clearpage

\clearpage

\begin{thebibliography}{}

\bibitem[Aihara et al(2011)]{sdss_dr8}
Aihara, H. et al, 2011, \apjs, 193, 29

\bibitem[Andersen, Hansen \& Nordstr\"om(2013)]{andersen13}
Andersen, J.,  Hansen, T. \& Nordstr\"om, B., 2013, in preparation

\bibitem[Andrievsky et al(2011)]{nonlte_sr}
Andrievsky, S.~M., Spite, F., Korotin, S.~A., Francois, P.,
Spite, M., Bonifacio, P., Cayrel, R. \& Hill, V., 2011,
\aap, 530, A105

\bibitem[Andrievsky et al(2010)]{nonlte_mgk}
Andrievsky, S.~M., Spite, M., Korotin, S.~A., Spite, F.,
Bonifacio, P., Cayrel, R., Francois, P. \& Hill, V.,
2010, \aap, 509, 88

\bibitem[Aoki et al(2002)]{aoki02}
Aoki, W., Norris, J.~E., Ryan, S.~G., Beers, T.~C. \& Ando, H.,
2002, \apjl, 576, L141

\bibitem[Aoki et al(2010)]{aoki10}
Aoki, W., Beers, T.~C., Honda, S. \& Carollo, M., \apj, 723, L201


\bibitem[Asplund(2005)]{asplund05}
Asplund, M., 2005, \araa, 43, 481


\bibitem[Asplund et al(2009)]{asplund09}
Asplund, M., Grevesse, N., Sauval, A.~J. \& Scott, P., 2009,
\araa, 47, 481

\bibitem[Barklem et al(2005)]{barklem05}
Barklem, P.S., Christlieb, N.,  Beers, T.~C., et al, 2005, \aap, 439, 129

\bibitem[Baum{\"u}ller \& Gehren(1997)]{al_nonlte}
Baum{\"u}ller, D.~G. \& Gehren, T., 1997, \aap, 325, 108


\bibitem[Beers, Preston \& Shectman(1985)]{hk1}
Beers, T.C.,  Preston, G.W. \& Shectman, S., 1985, \aj, 90, 2089

\bibitem[Beers, Preston \& Shectman(1992)]{hk2}
Beers, T.C.,  Preston, G.W. \& Shectman, S., 1992, \aj, 103, 1987

\bibitem[Beers et al.(1998)]{beers98}
Beers, T.~C., Rossi, S., Norris, J.~E., Ryan, S., Molaro, P. \& Rebolo, R.,
1998, Space Science Reviews,  84, 139 

\bibitem[Beers \& Christlieb(2005)]{beers05}
Beers, T.~C. \& Christlieb, N., 2005, \araa, 43, 531


\bibitem[Bensby et al(2005)]{bensby05}
Bensby, T., Feltzing, S.. Lundstr\"{o}m, I. \& Ilyin, I., 2005,
\aap, 433, 185


\bibitem[Bernstein et al.(2003)]{bernstein_mike}
Bernstein, R., Shectman, S.~A., Gunnels, S.~M.,
Mochnacki, S. \& Athey, A.~E., 2003, Society of Photo-Optical
Instrumentation Engineers (SPIE) Conference Series, 4841, ed.
M.~Iye \& A.~F.~M.~Moorhead, 1694

\bibitem[Bergemann \& Gehren(2008)]{bergemann08}
Bergemann, M. \& Gehren, T., 2008, \aap, 492, 823

\bibitem[Bergemann et al(2012)]{nonlte_fe}
Bergemann, M., Lind, K., Collet, R., Magic, Z. \& Asplund, M.,
2012, \mnras, in press

\bibitem[Bisterzo et al(2011)]{bisterzo11}
Bisterzo, S., Gallino, R., Straniero, O., Cristallo, S. \&
 Kappeler, F., 2011, \mnras, 418, 284
 
\bibitem[Bonifacio et al(2009)]{first_stars_xii}
Bonifacio, P. et al, 2009, \aap, 501, 519

\bibitem[Brown, Wallerstein \& Zucker(1997)]{brown97}
Brown, J.~A., Wallerstein, G. \& Zucker, D., 1997, \aj, 114, 180

\bibitem[Bullock \& Johnston(2005)]{bullock05}
Bullock, J. \& Johnston, K.~V., 2005, \apj, 635, 931


\bibitem[Busso, Gallino \& Wasserburg(1999)]{busso_araa}
Busso, M., Gallino, R. \& Wasserburg, G.~J., 1999, \araa, 37, 239

\bibitem[Caffau et al(2011)]{caffau11}
Caffau, E. et al, 2011, Nature, 477, 67

\bibitem[Carney et al.(1997)]{carney97}
Carney, B.~W., Wright, J.~S., Sneden, C., Laird, J.~B.,
Aguilar, L.~A. \& Latham, D.~W., 1997, \aj, 114, 363

\bibitem[Carollo et al.(2007)]{carollo07}
Carollo, D. et al, 2007, Nature, 450, 1020

\bibitem[Carretta et al.(2002)]{keck_pilot_2}
Carretta, E., Gratton,~R., Cohen J.~G., Beers, T.~C. \&  Christlieb,~N.,  
2002, \aj, 124, 481

\bibitem[Castelli \& Kurucz(2004)]{castelli04} Castelli, F., \& Kurucz,
  R.~L.\ 2004, arXiv:astro-ph/0405087

\bibitem[Cayrel et al.(2004)]{cayrel04} 
Cayrel, R. et al, 2004, \aap, 416, 1117


\bibitem[Christlieb et al.(2002)]{christlieb02}
Christlieb, N. et al, 2002, Nature, 419, 904


\bibitem[Christlieb et al.(2008)]{christlieb08}
Christlieb, N. et al, 2008, \aap, 721

\bibitem[Christlieb et al.(2013)]{christlieb13}
Christlieb, N. et al, 2013, in preparation
 

\bibitem[Cohen et al.(2002)]{keck_pilot_1}
Cohen, J.~G., Christlieb, N.,  Beers, T.~C., Gratton, R. \& Carretta,~E.,
2002, \aj, 124, 470 

\bibitem[Cohen et al.(2003)]{cohen03}
Cohen, J.~G., Christlieb,  N., Qian, Y. \& Wasserburg, G., 
2003, \apj, 588, 1082 

\bibitem[Cohen(2004)]{cohen_pal12}
 Cohen, J.~G., 2004, \aj, 127, 1545

\bibitem[Cohen et al.(2004)]{cohen04}
Cohen, J.~G., Christlieb, N.,  McWilliam, A., Shectman, S.,
Thompson, I., Wasserburg, G., Ivans, I., Dehn,  Karlsson, T. \& 
Melendez, J., 2004, \apj, 612, 1107

\bibitem[Cohen et al.(2005)]{feh_cstars}
Cohen, J.~G., Shectman,~S., Thompson,~I., McWilliam,~A., Christlieb,~N.,
Melendez,~J., Zickgraf,~F.~J., Ram\'{\i}rez, S. \& Swenson,~A., 2005,
\apjl, 633, L109

\bibitem[Cohen \& Melendez(2005)]{cohen05}
Cohen, J.~G. \& Melendez, J., 2005, \aj, 129, 303

\bibitem[Cohen et al.(2006)]{cohen06}
Cohen, J.,  McWilliam, A., Shectman, S., Thompson, I., Christlieb, N.,
Ram\'irez, S., Swenson, A. \& Zickgraf, F.~J., 2006,
\aj, 132, 137

\bibitem[Cohen et al.(2007)]{cohen_1424}
Cohen, J.~G., McWilliam,~A., Christlieb,~N., 
 Shectman,~S., Thompson,~I., Melendez, J., Wisotzki, L, \& Reimers, D.,
 2007, \apjl, 659, L161


\bibitem[Cohen et al.(2008)]{cohen08}
Cohen, J.~G., Christlieb, N., McWilliam, A.,
Shectman, S., Thompson, I., Melendez, J.,
Wisotzki, L. \& Reimers, D., 2008, \apj, 672, 320

\bibitem[Cohen \& Huang(2009)]{cohen_draco}
Cohen, J.~G. \& Huang, W., 2009, \apj, 701, 1053

\bibitem[Cohen \& Kirby(2012)]{ngc2419}
Cohen, J.~G. \& Kirby, E., 2012, \apj, 760, 86

\bibitem[Cohen et al.(2012)]{cohen12}
Cohen, J.~G., Christlieb, N., Thompson, I., McWilliam, A. \& 
Shectman, S., 2012, ``Galactic Archeology: Near-Field Cosmology and the 
Formation of the
Milky Way'', PASP Conference Series, volume 458, edited W. Aoki, M. Ishigaki, T. Suda,
T. Tsuijimoto \& N. Arimoto, pg. 61


\bibitem[Cooke, Pettini \& Murphy(2011)]{cooke12}
Cooke, R., Pettini, M. \& Murphy, M.~T., 2012, \mnras, 425, 347

\bibitem[Cutri et al.(2003)]{2mass2}
Cutri, R.~M. et al, 2003,
``Explanatory Supplement to the 2MASS All-Sky Data Release,
http://www.ipac.caltech.edu/2mass/releases/allsky/doc/explsup.html

\bibitem[Depagne et al.(2002)]{first_stars_2}
Depagne, E. et al, 2002, \aap, 390, 187

\bibitem[de Jong et al.(2010)]{dejong10}
de Jong, J.~T.~A., Yanny, B., Rix, H.~W., Dolphin, A.,
Martin, N~.F. \& Beers, T.~C., 2010, \apj, 714, 663 

\bibitem[Dobrovolskas et al.(2012)]{nonlte_ba}
Dobrovolskas, V., Kucinskas, A., Andrievsky, S.~M.,
Korotin, S.~A., Mishenina, T.~V., Bonifacio, P.,
Ludwig, H.~G. \& Caffau, E., 2012, \aap, 540, A128

\bibitem[Dominy(1984)]{dominy84}
Dominy,J.~F., 1984, \apjs, 55, 27

\bibitem[Edvardsson et al.(1993)]{edvard93}
Edvardsson, B., Andersen, J., Gustaffson, B., Lambert, D.~L.,
 Nissen, P.~E. \& Tomkin, J., 1993, \aap, 275, 101
 
\bibitem[Francois et al.(2008)]{first_stars_viii}
Francois, P. et al, 2008, \aap, 476, 935
 
\bibitem[Frebel et al.(2005)]{frebel05}
Frebel, A. et al, 2005, Nature, 434, 871
 

\bibitem[Frebel et al.(2007)]{frebel07}
Frebel, A., Norris, J.~E., Aoki, W., Honda, S.,
Bessell, M.~S., Takada-Hidai, M., Beers, T.C. \&
Christlieb, N., 2007, \apj, 658, 534



\bibitem[Fulbright(2002)]{fulbright02} 
Fulbright, J.~P., 2002, \aj, 123, 404




\bibitem[Fulbright, Rich \& Castro(2004)]{draco119}
Fulbright, J., Rich, R.~M. \& Castro, S., 2004, \apj, 612, 447

\bibitem[Fumagalli, O'Meara \& Prochaska(2011)]{fumagalli11}
Fumagalli, M., O'Meara, J.~M., Prochaska, J.~X., 2011 Science, 334, 1245

\bibitem[Gratton et al.(2003)]{gratton03}
Gratton, R.~G., Carretta, E., Desidera, S., Lucatello, S.,
Mazzei, P. \& Barbieri, M., 2003, \aap, 406, 131 

\bibitem[Gratton, Sneden \& Carretta(2004)]{gratton04} 
Gratton, R., Sneden, C. \& Carretta, E., 2004, \araa, 42, 385

\bibitem[Greggio, Renzini \& Daddi(2008)]{greggio08}
Greggio, L., Renzini, A. \& Daddi, E., 2008, \mnras, 388, 829

\bibitem[Grevesse \& Sauval(1998)]{grevesse98}
Grevesse, N. \& Sauval, A.~J., 1998, Space Science Reviews, 85, 161

\bibitem[Gunn \& Griffin(1979)]{gunn79}
Gunn, J.E. \& Griffin, R.~F., 1979, \aj, 84, 752

\bibitem[Gustaffson etal.(1975)]{marcs}
Gustaffson, B., Bell, R.~A., Eriksson, K. \& Nordlund, B.,
1975, \aap, 42, 407

\bibitem[Hansen et al.(2013)]{hansen13}
Hansen, C.~J., Bergemann, M., Cescutti, G., Arcones, A., Karakas, A.~I.
\& Chiappini, C., 2013, \aap, 551, A57 

\bibitem[Heger \& Woosley(2010)]{heger10}
Heger, A. \& Woosley, S.~E., 2010, \apj, 724, 341 

\bibitem[Honda et al.(2011)]{honda11}
Honda, S., Aoki, W., Beers, T.~C. \& Takada-Hidei, M., 2011,
\apj, 730, A77

\bibitem[Houdashelt, Bell \& Sweigart(2000)]{houdashelt00}
Houdashelt, M.~L., Bell, R.~A. \& Sweigart, A.~V., 2000, \aj, 119, 1448

\bibitem[Ivanova \& Shimanski{\u i}(2000)]{ivanova_k}
Ivanova, D.~V. \& Shimanski{\u i}, V.~V., 2000, Astronomy Reports,
44, 376

\bibitem[Ivans et al.(2003)]{ivans}
Ivans, I.~I., Sneden, C., Renee James, C., Preston, G.~W.,
   Fulbright, J.~P., Hoflich, P.~A., Carney, B.~W. \&
   Wheeler, J.~C., 2003, \apj, 592, 906

\bibitem[Iwamoto et al(1999)]{iwamoto99}
Iwamoto, K., Brachwitz, F., Nomoto, I.~I., Kishimoto, N.,
   Umeda, H., Hix, W.~R. \& Thielemann, F.~K., 1999, \apjs, 125, 439

\bibitem[Izzard et al(2009)]{izzard09}
Izzard, R.~G., Glebbeek, E., Stancliffe, R.~J. \& Pols, O.,
\aap, 508, 1359-U332

\bibitem[Johnson \& Bolte(2002)]{johnson02}
Johnson, J.~A. \& Bolte,~M., \apjl, 2002,  579, L87

\bibitem[Kelson(2003)]{kelson03}
Kelson, D.~D., 2003, \pasp, 115, 688

\bibitem[Kirby et al(2011)]{kirby11}
Kirby, E.~N., Cohen, J.G., Smith, G.~H., Majewski, S.~R., Sohn, S.~T.
\& Guhathakurta, P., 2011, \apj, 727, 79


\bibitem[Kobayashi et al(2006)]{kobayashi06}
Kobayashi, C., Umeda, K., Nomoto K.,
  Tominaga, N. \& Ohkubo, T., 2006, \apj, 653, 1145


\bibitem[Kurucz(1993)]{kurucz93} Kurucz, R. L., 1993, ATLAS9 Stellar
Atmosphere Programs and 2 km/s Grid, (Kurucz CD-ROM No. 13)

\bibitem[Lai et al(2009)]{lai09}
Lai, D. et al, 2009, \apjl, 697, L63


\bibitem[Latham et al(2002)]{latham02}
Latham, D.~W., Stefanik, R.~P., Torres, G., Davis, R.~J.,
Mazeh, T., Carney, B.~W., Laird, J.~B. \& Morse, J.~A.,
2002,\aj, 124, 1144

\bibitem[Lau, Stancliffe \& Tout(2007)]{lau07}
Lau, H., Stancliffe, R.~J. \& Tout, C.~A., 2007, \mnras, 385, 301

 

\bibitem[Lee et al(2008)]{segue1}
Lee, Y.~S. et al, 2008, \aj, 136, 2022

\bibitem[Lind et al(2011)]{nonlte_na}
Lind, K., Asplund, M., Barklem, P.~S. \& Belyaev, A,~K.,
2011, \aap, 528, A103

\bibitem[Lucatello et al(2003)]{keck_pilot_3}
Lucatello,~S., Gratton,~R., Cohen,~J.~G., 
Beers,~T., Christlieb,~N.,  Carretta,~E. \& Ram\'{\i}rez,~S.,  
2003, \aj, 125, 875 

\bibitem[Lucatello et al(2007)]{lucatello_bin}
Lucatello,~S.,  Tsangarides,~S., Beers,~T.~C., Carretta,~E., Gratton,~R.~G. \& Ryan,~S.~G.,
2005, \apj, 625, 825



\bibitem[Lucatello et al(2006)]{lucatello06}
Lucatello, S., Beers, T.C., Christlieb, N., Barklem, P.~S.,
Rossi, S., Barsteller, B., Sivarani, T. \& Lee, Y.~S., 2006, \apj, 652, L37

\bibitem[Lugaro et al(2012)]{lugaro12}
Lugaro, M., Karakas, A.~I., Stancliffe, R.~J. \& Rijs, C., 2012,
\apj, 747, A2

\bibitem[Magic et al(2013)]{stagger2_13}
Magic, Z., Collet, R., Hayek, W. \& Asplund, M., 2013, \aap, submitted
  (see arXiv:1307.3273)

\bibitem[McCarthy(1988)]{mccarthy88}
McCarthy, J.~K., 1988, PhD thesis, California Institute of Technology

\bibitem[McClure(1997)]{mcclure97}
McClure, R.~D., 1997, \pasp, 109, 256


\bibitem[McWilliam et al(1995a)]{mcwilliam95a}
McWilliam, A., Preston, G.~W., Sneden, C. \& Shectman, S., 1995,
\aj, 109, 2736

\bibitem[McWilliam et al(1995b)]{mcwilliam95b}
McWilliam, A., Preston, G.~W., Sneden, C. \& Searle, L., 1995,
\aj, 109, 2757

\bibitem[McWilliam(1998)]{mcwilliam98}
McWilliam, A., 1998, \aj, 115, 1640

\bibitem[Nissen \& Schuster(2010)]{nissen1}
Nissen, P.~E. \& Schuster, W.~J., 2010, \aap, 511, L10

\bibitem[Nissen \& Schuster(2011)]{nissen2}
Nissen, P.~E. \& Schuster, W.~J., 2011, \aap, 530, A15

\bibitem[Nomoto, Thielemann \& Yokoi(1984)]{nomoto84}
Nomoto, K., Thielemnn, F.~K. \& Yokoi, K., 1984, \apj, 286, 644

\bibitem[Nomoto et al(2006)]{nomoto06}
Nomoto, K., Tominaga, N., Umeda, H., Kobayashi, C. \& Maeda, K.,
2006, Nucl. Phys. A777, 424

\bibitem[Norris, Ryan \& Beers(2001)]{norris01}
Norris, J.~E., Ryan, S.~G. \& Beers, T.~C., 2001, \apj, 561, 1034

\bibitem[Norris et al(2007)]{norris07}
Norris, J.~E., Christlieb, N., Korn, A.~J., Eriksson, K., Bessell, M.~S.,
Beers, T.~C., Wisotzki, L. \& Reimers, D., \apj, 670, 774

\bibitem[Norris et al(2013a)]{norris12a}
Norris, J.~E. et al, 2013a, \apj, 762, A25


\bibitem[Norris et al(2013b)]{norris12b}
Norris, J.~E. et al, 2013b, \apj, 762, A28

\bibitem[Oke \& Gunn(1982)]{dbsp}
Oke, J.~B. \& Gunn, J.~E., 1982, \pasp, 94, 586

\bibitem[Preston \& Sneden(2001)]{preston01}
Preston, G.~W. \& Sneden, C., 2001, \aj, 122, 1545

\bibitem[Preston et al(2006)]{preston06}
Preston, G.~W., Sneden, C., Thompson, I.~B., Shectman, S.~A.
\& Burley, G.~S., 2006, \aj, 132, 85

\bibitem[Qian \& Wasserburg(2008)]{qian08}
Qian, Y.~Z. \& Wasserburg, G.~J., 2008, \apj, 687, 272

\bibitem[Ralchenko et al(2010)]{nist}
Ralchencko, Yu., Kramida, A.~E., Reader, J. \& the NIST ASD Team,
2010, NIST Atomic Spectra Database (version 4.0),
http://physics.nist.gov/asd


\bibitem[Ram\'{\i}rez et al.(2001)]{ramirez_m71_fe}
Ram\'{\i}rez, S.~V., Cohen, J.~G., Buss, J. \& Briley, M.~M.,
2001, \aj, 122, 1429

\bibitem[Reddy et al.(2006)]{reddy06}
Reddy, B.~E., Tomkin, J., Lambert, D.~L. \& Allende Prieto, C.,
2006, \mnras, 367, 1329


\bibitem[Rafelski et al(2012)]{rafelski12}
Rafelski, M., Wolfe, A., Prochaska, J.~X.,
 Neeleman, M. \& Mendez, A.~J., 2012, \apj, 755, A89

\bibitem[Roederer(2009)]{roederer09}
Roederer I.~U., 2009, \aj, 137, 272

\bibitem[Roederer et al(2010)]{roederer10}
Roederer, I.~U., Sneden, C., Lawler, J.~W. \& Cowan, J.~J., 2010
\apjl, 714, L123


\bibitem[Schlaufman et al(2012)]{schlaufman12}
Schlaufman, K.~C. et al, 2012, \apj, 749, A77


\bibitem[Schlegel, Finkbeiner \& Davis(1998)]{schlegel}
Schlegel, D.~J., Finkbeiner, D.~P. \& Davis, M., 1998, \apj, 500, 525

\bibitem[Sch\"{o}rck et al(2009)]{hes_mdf}
Sch\"{o}rck, T., Christlieb, N., Cohen, J.G.  et al, 2009,
\aap, 507, 817


\bibitem[Schuster et al(2012)]{nissen3}
Schuster, W.~J., Moreno, E., Nissen, P.~E. \& Pichardo, B.,
2012, \aap, 538, A21

\bibitem[Sesar, Juric \& Ivezic(2011)]{sesar11}
Sesar, B., Juric, M. \& Ivezic, Z., 2011, \apj, 731, 4


\bibitem[Shi et al.(2009)]{shi09}
Shi, J.~R., Gehren, T., Mashonkina, L. \& Zhao, G., 2009,
\aap, 503, 533

\bibitem[Shortridge(1993)]{figaro}
Shortridge K. 1993, in {\it{Astronomical Data Analysis Software and
      Systems II}}, A.S.P. Conf. Ser., Vol 52, eds. R.J. Hannisch, 
      R.J.V. Brissenden, \& J. Barnes, 21
      
\bibitem[Skrutskie et al.(2006)]{2mass1}
Skrutskie, M.~F. et al, 2006, \aj, 131, 1163

\bibitem[Smith et al.(2002)]{sdss_convert}
Smith, J.~A., Tucker, D.~L., Kent, S. et al, 2002, \aj, 123, 2121

\bibitem[Sneden(1973)]{moog} Sneden, C., 1973, Ph.D. thesis, Univ.
of Texas

\bibitem[Sneden et al.(2003)]{sneden03}
Sneden, C. et al, 2003, \apj, 591, 936

\bibitem[Sneden et al.(2009)]{sneden09}
Sneden, C., Lawler, J.~E., Cowan, J.J., Ivans, I.~I.
\& Den Hartog, E.~A., 2009, \apjs, 182, 80

\bibitem[Sneden, Cowan \& Gallino(2008)]{r_process}
Sneden, C., Cowan, J.~J. \& Gallino, R., 2008, \araa, 46, 241
      
\bibitem[Sobeck et al.(2011)]{sobeck_moog}
Sobeck, J.~S. et al, 2011, \aj, 141, 175

\bibitem[Spite et al.(2006)]{spite_fs9}
Spite, M. et al, 2006, \aap, 455, 291

\bibitem[Spite et al.(2012)]{nonlte_ca}
Spite~M. et al, 2012, \aap, 541, A143

\bibitem[Stephens \& Boesgaard(2002)]{stephens02}
Stephens, A. \& Boesgaard, A.~M., 2002, \aj, 123, 1647

\bibitem[Takeda et al.(2002)]{takeda_knonlte}
Takeda, Y., Zhao, G., Chen, Y.~Q., Qui, H.~M.
\& Takada-Hidai, M., 2002, PASJ, 54, 275 

\bibitem[Thevenin \& Idiart(1999)]{thevenin99}
Thevenin, F. \& Idiart, T.~P., 1999, \apj, 521, 753


\bibitem[Tissera, White \&  Scannapieco(2012)]{tissera12}
Tissera, P.~B., White, S.~D.~M. \& Scannapieco, C., 2012,
   \mnras, 420, 255

\bibitem[Tolstoy, Hill \& Tosi(2009)]{tolstoy09}
Tolstoy, E., Hill, V. \& Tosi, M., 2009, \araa, 47, 371

\bibitem[Tominaga, Umeda \& Nomoto(2007)]{tominaga07}
Tominaga, N., Umeda, H. \& Nomoto, K., 2007, \apj, 660, 516

\bibitem[Travaglio et al(2004a)]{travaglio04a}
Travaglio, C.~D., Gallino, R., Arnone, E., Cowan, J., Jordan, F. \&
Sneden, C., 2004, \apj, 601, 864

\bibitem[Travaglio et al(2004b)]{travaglio04b}
Travaglio, C., Hillebrandt, W., Reinicke, M. \&
 Thielemann, F.~K., 2004, \aap, 425, 1029

\bibitem[Tsangarides et al(2004)]{tsangarides04}
Tsangarides, S., Ryan, S.~G., \& Beers, T.~C., 2004, 
Mem Soc Astron Italy, 75, 772

\bibitem[VandenBerg et al(2012)]{vandenberg12}
VandenBerg, D.~A., Bergbusch, P.~A., Dotter, A., Ferguson, J.,
Michaud, G., Richer, J. \& Profitt, C.~R., 2012 \apj, 755, A15

\bibitem[Ventura \& d'Antona(2009)]{ventura09}
Ventura, P. \& d'Antona, F., 2009, \aap, 499, 835

      
\bibitem[Vogt et al.(1994)]{vogt94} Vogt, S.~E. et al.\, 1994, SPIE, 2198, 362

\bibitem[Wallerstein \& Knapp(1998)]{araa_chstars}
Wallerstein G. \& Knapp, G.~R., 1998, \araa, 36, 369

\bibitem[Wisotzki et al(2000)]{wis00} 
Wisotzki, L., Christlieb, N., 
Bade, N.,Beckmann, V., K\"ohler, T., Vanelle, C. \& Reimers, D., 2000, 
\aap, 358, 77

\bibitem[Woosley \& Weaver(1995)]{woosley95}
Woosley, S.~E. \& Weaver, T.~A., 1995, \apjs, 101, 181

\bibitem[Yi et al.(2003)]{yi03}
Yi, S.,  Kim, Y.-C., Demarque, P. \& Alexander, D.~R., 2003, \apjs, 143, 499

\bibitem[Yong et al(2013)]{yong13}
Yong, D. et al, 2013, \apj, 762, A26

\bibitem[York et al.(2000)]{york_sdss}
York, D.~G. et al, 2000, \aj, 120, 1579

\bibitem[Zhang et al(2010)]{zhang11}
Zhang, L., Karlsson, T., Christlieb, N., Korn, A.~J., Barklem, P.~S.
\& Zhao, G., 2011, \aap, 528, A92

\bibitem[Zolotov et al(2011)]{zolotov10}
Zolotov, A., Willman, B., Brooks, A., Governato, F., Hogg, D.~W.,
   Shen, S. \& Wadsley, J., 2010, \apj, 721, 783


\end{thebibliography}
\end{document}